\documentclass{aa}
\usepackage{graphicx}
\usepackage{txfonts}
\usepackage{hyperref}
\usepackage{multirow}
\usepackage{xcolor}
\usepackage{soul} 
\hypersetup{
     colorlinks   = true,
     citecolor    = blue
}
\usepackage{float}
\DeclareUnicodeCharacter{2212}{\textendash}
\begin{document}

    \def \aCO{$\alpha_{\rm CO}\,$}
    \def \aCII{$\alpha_{\rm [CII]}\,$}
    \def \XCI{$\rm X_{CI}\,$}
    \def \gdr{$\delta_{\rm GDR}\,$}

\title{Observations of neutral carbon in 29 high-$z$ lensed dusty star forming galaxies and the comparison of gas mass tracers}

    \authorrunning{Gururajan et al.}
	\titlerunning{[CI] observations of high-$z$ DSFGs and cross-calibration of molecular gas mass tracers.}

\author{G. Gururajan
          \inst{1,2,3},
          M. Béthermin
          \inst{1,4},
          N. Sulzenauer
          \inst{5},
          P. Theulé 
          \inst{1},
          J. S. Spilker
          \inst{6},
          M. Aravena,
          \inst{7}
          S. C. Chapman
          \inst{8,9,10},
          A. Gonzalez
          \inst{11},
          T. R. Greve
          \inst{12,13},
          D. Narayanan
          \inst{11,14,12},
          C. Reuter
          \inst{15},
          J. D. Vieira
          \inst{15,16,17},
          \and
          A. Weiss
          \inst{5}
          }

   \institute{
            Aix Marseille Univ, CNRS, CNES, LAM, Marseille, France
        \and
            University of Bologna - Department of Physics and        Astronomy “Augusto Righi” (DIFA), Via Gobetti 93/2,        I-40129, Bologna, Italy, \email{gayathri.gururajan@unibo.it}
        \and
            INAF - Osservatorio di Astrofisica e Scienza dello Spazio, Via Gobetti 93/3, I-40129, Bologna, Italy
        \and
            Universit\'e de Strasbourg, CNRS, Observatoire astronomique de Strasbourg, UMR 7550, 67000 Strasbourg, France
        \and
            Max-Planck-Institut für Radioastronomie, Auf dem Hügel 69, D-53121 Bonn, Germany
        \and
             Department of Physics and Astronomy and George P. and Cynthia Woods Mitchell Institute for Fundamental Physics and Astronomy, Texas A$\&$M University, 4242 TAMU, College Station, TX 77843-4242, US
        \and 
            Instituto de Estudios Astrof\'{\i}sicos, Facultad de Ingeniería y Ciencias, Universidad Diego Portales, Av. Ejército 441, Santiago, Chile
        \and
            Department of Physics and Astronomy, University of British Columbia, 6225 Agricultural Road, Vancouver, V6T 1Z1, Canada
        \and
            National Research Council, Herzberg Astronomy and Astrophysics, 5071 West Saanich Road, Victoria, V9E 2E7, Canada
        \and
            Department of Physics and Atmospheric Science, Dalhousie University, 6310 Coburg Road, B3H 4R2, Halifax, Canada
        \and
            Department of Astronomy, University of Florida, 211 Bryant Space Sciences Center, Gainesville, FL 32611 USA
        \and
            Cosmic Dawn Center (DAWN), DTU-Space, Technical University of Denmark, Elektrovej 327, DK-2800 Kgs. Lyngby, Denmark
        \and
            Department of Physics and Astronomy, University College London, Gower Place, WC1E 6BT London, UK
        \and
            University of Florida Informatics Institute, 432 Newell Drive, CISE Bldg E251, Gainesville, FL 32611, USA
        \and
            Department of Astronomy, University of Illinois at Urbana-Champaign, 1002 West Green St., Urbana, IL 61801, USA
        \and
            Department of Physics, University of Illinois Urbana-Champaign, 1110 West Green Street, Urbana, IL, 61801, USA
        \and
            Center for AstroPhysical Surveys, National Center for Supercomputing Applications, Urbana, IL, 61801, USA
         }
         
\abstract {The nature and evolution of high-redshift dusty star-forming galaxies (high-z DSFGs) remain an open question. Their massive gas reservoirs play an important role in driving the intense star-formation rates hosted in these galaxies.}
{We aim to estimate the molecular gas content of high-$z$ DSFGs by using various gas mass tracers such as the [CI], CO, [CII] emission lines and the dust content. These tracers need to be well calibrated as they are all limited by uncertainties on factors such as \aCO, \XCI, \aCII and \gdr, thereby affecting the determination of the gas mass accurately. The main goal of our work is to check the consistency between the gas mass tracers and cross-calibrate the uncertain factors. }
{We observe the two [CI] line transitions for 29 South Pole telescope - submillimeter galaxies (SPT-SMGs) with the Atacama large millimeter/sub-millimeter array - Atacama compact array (ALMA-ACA). Additionally, we also present new Atacama pathfinder experiment (APEX) observations of [CII] line for 9 of these galaxies. We combine our observations with the rich ancillary data of low and mid-J CO lines, ancillary [CII] line, and dust mass estimations for these galaxies.}
{We find a nearly linear relation between the infrared luminosity and [CI] luminosity if we fit the starbursts and main-sequence galaxies separately. We measure a median [CI]-derived excitation temperature of 34.5$\pm$2.1\,K. We probe the properties of the interstellar medium (ISM) such as density and radiation field intensity using [CI] to mid- or high-J CO lines and [CI] to infrared luminosity ratio, and find similar values to the SMG populations in literature. Finally, the gas masses estimated from [CI], CO, dust, and [CII] do not exhibit any significant trend with the infrared luminosity or the dust temperature. We provide the various cross-calibrations between these tracers.}{Our study confirms that [CI] is a suitable tracer of the molecular gas content, and shows an overall agreement between all the classical gas tracers used at high redshift. However, their absolute calibration and thus the gas depletion timescale measurements remain uncertain.}

\keywords{Galaxies:high-redshift -- Galaxies:evolution -- Galaxies:ISM --  Galaxies:star formation -- Submillimeter:galaxies }

\maketitle
\section{Introduction}


The formation and evolution of galaxies across cosmic time plays an important role in understanding the evolution of the Universe. The cosmic star-formation rate density (SFRD) when traced as a function of the redshift ($z$), shows a peak in star-formation at $1<z<3$ \citep[e.g.][]{Madau96, Lefloch05, Hopkins06, Madau14}, illustrating that more than half the stars we see today were already formed at $z\sim 1$ \citep{Walter11}. At these redshifts, the most massive galaxies tend to be heavily dust obscured \citep[e.g.][]{Heinis14, Fudamoto20}. These objects can host prodigious star-formation rates (SFR, $>$500 M$_{\odot}$\,yr$^{-1}$), and have a large contribution to the high-$z$ SFRD \citep[e.g.,][]{Madau14,Casey14}. Despite playing a crucial role in our understanding of early galaxy evolution, the nature and star-formation mechanisms of such high-$z$ dusty star-forming galaxies (DSFGs) remain unclear.

An important factor dictating the star-formation rates and mechanisms in galaxies is its molecular gas content. The evolution of the molecular gas density with redshift as shown by e.g. \citet{Decarli20} has a similar trend to that of the cosmic SFRD. Additionally, a rapid increase in the gas fraction (f$_{\rm gas} \equiv \rm M_{\rm gas}/(M_{\rm gas} + M_{*})$) in galaxies with increasing redshifts (from $\sim$5\,\% at z$\sim$0 to $\sim$50\,\% at z$\gtrsim$3) is also shown by various surveys \citep[e.g.][]{Tacconi10,Daddi10a,Saintonge13,Dessauges-Zavadsky15,Bethermin15,Dessauges-Zavadsky20}. This could be driven by rapid accretion of cold gas from the cosmic web \citep[e.g.][]{ Dekel09,Kleiner17,Kretschmer20,Chun20}. The molecular gas content in a galaxy is mainly dominated by the H$_2$ molecule. As the H$_2$ molecule is not polar, it is mostly observable through vibrational transitions which are currently inaccessible at high-$z$. Thus we rely on other line emissions that can trace the H$_2$ content of a galaxy.

The carbon monoxide ($^{12}$CO, CO henceforth) molecule is one of the commonly used tracer of molecular H$_2$ content in a galaxy. The J $=$ 1-0 transition of CO is assumed to be proportional to hydrogen column densities (n$_{\text{H}_2}$, \citealt{Solomon05,Bolatto13}). At high-$z$, particularly $z>3$, the CO(1-0) line becomes harder to detect due to its low excitation temperature and the increasing influence of cosmic microwave background (CMB) emission at $z>4$. We thus rely on mid- or high-J CO lines. Estimating the molecular gas mass from these lines require assumptions of CO line excitation \citep{Narayanan14,Tunnard16} and the CO-spectral line energy distribution for the various transitions have been modelled and constrained \citep[e.g][]{Weiss07, Harrington21,Jarugula21}. 

To estimate the molecular gas mass of a galaxy from the CO(1-0) luminosity, one needs to assume a CO-to-H$_2$ conversion factor - \aCO. The value of \aCO is known to increase with decreasing metallicities as the size of the CO emission reduces \citep[e.g.][]{Maloney97,Wolfire10,Krumholz11, Shetty11b, Feldmann12, Lagos12, Narayanan12}. The value of \aCO was also shown to vary within the Milky-Way \citep[e.g.][]{Oka98, Strong04, Sandstrom13} and between normal star-forming galaxies (or disk-dominated systems) and starbursts (ultra luminous infrared galaxies, ULIGRs, e.g. \citealt{DownesSolomon98,Daddi10b,Genzel10}). Furthermore, \aCO values can be affected by physical processes like mergers and radiation field intensity of the interstellar medium (ISM, e.g. \citealt{Leroy11, Magdis11}). Additionally, the destruction of CO molecules by cosmic rays in intensely star-forming environments can reduce their abundances, thereby underestimating the gas mass \citep{Bisbas15, Bisbas17}. Under typical ISM conditions, observations \citep[e.g.][]{Allen12,Langer14,Pineda17} and theoretical works \citep[e.g.][]{Wolfire10,Smith14,GloverSmith16} suggest that the CO may miss nearly 30 -- 70$\%$\, of the total H$_2$ mass as the H$_2$ and CO may not be co-spatial in the ISM. 

The molecular gas mass of a galaxy can also be estimated from its dust mass assuming a gas-to-dust ratio, \gdr \citep[e.g.][]{Magdis11, Leroy11,Eales12,Scoville14, Scoville16}. The properties as dust mass and dust temperature can be estimated by modelling far-infrared (FIR) and sub-millimetre emission of the spectral energy distribution (SED) of a galaxy. \gdr is also sensitive to metallicity, decreasing with increasing metallicity \citep[e.g.][]{Magdis11,Leroy11,Remyruyer14, Capak15, Popping22a,Popping22b}. Therefore, assuming a single value of \gdr for all the galaxies across a large range of redshifts might not be accurate. Additionally, \gdr is also dependent on dust grain properties and the models describing the creation or destruction of dust \citep{Bolatto13}. 

In the recent years, the ionised carbon fine structure emission line [CII] at 158\,$\mu$m has been advocated as an estimator of molecular gas mass \citep[e.g.][]{Hughes17,Zanella18,Dessauges-Zavadsky20,Vizgan22}. [CII] emission arises from multiple phases of the ISM, the photodissociation regions (PDRs), atomic, molecular and diffuse ionised regions \citep[e.g.][]{Stacey91,Kaufman99,Sargsyan12,Rigopoulou14,Croxall17}. This makes it difficult to disentangle the contributions of these various regions to the [CII] emission. However, $\sim 75\%$ of [CII] emission is suspected to arise from the PDRs \citep{Pineda14,Cormier15, Diaz-Santos17}, majority of which arises in the molecular phase \citep[e.g.][]{Pineda13, Velusamy14}. This has also been found in simulations of high-$z$ galaxies \citep[e.g.][]{Vallini15,Popping19}. \citet{Zanella18} describes an empirical relation between [CII] luminosity and molecular gas mass for a sample of main-sequence galaxies and ULIRGs, with a [CII]-to-H$_2$ conversion factor, \aCII. They found that, \aCII does not seem to vary with metallicity or the mode of star formation. 

An alternative, promising tracer of the molecular gas content is the atomic carbon fine structure lines. The two transitions C$_{\rm I}$($^3\rm P_2 - ^3\rm P_1$) and C$_{\rm I}$($^3\rm P_1 - ^3\rm P_0$), henceforth the [CI](2-1) and [CI](1-0) lines, respectively, can also be used to estimate the gas content of a galaxy \citep[e.g.][]{Keene85,Papadopoulos04a, Papadopoulos04b}. The [CI] emission in the ISM of a galaxy was thought to arise only in a small region of the PDR, between the CO and [CII] \citep{Langer76,Tielens85}. Theoretical works \citep{Tomassetti14} and the detection of [CI] in the Galactic molecular clouds along with CO, have suggested that [CI] emission is more extended in the ISM \citep{Keene85,Ojha01}, thereby making it a candidate to trace the molecular gas content. Studies have also shown [CI] to arise from the same volume as CO(1-0) \citep[e.g.][]{Plume99, Ikeda02, Schneider03, Perez15}. The accuracy of the [CI](1-0) line as tracer of molecular gas content has thus been advocated by many studies \citep[e.g.][]{Papadopoulos04, Weiss03, Dunne21}. Additionally, it is optically thin for the bulk of the H$_2$ gas \citep{Perez15} facilitating its detection in intense starbursts, and the excitation temperature of [CI] is sufficiently low ($\sim 24$ K, \citealt{Dunne22}) thereby tracing cold gas effectively. 

In order to estimate the molecular gas mass from the [CI](1-0) observations, one needs to assume a [CI] abundance - \XCI. It can also be affected by factors such as metallicity and densities \citep{Bisbas21, Heintz20}, but it is less sensitive to metallicity than \aCO \citep[e.g.][]{Leroy11, Genzel12,Schruba12, Shi15, Shi16, Heintz20, Bisbas21}. Similar to CO, the [CI] excitation must also be known to accurately derive the gas mass. The [CI] excitation factor/partition function - Q$_{10}$ can be constrained if the excitation temperature of [CI] in the ISM is known. 

Owing to its relatively simple two-level excitation, and assuming local thermodynamical equilibrium (LTE) conditions, the excitation temperature of [CI] in the ISM can be estimated as a function of the ratio of the two [CI] line luminosites \citep[e.g.][]{Stutzki97,Weiss03,Papadopoulos04,Walter11}. The excitation temperature of the ISM can also be compared with the dust temperature as they are expected to be linear under LTE, assuming an effective gas to dust coupling \citep[e.g.][]{CarilliWalter13,daCunha13,Valentino20}. Other parameters such as the radiation field intensity and the density of the ISM can also be constrained using various line ratios. Mid/high-J CO lines are expected to arise from warm and dense ISM, when compared to an extended gas tracer such as [CI], the line ratio could be a proxy to the density of the ISM \citep[e.g.][]{Alaghband-Zadeh13,Yang17,Andreani18,Valentino18,Valentino20}.

In this paper, we explore the properties of the [CI] line and its ability as a molecular gas tracer using a sample of 29 lensed DSFGs from the SPT-SMG sample. Gravitational lensing can boost the apparent flux, thereby making these DSFGs bright enough to detect, and allowing us to build a statistical sample of [CI] detection at high-$z$. With the data in hand, we study the relation between the [CI] line luminosity and the infrared luminosity. We compare the [CI] excitation temperature with the dust temperature. We use the combination of [CI] and other observables to constrain ISM properties of our sample. Finally, we compare [CI] to other molecular gas tracers such as CO, [CII], and dust to cross-calibrate the unknown factors such as \XCI, \aCO, \aCII and \gdr in these tracers.

The structure of this paper is as follows. The sample selection, the ALMA/ACA observations and the APEX-[CII] observations are presented in Sect.\,\ref{sec2}. In Sect.\,\ref{sec3}, we describe the data processing, imaging and the flux extraction for our sample. We analyse the properties of the [CI] line in Sect.\,\ref{CI line properties}, where we probe the relation between the infrared and [CI] luminosities (Sect.\,\ref{CI IR luminosities}), the [CI] excitation temperature (Sect.\,\ref{exc temp}), and the properties of the ISM as traced by the line and line-to-continuum ratios (Sect.\,\ref{ism diagnostics}). We then compare the gas mass estimated with [CI] line with other tracers such as CO, dust continuum, and [CII] emission line and cross-calibrate these tracers in Sect.\,\ref{gas mass estimation}. In Sect.\,\ref{discussion cross calib}, we discuss the results of our cross-calibration of the various gas mass tracers and probe the gas depletion timescales for our sample as a function of their redshifts (Sect.\,\ref{depletion timescale}). Finally, we conclude in Sect.\,\ref{conclusions}.  

Throughout this paper, we adopt a $\Lambda$CDM cosmology with $\Omega_m = 0.3$, $\Omega_{\Lambda} = 0.7$ and $H_0 = 70$ km\,s$^{-1}$\,Mpc$^{-1}$ and a \citet{2003PASP..115..763C} initial mass function (IMF). 

\section{Sample and ALMA observations}\label{sec2}


\subsection{Sample selection}\label{sec2.1}

We build a statistical sample of 30 galaxies targeting both the [CI] lines from the SPT-SMG sample \citep{Vieira13, Weiss13, Strandet16, Reuter20}. These galaxies are in the ideal redshift range ($1.87<z<4.8$) to allow the observations of both the [CI] transitions with ALMA (with the bands available in the cycle and in frequencies unaffected by atmospheric absorption). For this sample, 21 ancillary [CI](1-0) observations exist from the observations presented in \citet{Bothwell17} and \citet{Reuter20}. In this work, we present 39 observations of [CI] lines (30 observations of [CI](2-1) line and 9 observations of [CI](1-0) lines) with ALMA-ACA (PI: Bethermin, 2019.1.00297.S) to build a complete sample with the ancillary data for these 30 galaxies. We also present the APEX-[CII] observations for 9 of these galaxies (M-090.F-0016-2012, M-093.F-0012-2014, M-095.F-0028-2015, M-097.F-0019-2016). Additionally, 9 more galaxies have ancillary [CII] observations from \citet{Gullberg15}. The ACA observations are described in Sect.\,\ref{sec:ACA observations} and the APEX-[CII] observations are described in Sect.\,\ref{APEX CII observations}. For this sample, ancillary low-J CO observations from \citet{Aravena16} and mid-J CO observations from \citet{Reuter20} are also available. Unfortunately, most of these sources lack an estimate of the stellar masses as the near-infrared bright foreground lens usually outshines the DSFG. However, for 6 SPT-SMGs, the stellar masses have been estimated based on IRAC photometry, but remain uncertain \citep{Ma2015}. These sources were found above the main sequence of star-forming galaxies, but because of the uncertainties it is not fully clear if they are starbursts or just on the upper envelop of the relation.

For the [CI](1-0) transition, the observed frequency was required to be higher than 84\,GHz ($z<4.8$) to be within the observation limit of band 3 and not in the range 116--125\,GHz ($2.9<z<3.2$), as it corresponds to the gap between band 3 and band 4. Additionally, we exclude the frequency range 175--190\,GHz ($3.2<z<3.6$) for [CI](2-1) as it can fall in a strong atmospheric water absorption feature in band 5, and thus would require a very long integration with excellent weather for a detection, which could lead to bad quality data. Thus we present 21 [CI](2-1) observations in $3.6<z<4.8$ for which there is existing [CI](1-0) observations from the ancillary data and 10 galaxies at $z<2.9$ for which both the [CI] lines can be observed.

Together with the [CI](2-1) line, we also observe simultaneously the CO(7-6), since they are close in frequency. For three of our sources, we also have CO(4-3) line imaged on the lower sideband of [CI](1-0) observational setup. All these galaxies are strongly gravitationally lensed and 18 of them have detailed lens modelling presented in \citet{Spilker16}. For the sources without lens modelling, we assume a median magnification of 5.5 \citep{Reuter20}. This median value of magnification is adopted based on the available lens models for a sample of 39 SPT-SMGs presented in \citet{Spilker16}. As the sources presented in \citet{Spilker16} for lens modelling were drawn at random from a larger sample of SPT-SMGs, the median value of magnification adopted is a reasonable assumption. Our sample is presented in Table\,\ref{tab:1_observation_details_ACA}.

The initial ACA proposal consisted of 39 observations and one of our source - SPT0452-52 was not detected in both [CI](2-1) and [CI](1-0). This source had not been detected clearly in the earlier band-3 observations presented in \citet{Reuter20}, hence we suspect an ambiguous redshift to be the cause of the non-detection. We thus chose to discard this source from our analysis. Hence, we proceed with a final sample of 29\,galaxies with both [CI] transitions, 18\, of the 29\,galaxies also have [CII] line observations using APEX (see Sect.\,\ref{APEX CII observations}) and 5\, galaxies with [CI](2-1) from the APEX observations.


\begin{table*}[h]
\centering
\caption{\label{tab:1_observation_details_ACA} ALMA/ACA observation details of our sample.}

\begin{tabular}{cccccccc}
\hline
\hline
&&&&&&&\\
Source & Target  & Observed & Observation  & Time on source & PWV & $\sigma_{\rm channel}$ &  Resolution  \\
 & line & frequency (GHz) & date(s) & (min) & (mm) & (mJy) &  $^{\prime\prime} \times ^{\prime\prime}$ \\
&&&&&&&\\
\hline
&&&&&&&\\
SPT0002-52 & [CI](1-0) & 146.87 & 29/10/2019 & 17 & 0.5 & 2.4 & 12.6 $\times$ 8.8 \\
 & [CI](2-1) & 241.52 & 08/10/2019 & 8 & 0.9 & 4.4 & 7.3 $\times$ 4.9 \\
SPT0020-51 & [CI](2-1) & 157.98 & 13/10/2019 & 18 & 1.8 & 2.8 & 11.1 $\times$ 7.6 \\
SPT0113-46 & [CI](2-1) & 154.66 & 13/10/2019 & 49 & 1.6 & 1.7 & 11.3 $\times$ 7.4 \\
SPT0125-50 & [CI](2-1) & 163.27 & 09/11/2019 & 24 & 1.4 & 2.4 & 11.5 $\times$ 6.9 \\
SPT0136-63 & [CI](2-1) & 152.73 & 15/11/2019 & 29 & 2.0 & 2.0 & 12.5 $\times$ 7.7 \\
SPT0150-59$^*$ & [CI](1-0) & 129.93 & 03/11/2019 & 28 & 4.8 & 2.4 & 12.8 $\times$ 10.2 \\
 & [CI](2-1) & 213.66 & 12/10/2019 & 20 & 2.1 & 3.7 & 9.1 $\times$ 5.6 \\
SPT0155-62 & [CI](2-1) & 151.31 & 01/11/2019 & 19 & 2.0 & 3.4 & 11.1  $\times$8.8 \\
SPT0319-47 & [CI](2-1) & 146.89 & 31/10/2019 & 17 & 3.4 & 3.6 & 10.4 $\times$ 8.9 \\
SPT0345-47 & [CI](2-1) & 152.82 & 02/11/2019 & 17 & 0.01 & 3.3 & 10.3 $\times$ 8.5 \\
SPT0418-47 & [CI](2-1) & 154.9 & 31/10/2019 & 17 & 3.5 & 3.7 & 10.1 $\times$ 8.6 \\
SPT0441-46 & [CI](2-1) & 147.77 & 31/10/2019 & 17 & 3.4 & 3.1 & 10.6 $\times$ 8.7 \\
SPT0459-59 & [CI](2-1) & 139.57 & 02/11/2019 & 24 & 3.2 & 2.5 & 12.0 $\times$ 10.1 \\
SPT0512-59 & [CI](1-0) & 152.23 & 31/10/2019 & 18 & 3.5 & 3.1 & 10.2 $\times$ 9.8 \\
 & [CI](2-1) & 250.34 & 13/11/2019 & 21 & 1.6 & 6.0 & 6.5 $\times$ 5.6 \\
SPT0544-40 & [CI](2-1) & 153.6 & 13/10/2019 & 17 & 0.01 & 3.6 & 12.6 $\times$ 7.6 \\
SPT0551-48$^*$ & [CI](1-0) & 137.36 & 01/10/2019 & 17 & 3.0 & 2.4 & 13.4 $\times$ 8.2 \\
 & [CI](2-1) & 225.88 & 03/10/2019 & 18 & 1.4 & 3.4 & 9.0 $\times$ 4.9 \\
SPT0552-42 & [CI](2-1) & 148.86 & 01/10/2019 & 38 & 2.6 & 2.0 & 11.9 $\times$ 8.0 \\
SPT0604-64 & [CI](1-0) & 141.38 & 01/10/2019 & 19 & 2.9 & 2.9 & 12.8 $\times$ 9.8 \\
 & [CI](2-1) & 232.5 & 10/11/2019 & 21 & 1.6 & 4.2 & 7.3 $\times$ 6.0 \\
SPT2037-65 & [CI](2-1) & 161.87 & 01/10/2019 & 17 & 1.6 & 3.0 & 9.5$\times$ 8.6 \\
SPT2048-55$^{a,*}$ & [CI](2-1) & 159.01 & 13/10/2019 & 29 & 3.0 & 1.6 & 11.6 $\times$ 7.6 \\
 & [CI](2-1) &  & 13/10/2019 & 29 & 2.8 &  &   \\
 & [CI](2-1) &  & 17/11/2019 & 29 & 3.4 &  &   \\
SPT2103-60 & [CI](2-1) & 148.89 & 02/10/2019 & 18 & 2.5 & 3.4 & 12.2 $\times$ 9.6 \\
SPT2132-58$^*$ & [CI](2-1) & 140.32 & 12/10/2019 & 21 & 2.1 & 1.8 & 12.9 $\times$ 9.0 \\
SPT2134-50$^*$ & [CI](1-0) & 130.2 & 02/11/2019 & 26 & 5.4 & 3.1 & 12.9 $\times$ 10.2 \\
 & [CI](2-1) & 214.11 & 04/10/2019 & 18 & 1.6 & 3.4 & 7.7 $\times$ 5.6 \\
SPT2146-55 & [CI](2-1) & 145.38 & 13/10/2019 & 21 & 2.0 & 1.9 & 11.8 $\times$ 8.7 \\
SPT2147-50 & [CI](2-1) & 170.03 & 01/10/2019 & 28 & 1.4 & 2.3 & 9.6 $\times$ 7.0 \\
SPT2311-54 & [CI](2-1) & 153.28 & 30/10/2019 & 19 & 0.8 & 2.8 & 10.3 $\times$ 8.9 \\
SPT2335-53 & [CI](2-1) & 140.58 & 13/10/2019 & 26 & 1.8 & 1.6 & 12.1 $\times$ 8.2 \\
SPT2349-50 & [CI](1-0) & 126.94 & 31/10/2019 & 29 & 3.9 & 2.6 & 13.2 $\times$ 9.5 \\
 & [CI](2-1) & 208.75 & 29/10/2019 & 20 & 0.7 & 3.8 & 9.5 $\times$ 5.9 \\
SPT2349-56$^{*,\dagger}$ & [CI](1-0) & 152.59 & 30/10/2019 & 41 & 3.7 & 1.7 & 11.7 $\times$ 8.2 \\
SPT2354-58 & [CI](1-0) & 171.66 & 23/10/2019 & 27 & 1.1 & 1.9 & 9.6 $\times$ 7.4 \\
 & [CI](2-1) & 282.3 & 11/10/2019 & 35 & 1.1 & 3.1 & 5.8 $\times$ 4.1 \\
\hline

\end{tabular}
\tablefoot{Some of our sources were observed for both [CI] transitions and thus have multiple observation dates. The precipitable water vapor (PWV) during the observation is given in mm. The channel sensitivity is the mean computed over every channel for a corresponding channel width of 0.031 GHz. The line resolution of the source is given in terms of the major axis $\times$ minor axis.\\
$\dagger$ : Protocluster candidate \\
$*$ : The sources for which the sensitivity is computed in a channel width of 0.063\,GHz. \\
$^{a}$: The source has been observed on multiple dates and the different datasets have been concatenated to proceed with the analysis. The channel sensitivity, width and resolution are for the concatenated dataset. 
}
\end{table*}

\begin{table*}[!htb]
\centering
\caption{\label{tab:2_flux_catalogue_ACA} [CI](1-0), [CI](2-1), CO(7-6), and [CII] line fluxes estimated for our sample. }

\begin{tabular}{cccccccc}
\hline
\hline
&&&&&&&\\
Source & $z$ & R.A. & Dec & [CI](1-0) & [CI](2-1) & CO(7-6) & [CII] \\
&&(J2000)&(J2000)&(Jy km s$^{-1}$)&(Jy km s$^{-1}$)&(Jy km s$^{-1}$)& (Jy km s$^{-1}$)\\
&&&&&&&\\
\hline
&&&&&&&\\
SPT0002-52 & 2.351 & 0:02:23.76 & -52:31:52.7 & 3.85$\pm$0.45 & 8.71$\pm$0.98 & 14.87$\pm$0.84& \\
SPT0020-51 & 4.123 & 0:20:23.58 & -51:46:36.4 & 1.98$\pm$0.47$^{b}$ & 6.32$\pm$0.89 & 8.12$\pm$1.08& \\
SPT0113-46 & 4.233 & 1:13:09.01 & -46:17:56.1 & 3.36$\pm$0.68$^{a}$ & 4.26$\pm$0.67 & 4.48$\pm$0.73& 82.1$\pm$13.3$^{c}$ \\
SPT0125-50 & 3.957 & 1:25:48.45 & -50:38:21.0 & 2.37$\pm$0.53$^{a}$ & 4.85$\pm$0.46 & 7.29$\pm$0.6 & <326.5\\
SPT0136-63 & 4.299 & 1:36:50.28 & -63:07:26.7 & 1.14$\pm$0.62$^{b}$ & $<$4.98 & 4.62$\pm$0.92 &40.8$\pm$6.3\\
SPT0150-59 & 2.788 & 1:50:09.26 & -59:23:57.1 & 3.29$\pm$0.84 & 6.76$\pm$0.81 & 12.03$\pm$1.25 &\\
SPT0155-62 & 4.349 & 1:55:47.75 & -62:50:50.0 & 5.06$\pm$0.58$^{b}$ & 7.08$\pm$0.85 & 11.61$\pm$1.12& \\
SPT0319-47 & 4.51 & 3:19:31.88 & -47:24:33.6 & <8.4$^{b}$ & 3.24$\pm$0.63 & 4.00$\pm$0.88 &32.1$\pm$7.6 \\
SPT0345-47 & 4.296 & 3:45:10.77 & -47:25:39.5 & $<$1.03$^{a}$ & $<$4.96 & 12.16$\pm$1.05 & 44.6$\pm$9.0$^{c}$\\
SPT0418-47 & 4.225 & 4:18:39.67 & -47:51:52.5 & 2.46$\pm$0.61$^{a}$ & 3.57$\pm$0.68 & 7.29$\pm$0.87& 116.0 $\pm$ 12.1$^{c}$ \\
SPT0441-46 & 4.477 & 4:41:44.08 & -46:05:25.6 & $<$2.22$^{a}$ & 4.01$\pm$1.04 & 6.57$\pm$0.88& 37.4$\pm$6.6$^{c}$\\
SPT0459-59 & 4.799 & 4:59:12.34 & -59:42:20.3 & 2.43$\pm$0.7$^{a}$ & 3.77$\pm$0.89 & 5.74$\pm$0.93& <57.5 \\
SPT0512-59 & 2.233 & 5:12:57.98 & -59:35:42.0 & 9.55$\pm$0.9 & 15.73$\pm$1.31 & 14.26$\pm$1.15 &\\
SPT0544-40 & 4.269 & 5:44:01.12 & -40:36:31.2 & $<$7.78$^{b}$ & 4.47$\pm$0.99 & 5.16$\pm$0.93 & 41.8$\pm$8.6\\
SPT0551-48 & 2.583& 5:51:54.65& -48:25:01.8&7.03$\pm$1.09& 12.08$\pm$0.62 & 34.09$\pm$0.65&\\
SPT0552-42 & 4.437 & 5:52:26.52 & -42:44:12.7 & $<$4.68$^{b}$ & 3.31$\pm$0.61 & 2.53$\pm$0.57&39.6$\pm$5.4 \\
SPT0604-64 & 2.481 & 6:04:57.57 & -64:47:22.0 & 9.80$\pm$0.5 & 13.03$\pm$0.84 & 18.11$\pm$0.86& \\
SPT2037-65 & 3.998 & 20:37:31.98 & -65:13:16.8 & $<$4.44$^{b}$ & 5.23 $\pm$0.87 & 7.53 $\pm$ 0.62 &\\
SPT2048-55 & 4.09 & 20:48:22.87 & -55:20:41.3 & 1.39$\pm$0.44$^{b}$ & 2.49$\pm$0.66 & 1.68$\pm$0.47 & <68.5 \\
SPT2103-60 & 4.436 & 21:03:30.90 & -60:32:39.9 & 3.07$\pm$0.76$^{a}$ & 4.59$\pm$1.23 & 5.57$\pm$0.89&52.0$\pm$17.2$^{c}$ \\
SPT2132-58 & 4.768 & 21:32:43.23 & -58:02:46.4 & $<$0.87 $^{a}$& 2.56$\pm$0.76 & 5.27$\pm$0.86 & 43.4$\pm$13.4$^{c}$\\
SPT2134-50 & 2.78 & 21:34:03.34 & -50:13:25.2 & 5.70$\pm$1.15 & 8.24$\pm$1.03 & 20.60$\pm$1.19 &\\
SPT2146-55 & 4.567 & 21:46:54.02 & -55:07:54.7 & 2.73$\pm$0.71$^{a}$ & 3.52$\pm$0.79 & 4.86$\pm$1.04 &33.4 $\pm$7.1 $^{c}$\\
SPT2147-50 & 3.76 & 21:47:19.05 & -50:35:53.5 & 2.01$\pm$0.6$^{a}$ & 4.90$\pm$0.7 & 6.37$\pm$0.94&101.9$\pm$14.3 $^{c}$\\
SPT2311-54 & 4.28 & 23:11:23.97 & -54:50:30.1 & $<$7.56$^{b}$ & 8.36$\pm$1.22 & 4.72$\pm$0.61&38.0$\pm$3.8 $^{c}$\\
SPT2335-53 & 4.757 & 23:35:13.96 & -53:24:21.0 & $<$2.34$^{b}$ & <3.08 & 1.75 $\pm$ 0.49 &12.4$\pm$3.3\\
SPT2349-50 & 2.877 & 23:49:42.20 & -50:53:30.9 & 3.58$\pm$0.68 & 4.14$\pm$0.63 & 3.74$\pm$0.68& \\
SPT2349-56 & 4.304 & 23:49:42.78 & -56:38:23.2 & $<$2.94$^{b}$ & <2.44 & <2.03&79.5$\pm$11.5 \\
SPT2354-58 & 1.867 & 23:54:34.31 & -58:15:08.3 & 5.06$\pm$0.56 & 11.62$\pm$0.75 & 23.58$\pm$0.82 &\\
\hline

\end{tabular}
\tablefoot{The redshift, $z$, the right ascension (R.A.) and declination (dec) are tabulated along with the [CI](1-0), [CI](2-1), CO(7-6), and [CII] fluxes. The fluxes presented in this table are not corrected for magnification. The errors given on the fluxes are 1\,$\sigma$ errorbars, and the upper limits are 3\,$\sigma$.\\   
$^a$: Ancillary [CI](1-0) fluxes from \citet{Bothwell17}\\
$^b$: Ancillary [CI](1-0) fluxes from \citet{Reuter20}\\
$^{c}:$ Ancillary [CII] fluxes from \citet{Gullberg15} \\}

\end{table*}

\subsection{ALMA/ACA observations} \label{sec:ACA observations}


Our target sources are observed with the Atacama compact array (ACA) of ALMA. We do not spatially resolve our sources and thus ACA is perfect as it is always in a very compact configuration contrary to the 12\,m configuration which changes. Owing to the low surface density of the SPT sources, we cannot share a calibrator since sources sharing a similar correlator setup are too distant from each other. Hence using 12\,m array would result in a shorter on-source observation time than the calibration. This would be an inefficient use of the telescope time and the observations would be mainly calibrations and overheads. We thus choose to observe in ACA configuration with median on-source observation time of 21.2\,minutes with a comparatively short calibration time. This would also require only ten to twelve 7\,m antennas and thus is a more optimal use of telescope time. 

To maximise the sensitivity of the integrated [CI] and CO fluxes, we require the entire source flux to be well-encompassed within a single synthesised beam. The median angular resolution reached by ACA for our sample is 8.85\,arcsec, while the sources are more compact than $\sim 2$\,arcsec. 

For all our targets, we use 4 spectral windows for every observation to secure the measurements of lines and the continuum. This also ensures maximal coverage of both the upper and the lower side bands. For the [CI](2-1) lines, we use one spectral window on frequency division mode (FDM) with a spectral resolution of 7.813\,MHz. This spectral window is centered on the [CI](2-1) observed frequency. As the observer-frame frequency separation between [CI](2-1) and CO(7-6) is $< 1$\,GHz for our sample, they can be imaged in the same spectral window. Additionally, we also place a second spectral window (with the same resolution) adjacent to it, in order to target any possible tails in the CO(7-6) emission. To measure the continuum emission, we use the other two spectral windows in the time division
mode (TDM) with a coarser resolution of 31.25\,MHz.

To observe the [CI](1-0) line, we place one spectral window (7.813\,MHz resolution), centered at its observed frequency. The other three spectral windows are used to measure the continuum emission. For four of our sources ($1.86<z<2.35$), we can also observe CO(4-3) by simultaneously shifting $\sim$ 1/4th of the width of the [CI](1-0) window to higher frequencies. This does not affect the [CI](1-0) measurements as the line width ($<$1000\,km\,s$^{-1}$) is lower than the velocity range of the spectral window ($\sim$3500\,km\,s$^{-1}$).

We aimed to achieve a sensitivity of 5\,$\sigma$ when integrated over the full line width ($\sim$ 400\,km\,s$^{-1}$) for both the lines for our sample. The target sensitivities for the [CI](1-0) line are estimated from their IR luminosities (L$_{\rm IR}$) assuming an L$_{\rm [CI](1-0)}$/L$_{\rm IR}$ ratio of $10^{-2.8}$. The L$_{\rm IR}$ measurements are well-constrained from the existing \textit{Herschel} and LABOCA observations \citep{Reuter20}. For the [CI](2-1) line, we assume a [CI](2-1)/[CI](1-0) flux ratio of 1.1 from the best-fit solution of the PDR modelling of \citet{Bothwell17}. 

The on-source integration time is significantly smaller than the calibration time, for a few bright sources, when the required sensitivity is $>$1.6\,mJy\,beam$^{-1}$ RMS in 400\,km\,s$^{-1}$. We thus fix the RMS goal to a maximum value of 1.6 \,mJy\,beam$^{-1}$, as a higher signal-to-noise could be a poor use of telescope time especially if the line is fainter than predicted and overhead dominated observations are not ideal. With the same sensitivity, the CO(7-6) line should also be detected as it is generally brighter than [CI](2-1). We present the observation details of our sample in Table\,\ref{tab:1_observation_details_ACA}. The sensitivity and resolution correspond to the achieved values and their estimation is described in Sect.\,\ref{sec:imaging and performance}. 

\begin{figure*}[h]
\centering

\begin{tabular}{ccc}
\includegraphics[width=6cm]{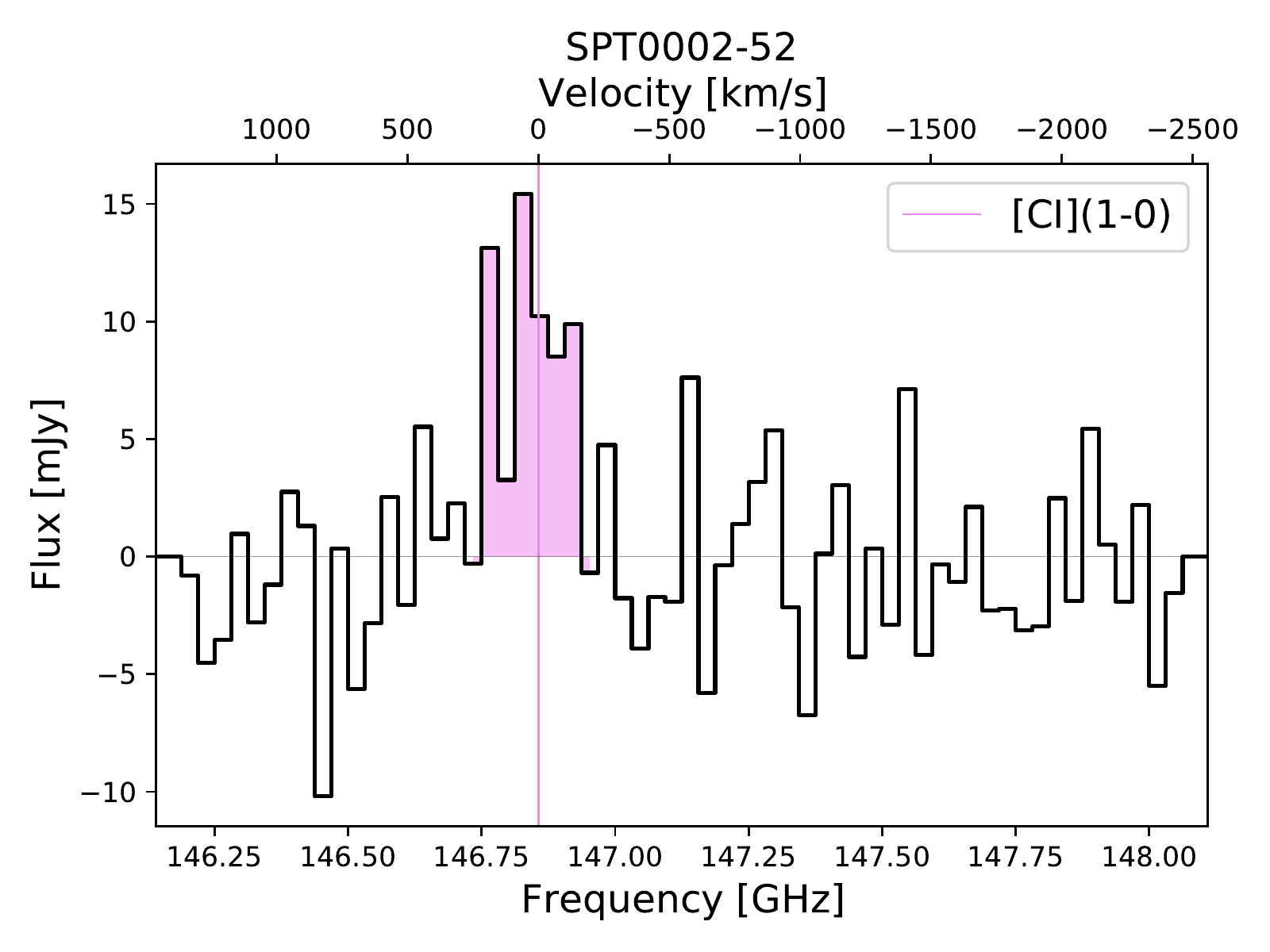} &  \includegraphics[width=6cm]{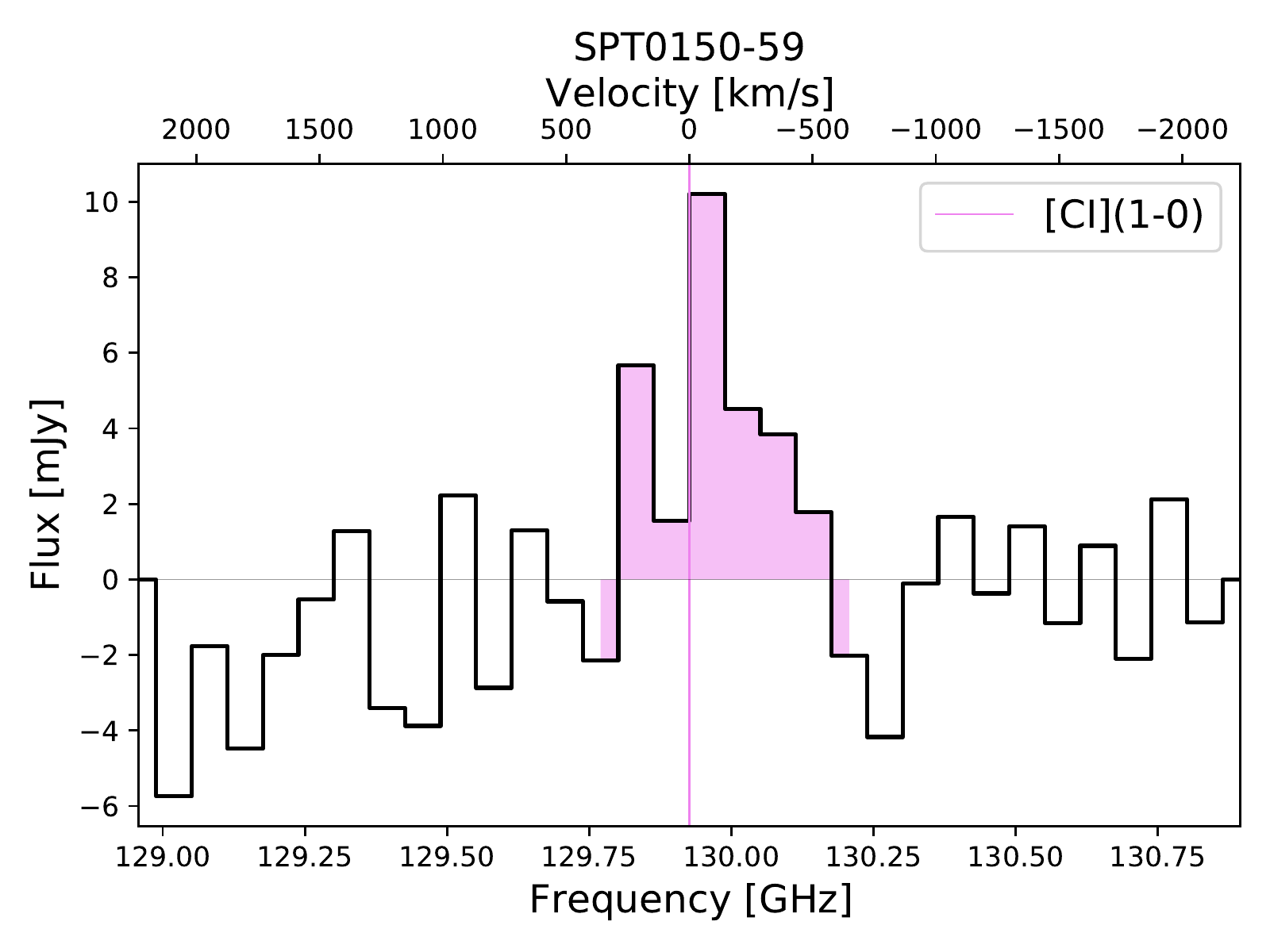} & \includegraphics[width=6cm]{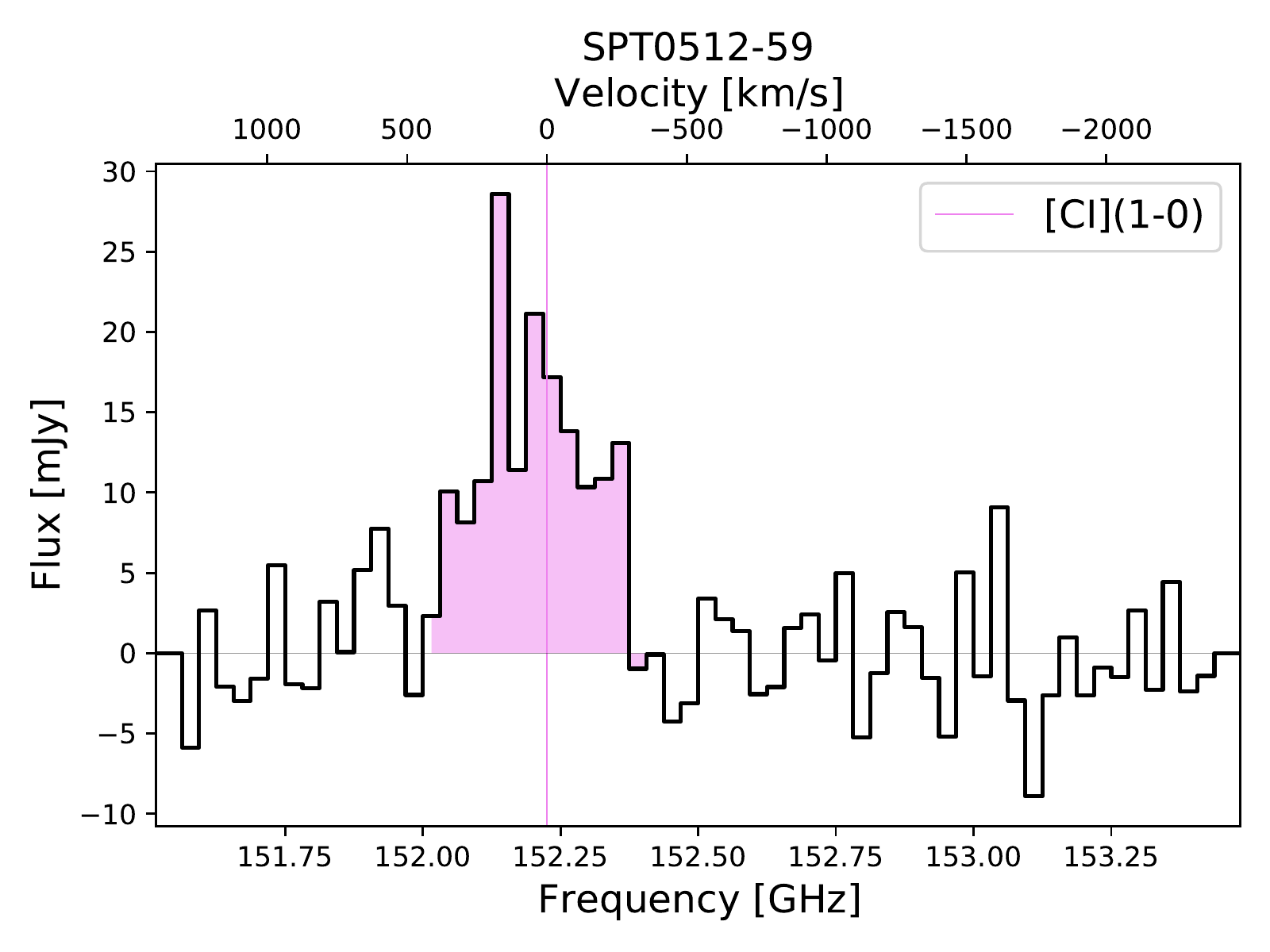} \\
\includegraphics[width=6cm]{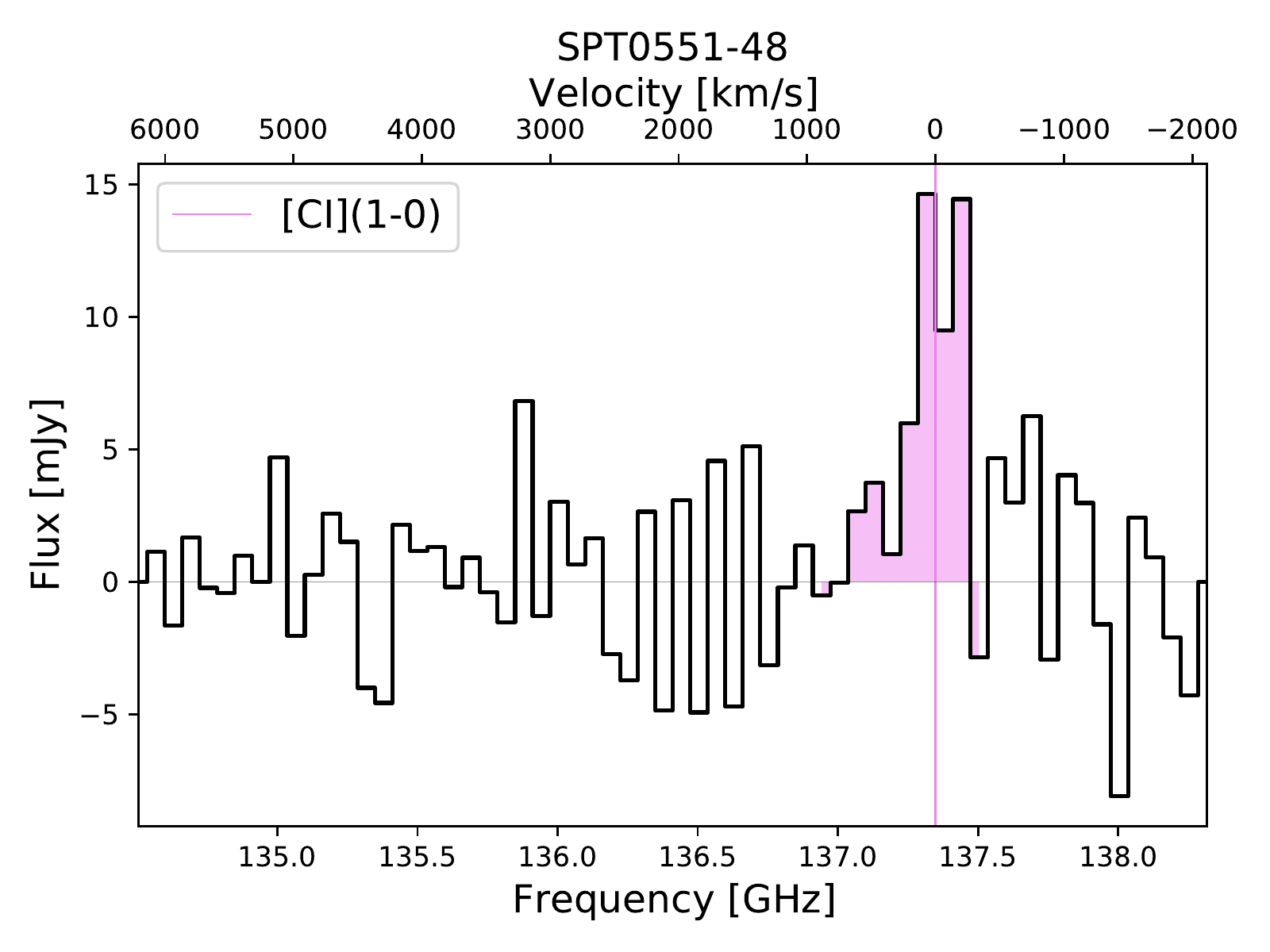} &  \includegraphics[width=6cm]{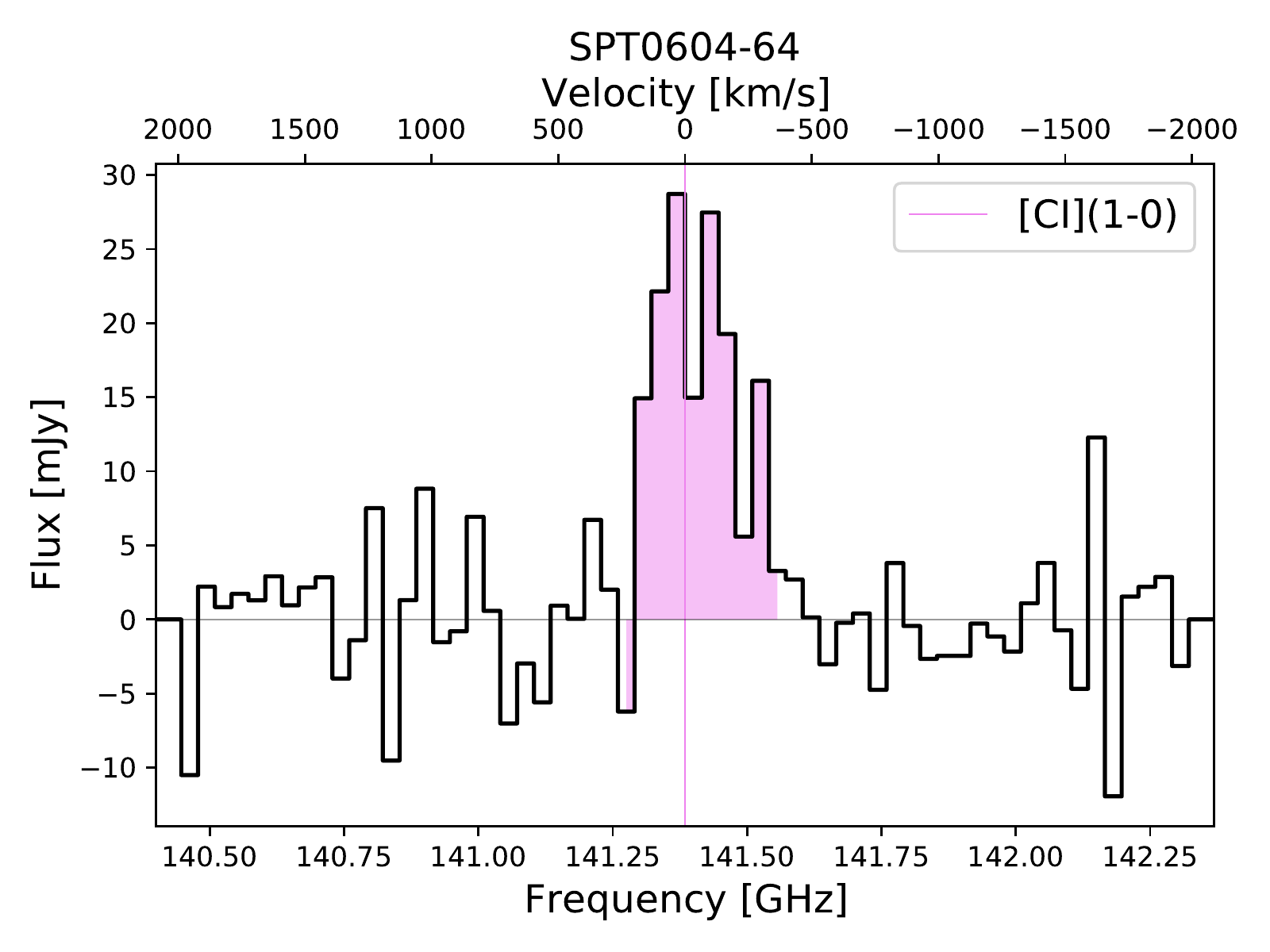} & \includegraphics[width=6cm]{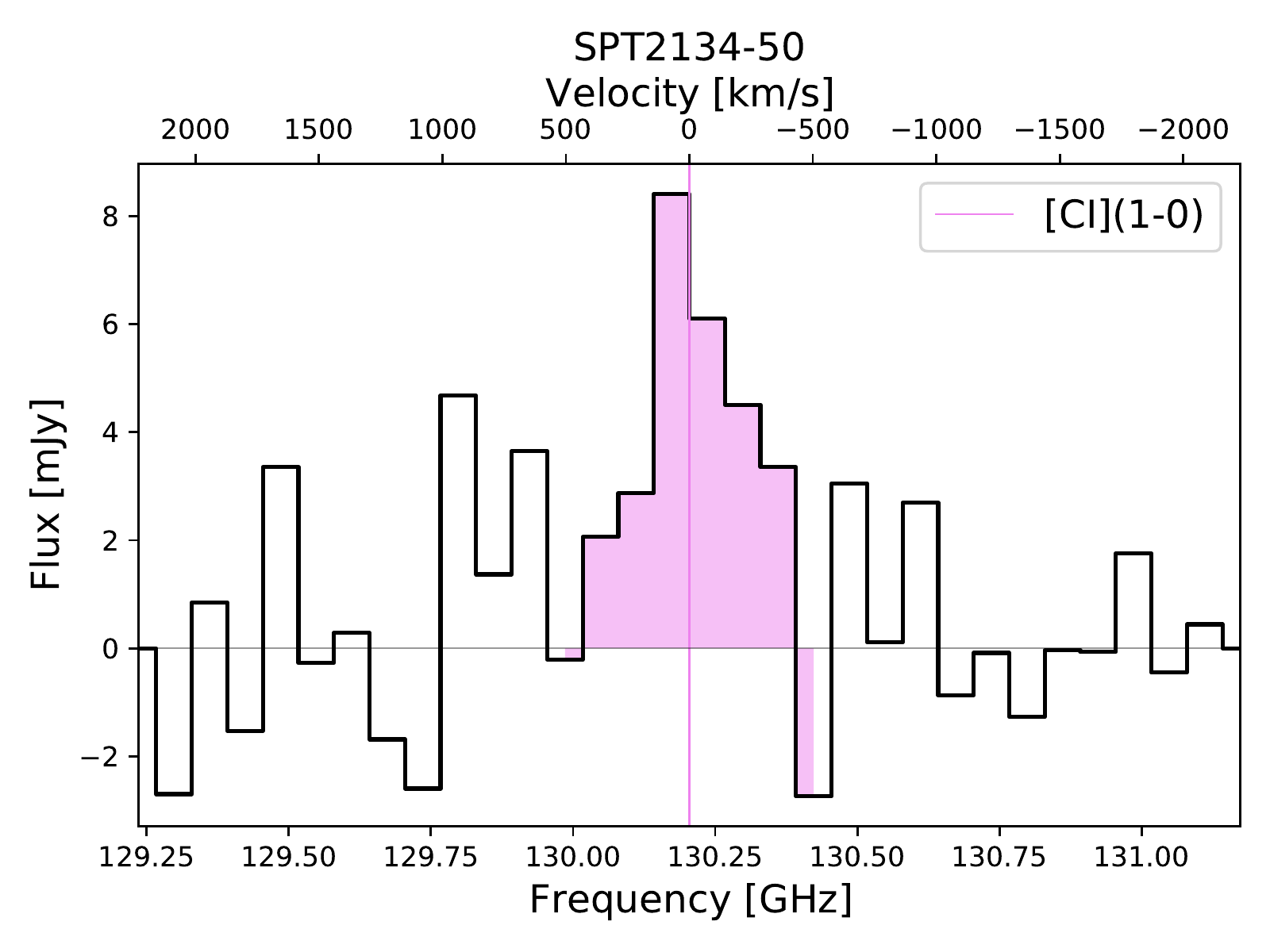} \\
\includegraphics[width=6cm]{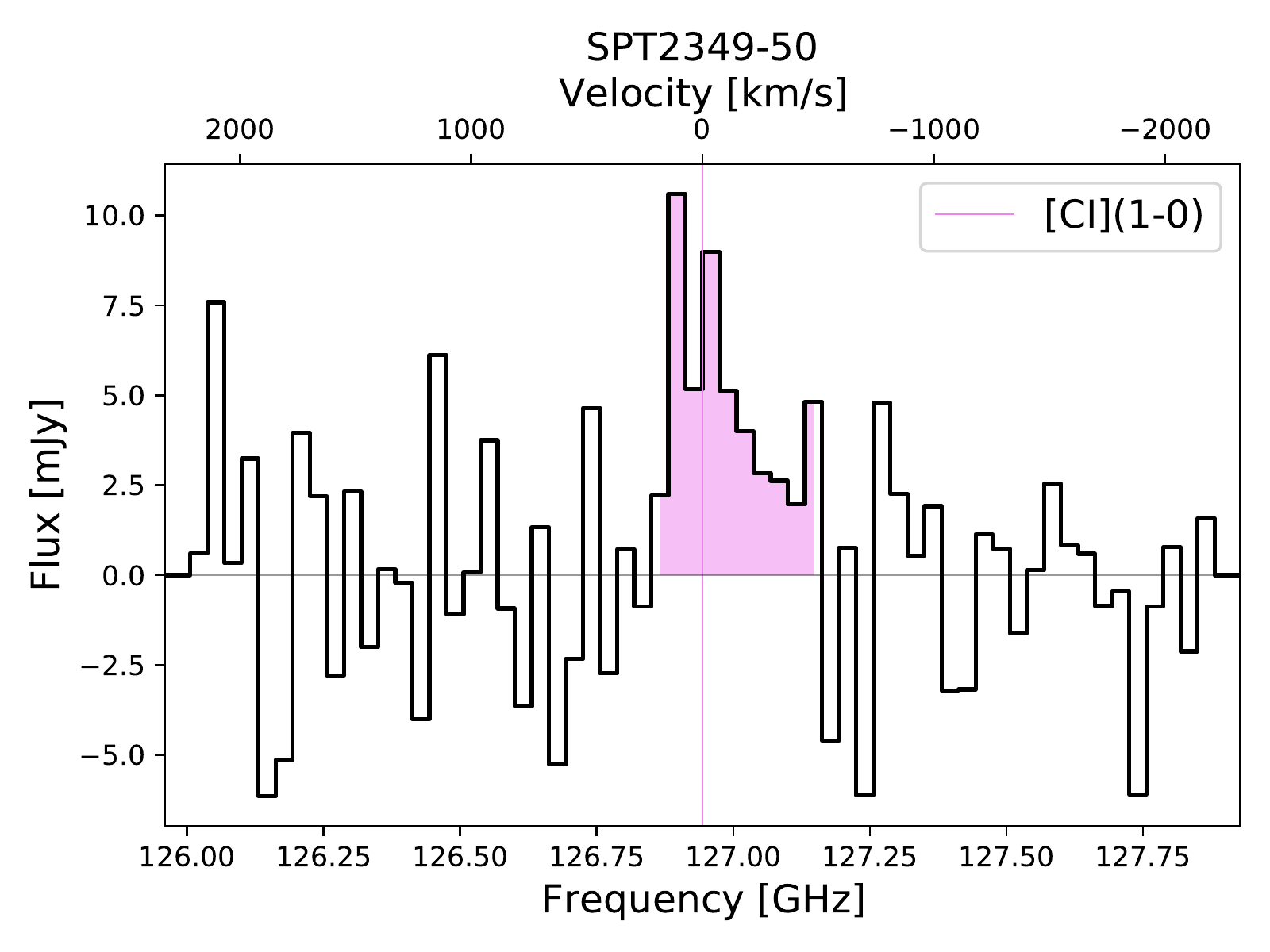} & 
\includegraphics[width=6cm]{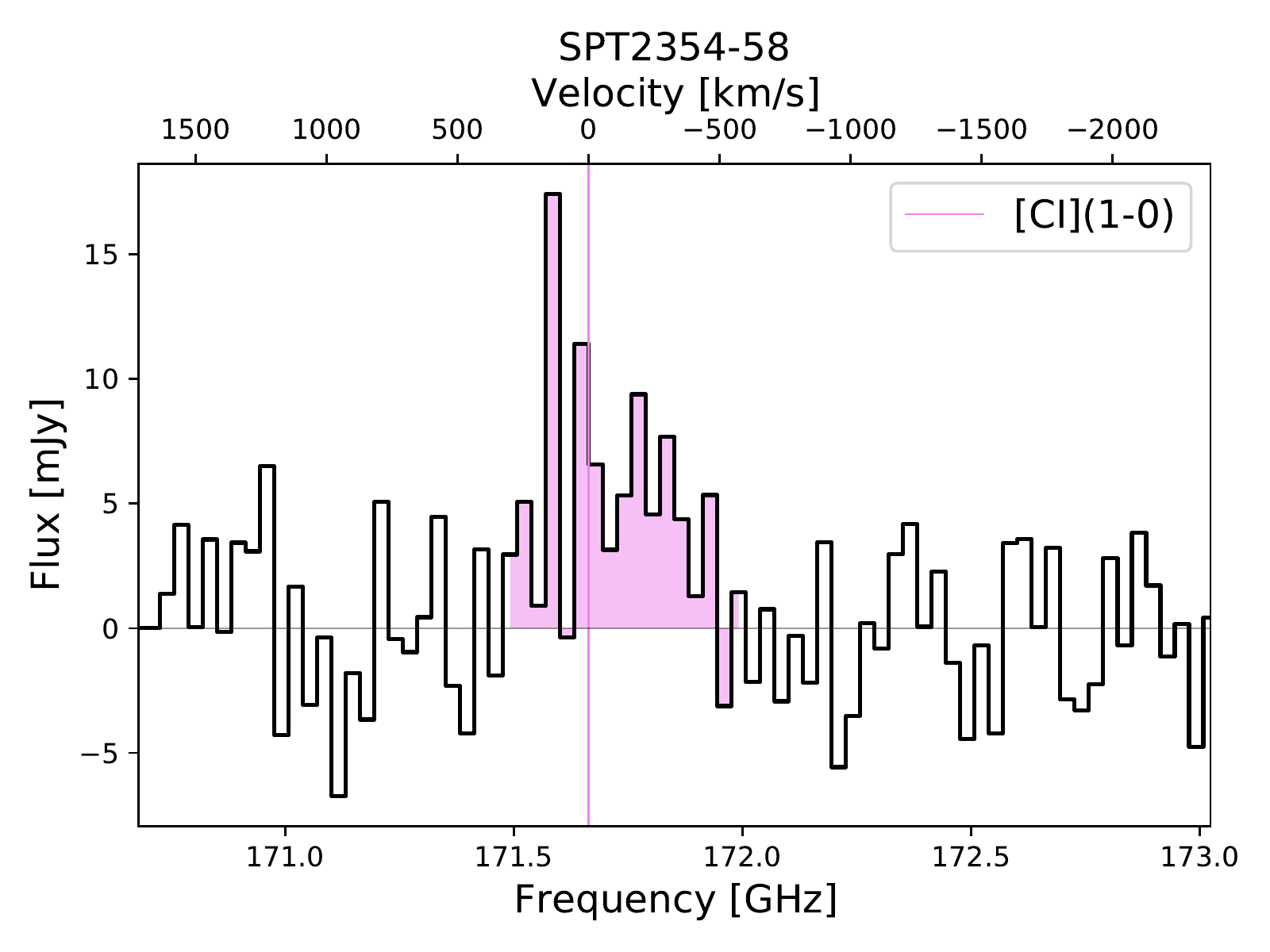} & \\

\end{tabular}
\caption{\label{fig:ACA CI spectra} The [CI](1-0) spectra of our sample. The violet-dotted line represents the [CI](1-0) line frequency based on the redshifts from \citet{Reuter20} and the velocity axis is centered at the this [CI](1-0) frequency. The violet shaded region represents the integrated limits to compute the integrated intensities of the [CI](1-0) line of our sources.}
\end{figure*}

\subsection{APEX [CII] observations}\label{APEX CII observations}
APEX [CII] observations were carried out using the First Light APEX Submillimetre Heterodyne receiver (FLASH, \citealt{Heyminck06}). 10 sources at 4.0 $< z <$ 4.8 ($\nu_{\rm obs}=327-388$\,GHz) were observed in the 345 GHz channel between 2014 July and 2017 September, during Max Planck time. All observations were done in good weather conditions with an average precipitable water vapour $<$1.1 mm, yielding typical system temperatures of 230K.  The beam sizes/antenna gains are 18.0 arcsec/40 Jy K$^{-1}$ and 15.0 arcsec/42 Jy $^{-1}$ for the lowest and highest observed frequencies of the [CII] line, respectively. The beam size is much larger than the observed Einstein radii of these sources and thus they are unresolved \citep{Spilker16}. The 92h of observations were done in wobbler switching mode, with switching frequency of 1.5\,Hz and a wobbler throw of 50 arcsec in azimuth. Pointing was checked frequently and was found to be stable to within 2.5 arcsec. Calibration was done every $\sim$10 min using the standard hot/cold load absorber measurements. The data were recorded with the MPIfR Fast Fourier Transform spectrometers (FFTS, \citealt{Klein06}) providing 4 $\times$ 2.5\,GHz of bandwidth to cover the full 4\,GHz bandwidth in each of the upper and lower sidebands of the sideband-separating FLASH receiver. 

The data were processed with the Continuum and Line Analysis Single-dish Software (CLASS). We visually inspected the individual scans and omitted scans with unstable baselines, resulting in $<10\%$ loss. We subtracted linear baselines from the individual 
spectra in each of the two FFTS units, and regridded to a velocity resolution of $\sim$90 km s$^{-1}$ in the averaged spectra. On-source integration times were between 2 and 15 h. The line intensities are summarised in Table\,\ref{tab:2_flux_catalogue_ACA}.

\begin{figure*}[h!]
\centering

\begin{tabular}{ccc}
\includegraphics[width=6cm]{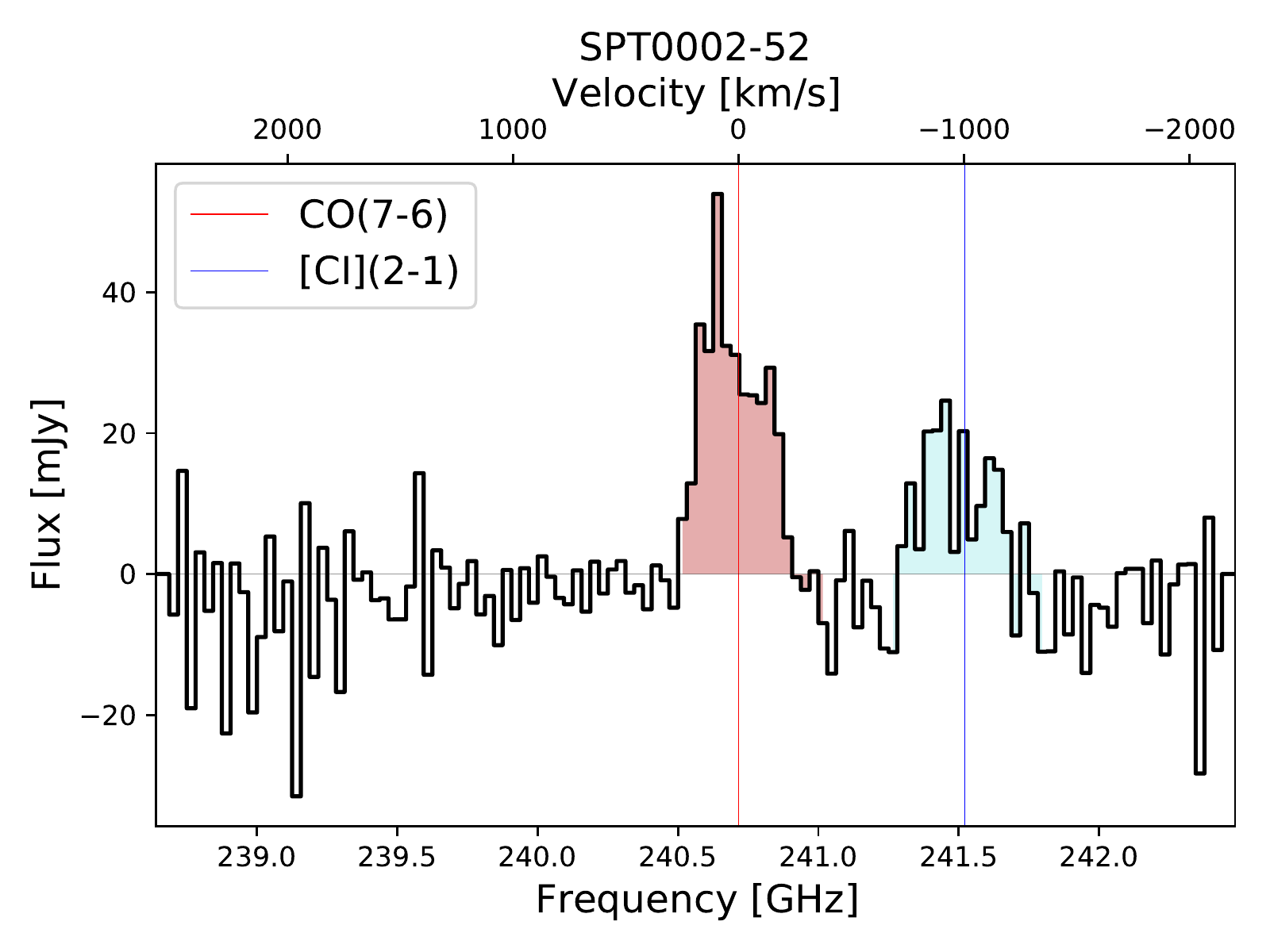} &  \includegraphics[width=6cm]{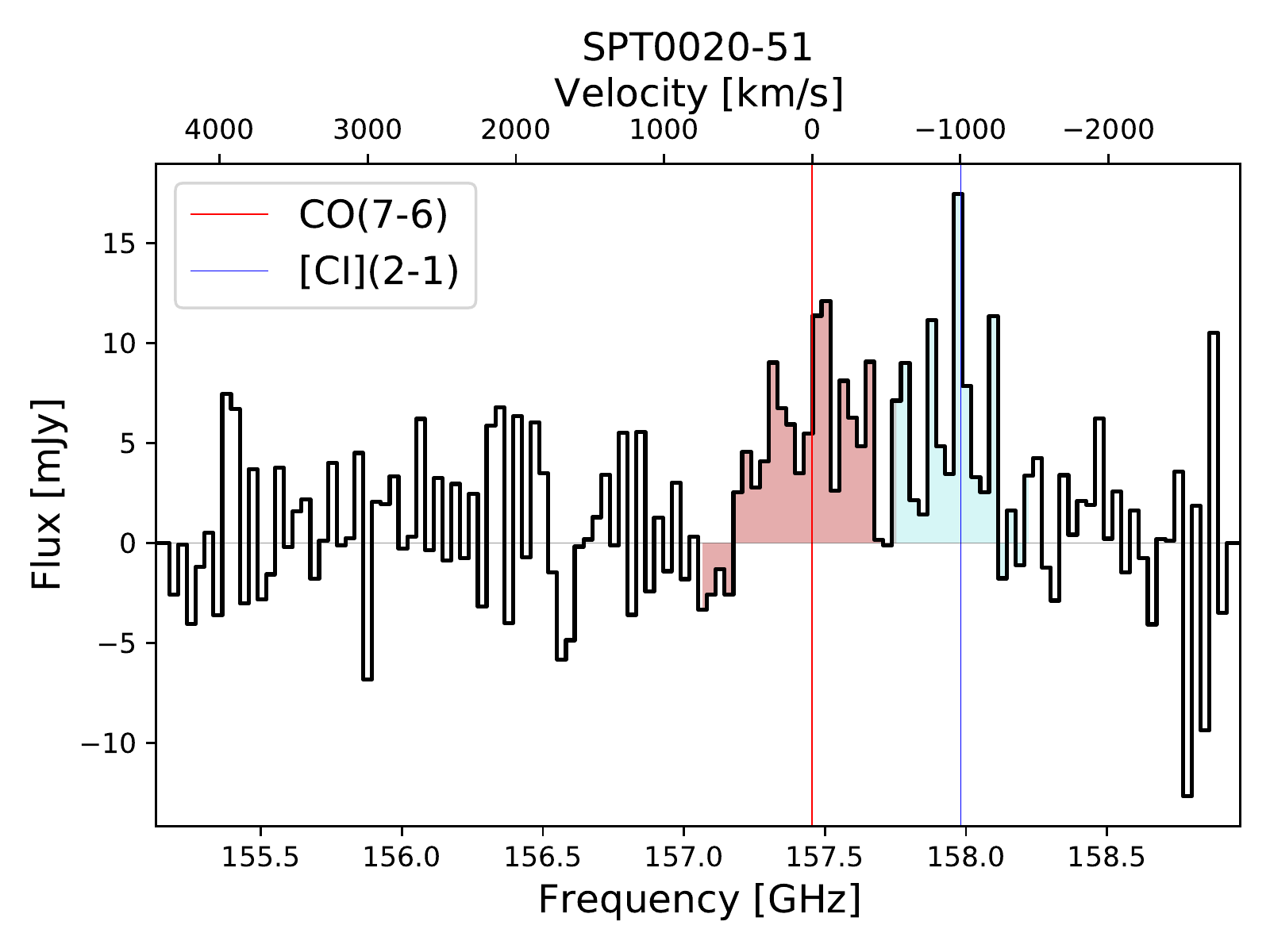} & \includegraphics[width=6cm]{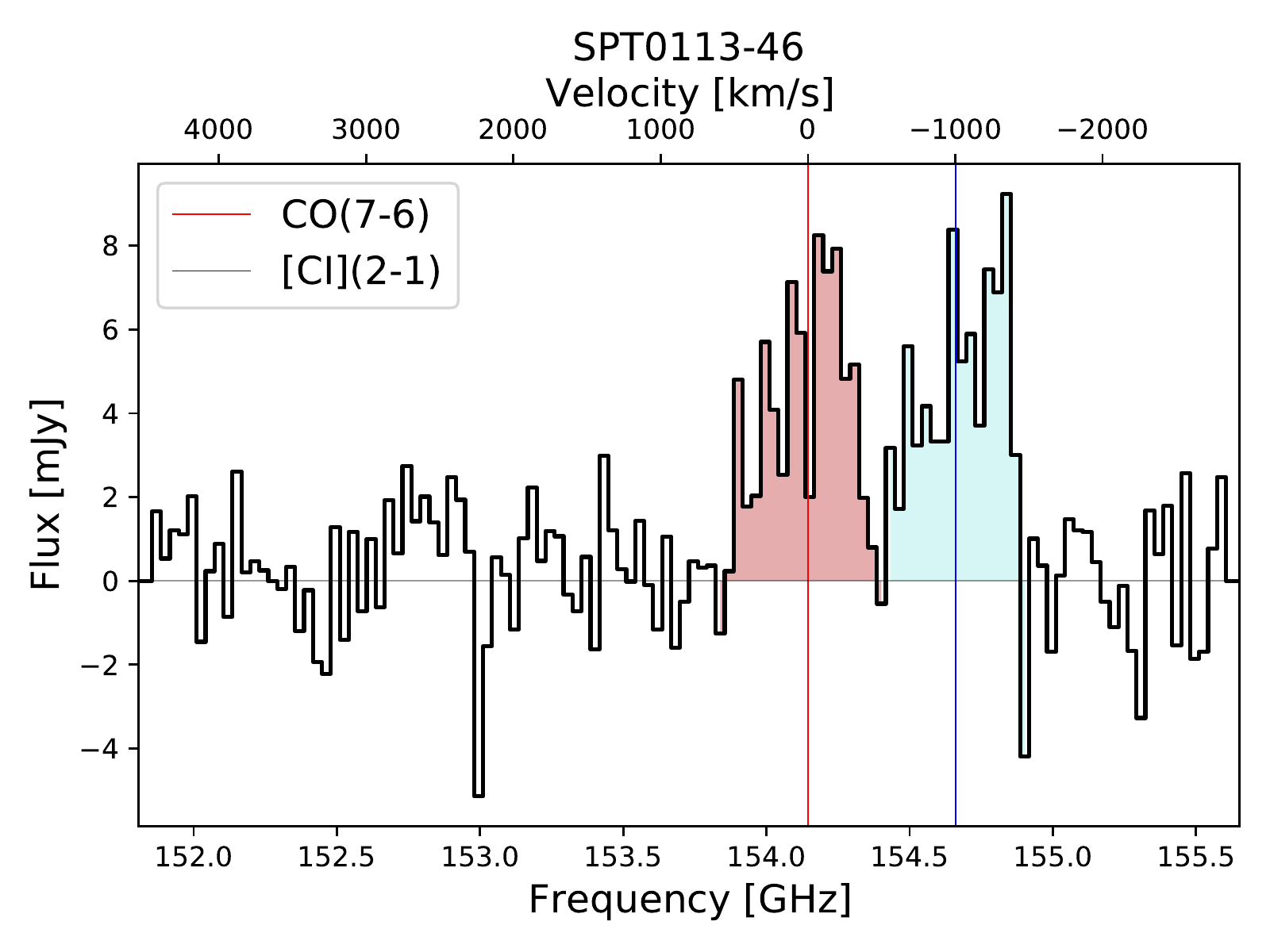} \\
\includegraphics[width=6cm]{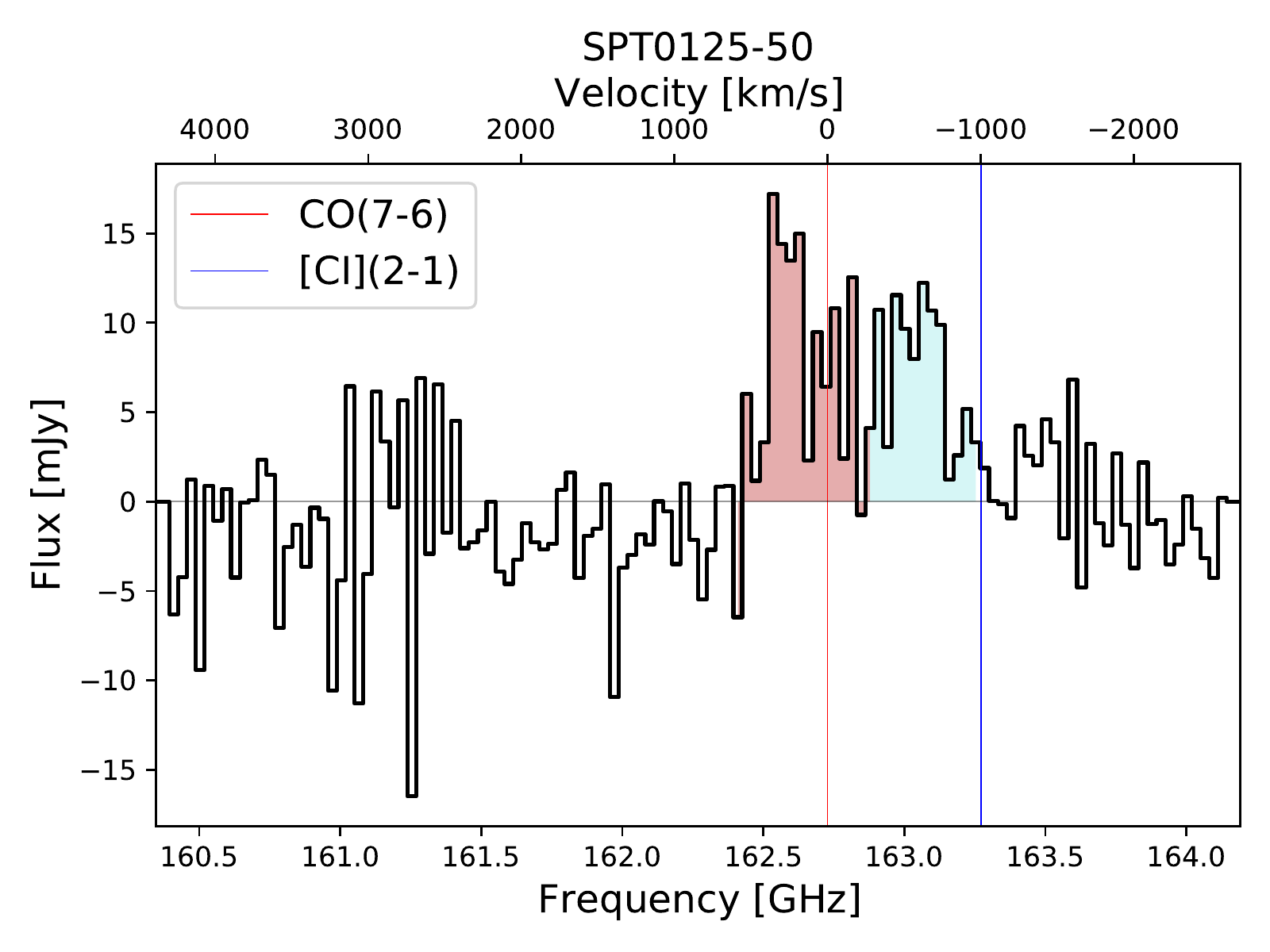} &  \includegraphics[width=6cm]{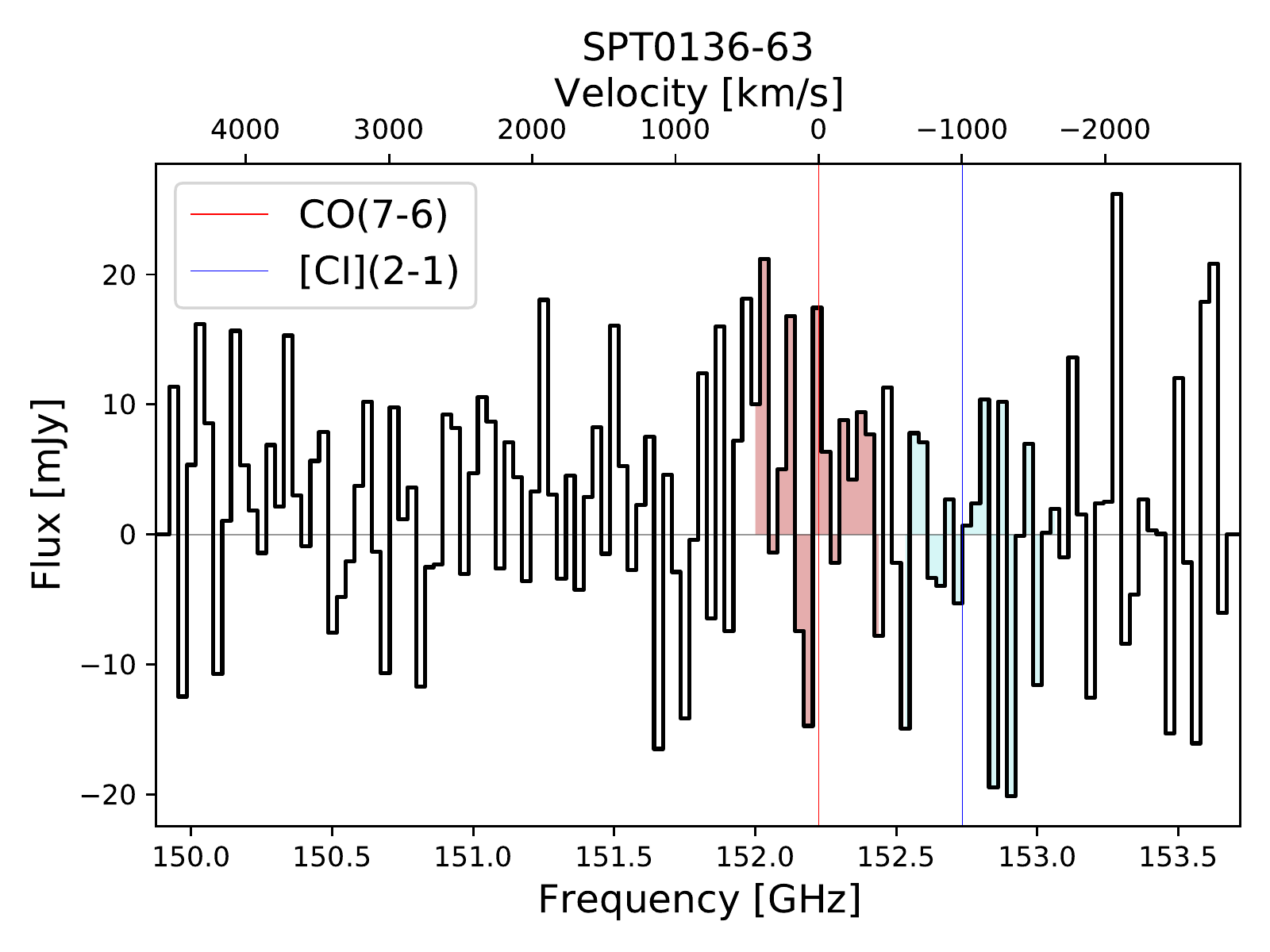} & \includegraphics[width=6cm]{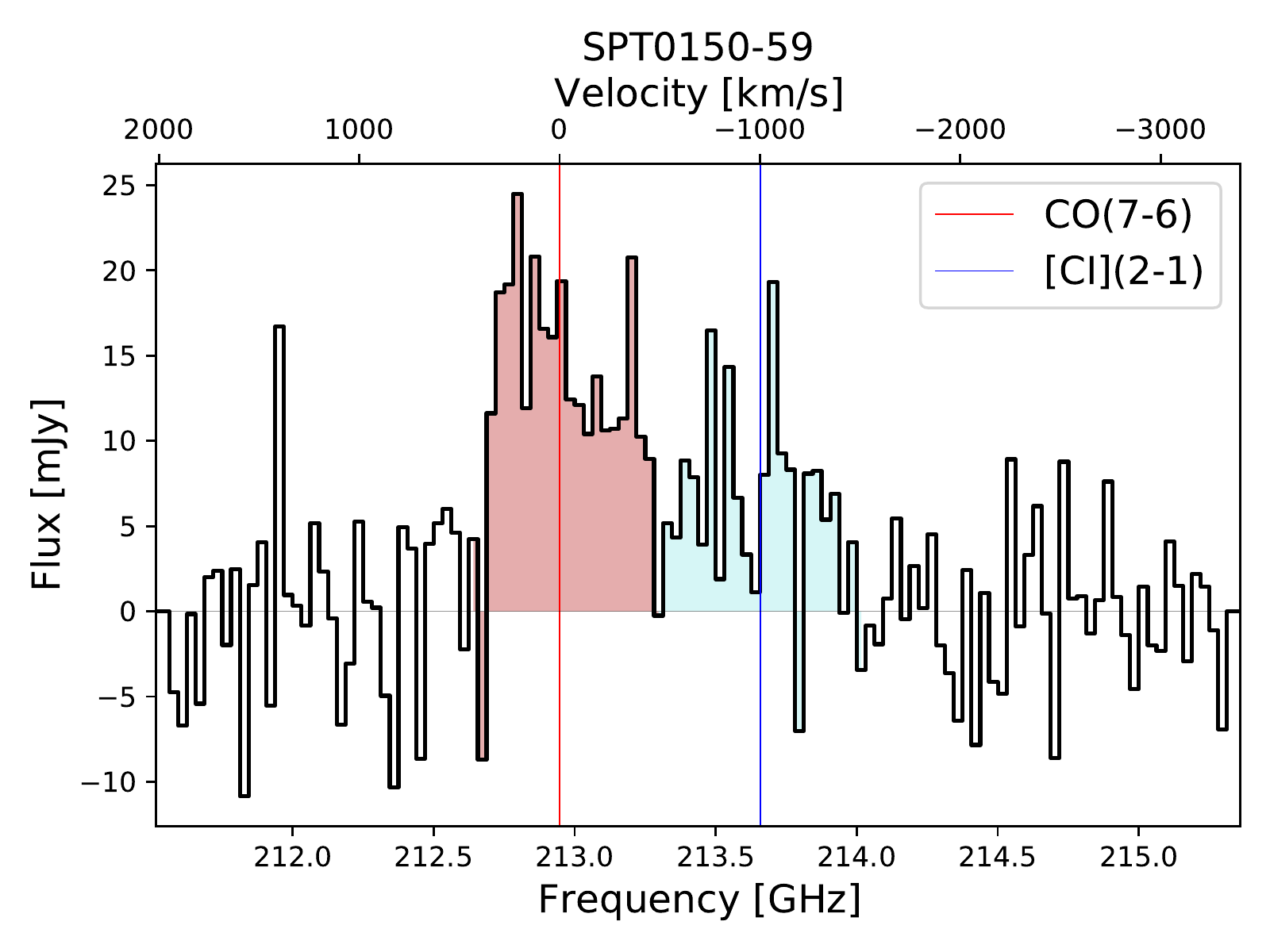} \\
\includegraphics[width=6cm]{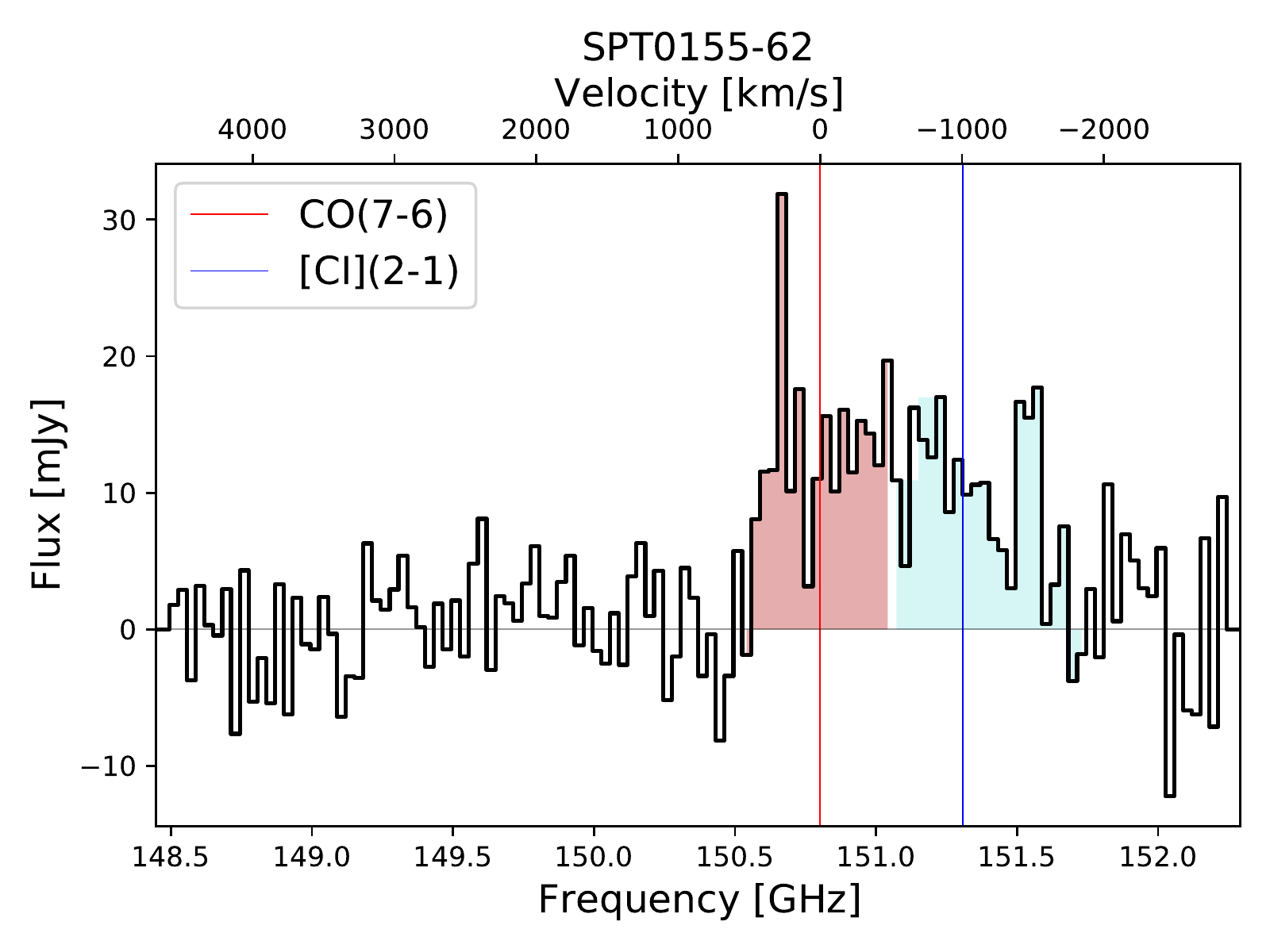} &  \includegraphics[width=6cm]{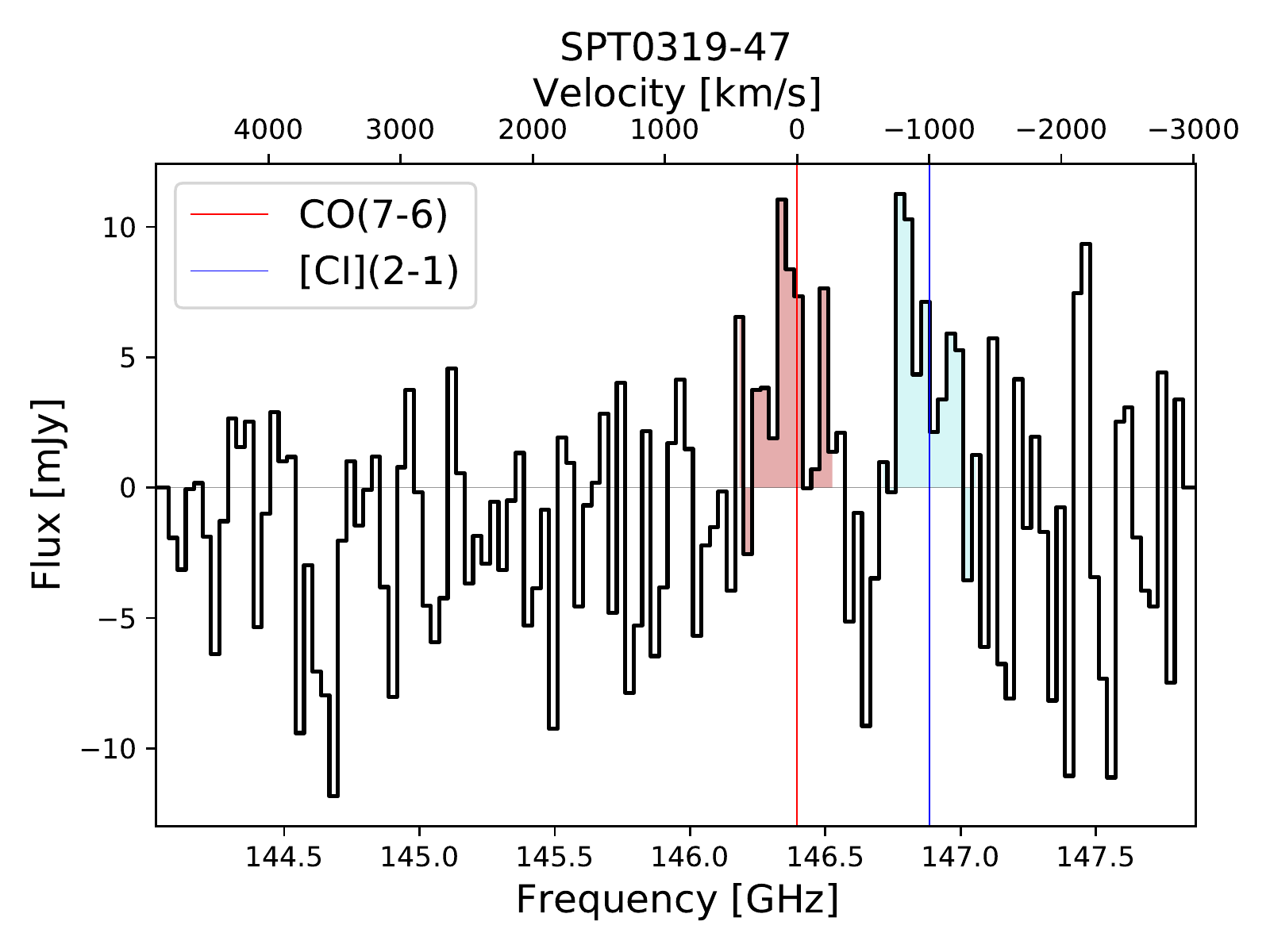} & \includegraphics[width=6cm]{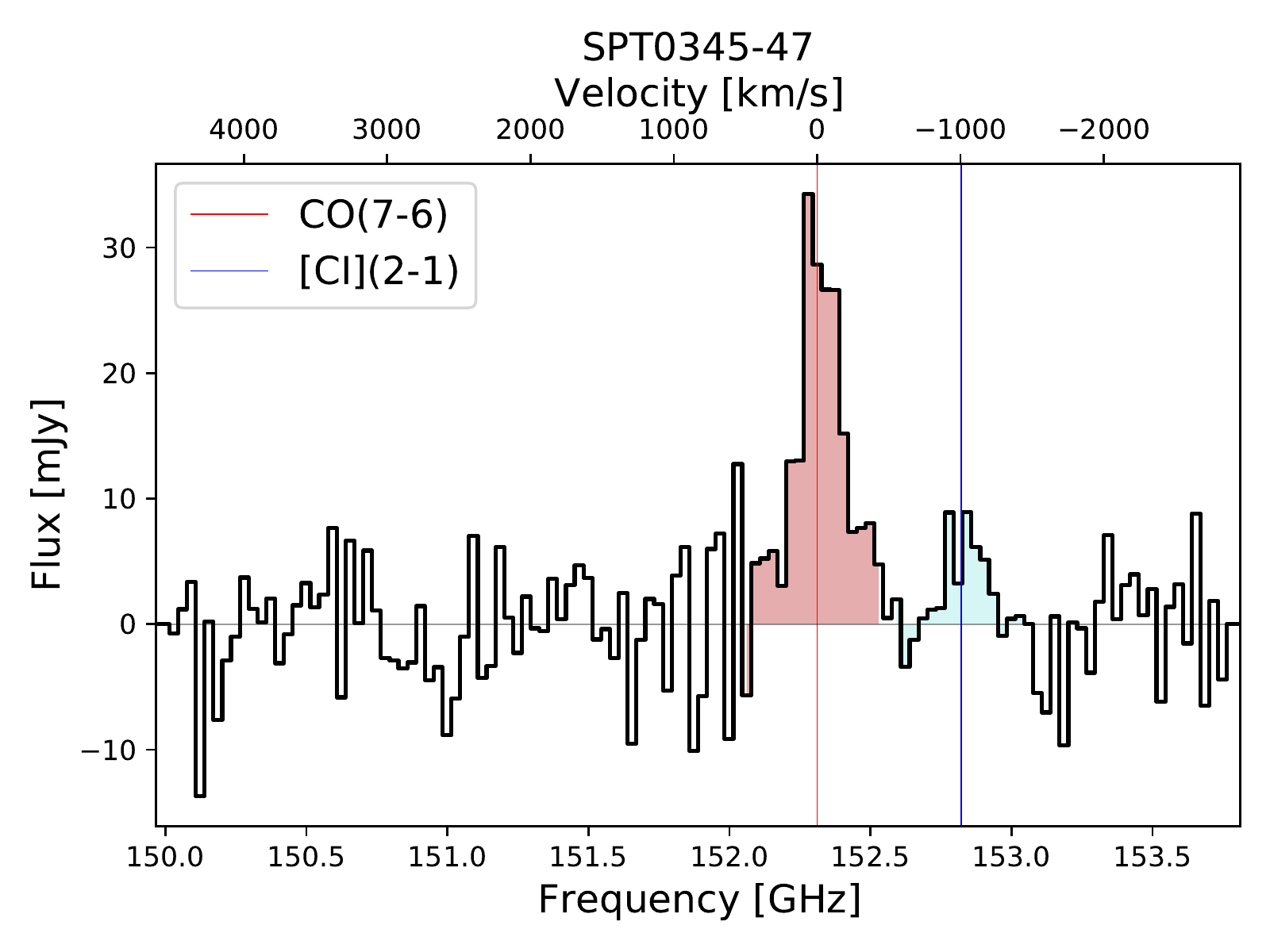} \\
\includegraphics[width=6cm]{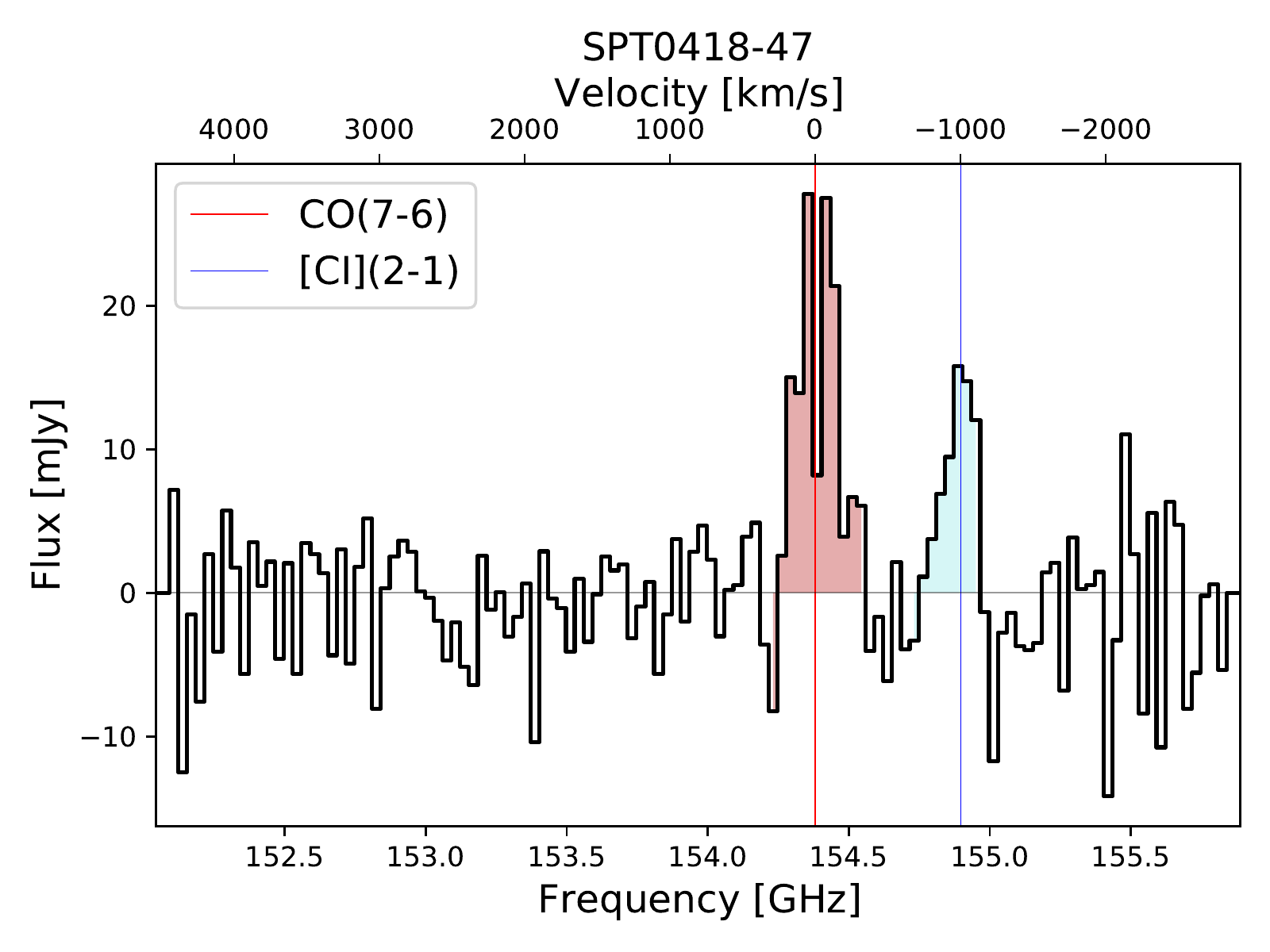} &  \includegraphics[width=6cm]{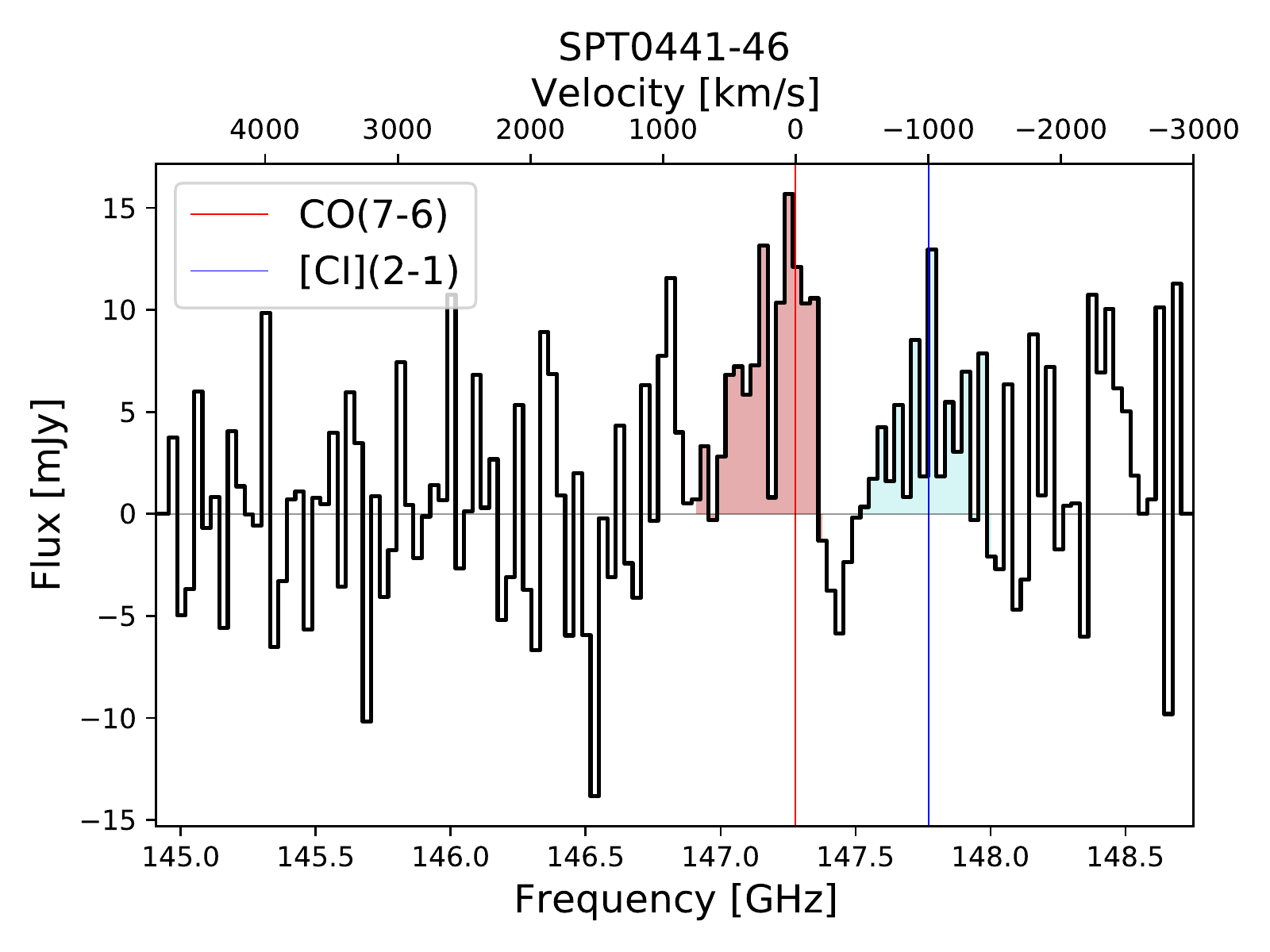} & \includegraphics[width=6cm]{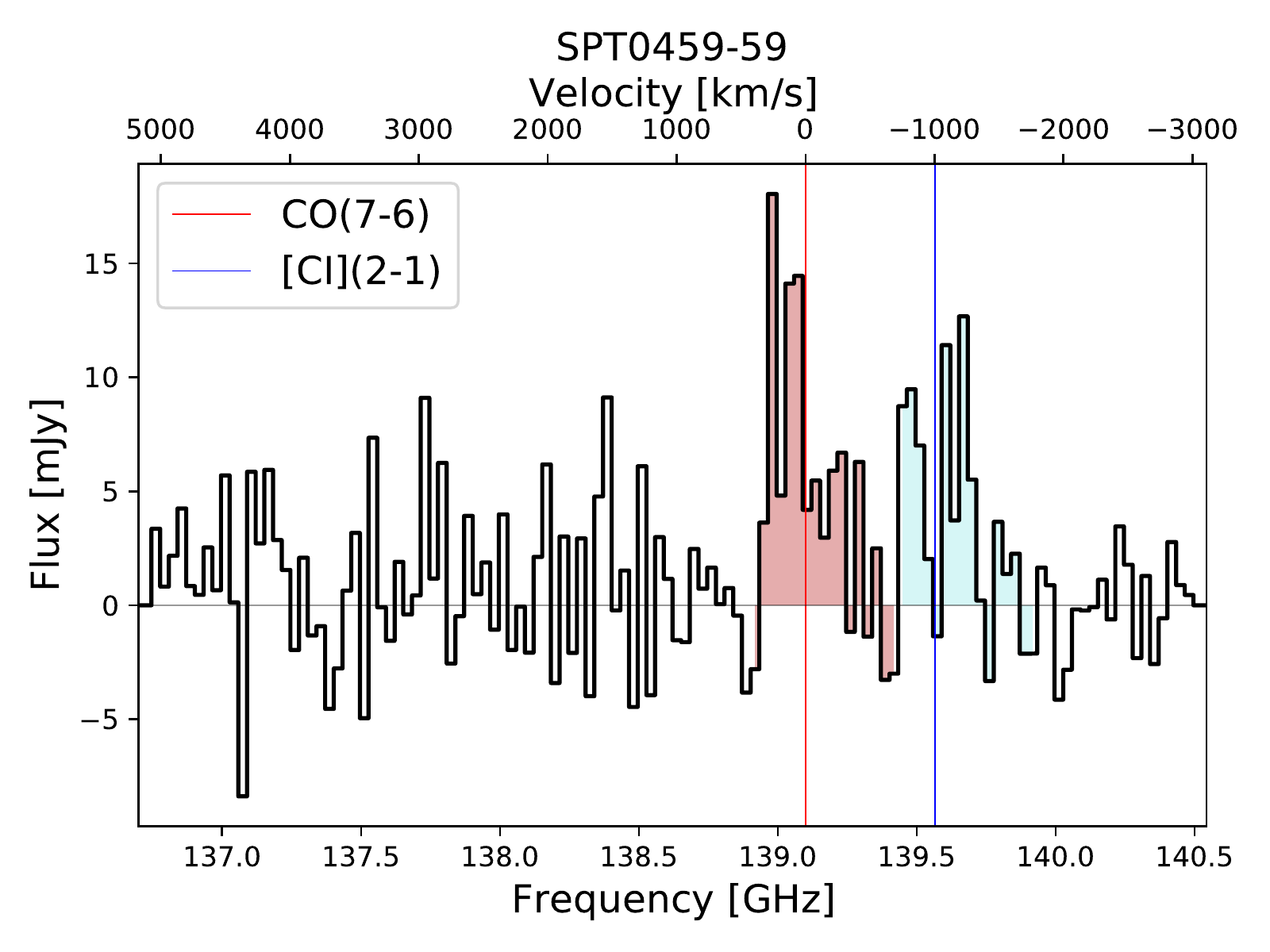} \\
\includegraphics[width=6cm]{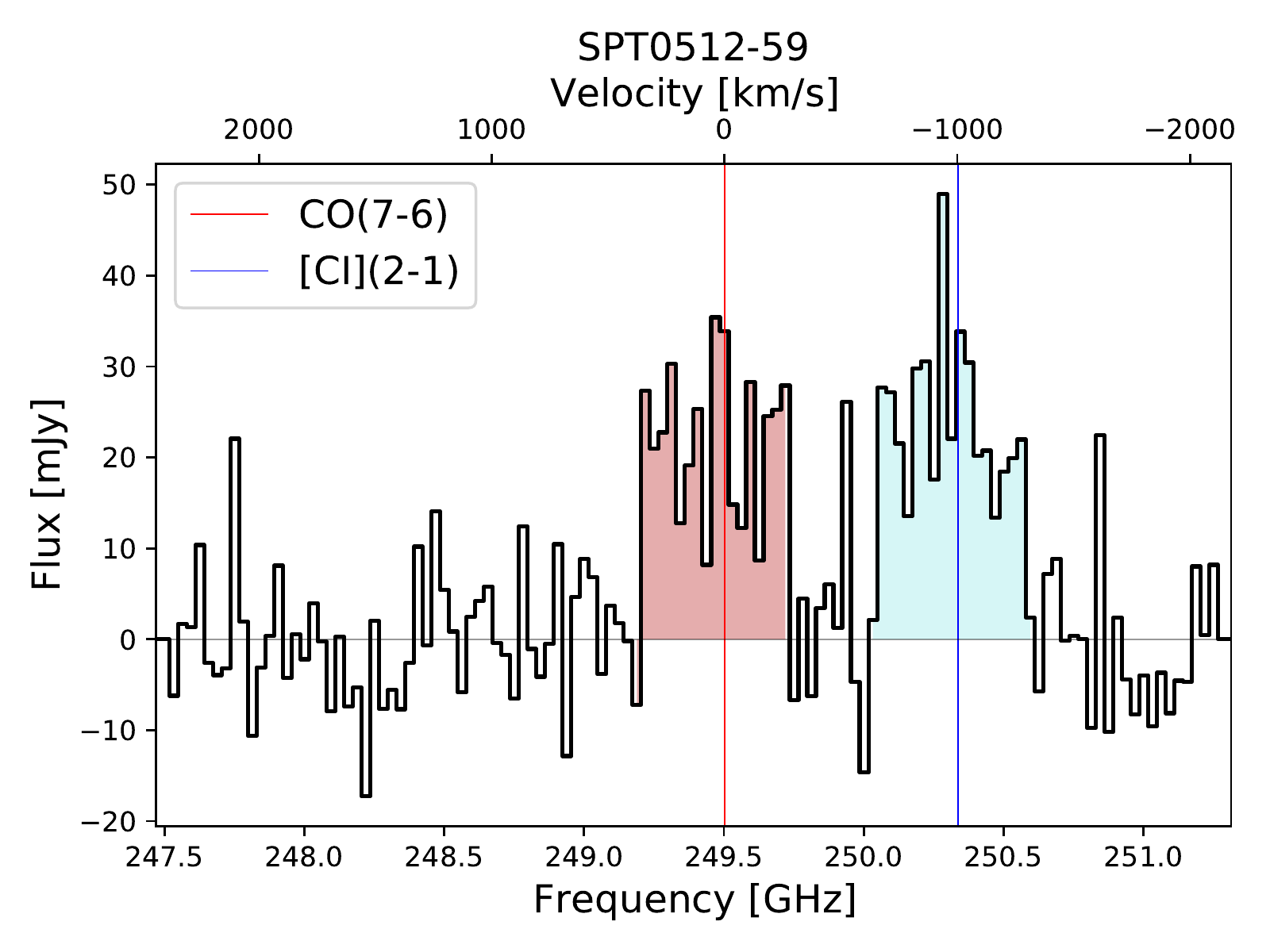} &  \includegraphics[width=6cm]{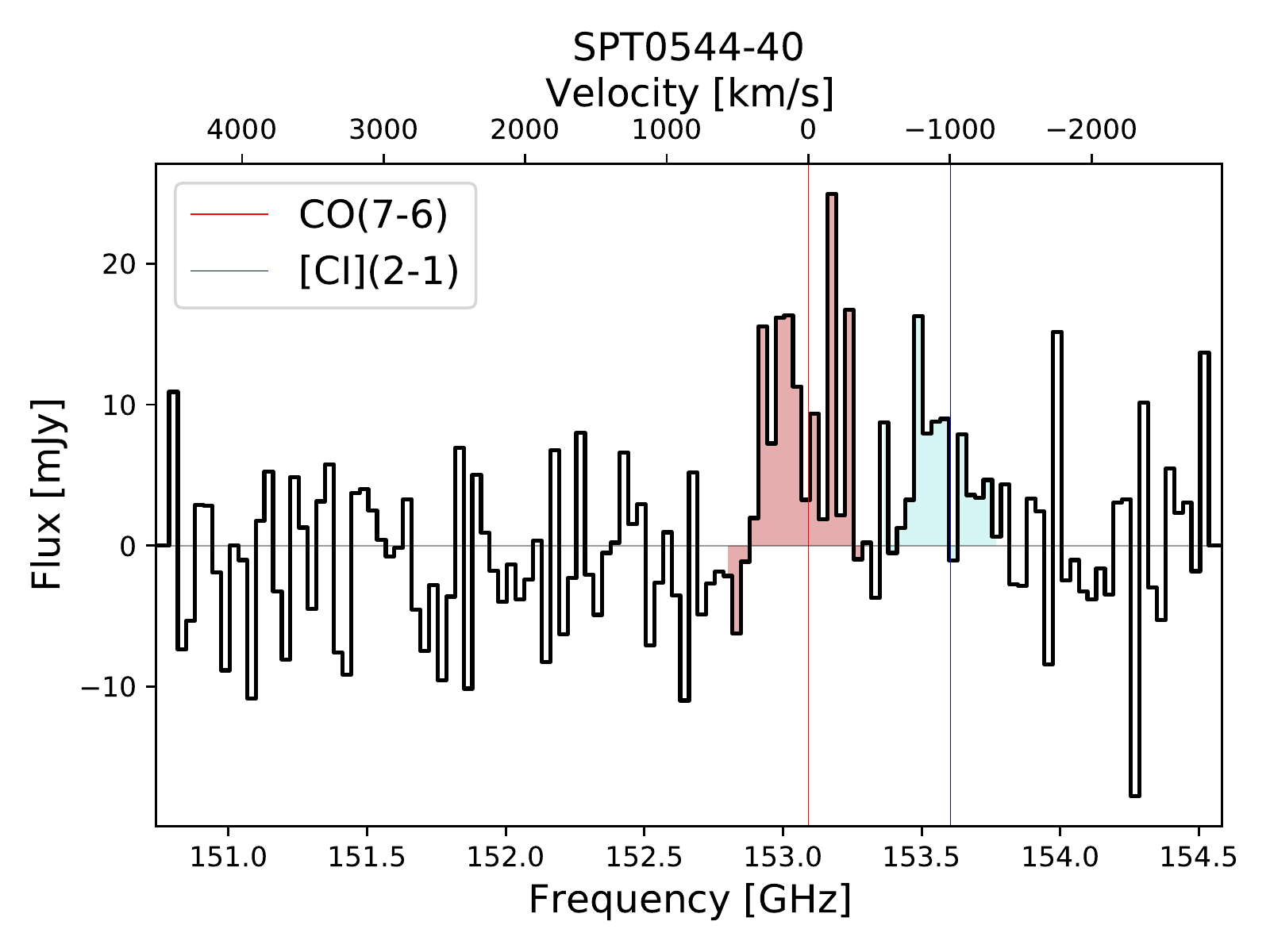} & \includegraphics[width=6cm]{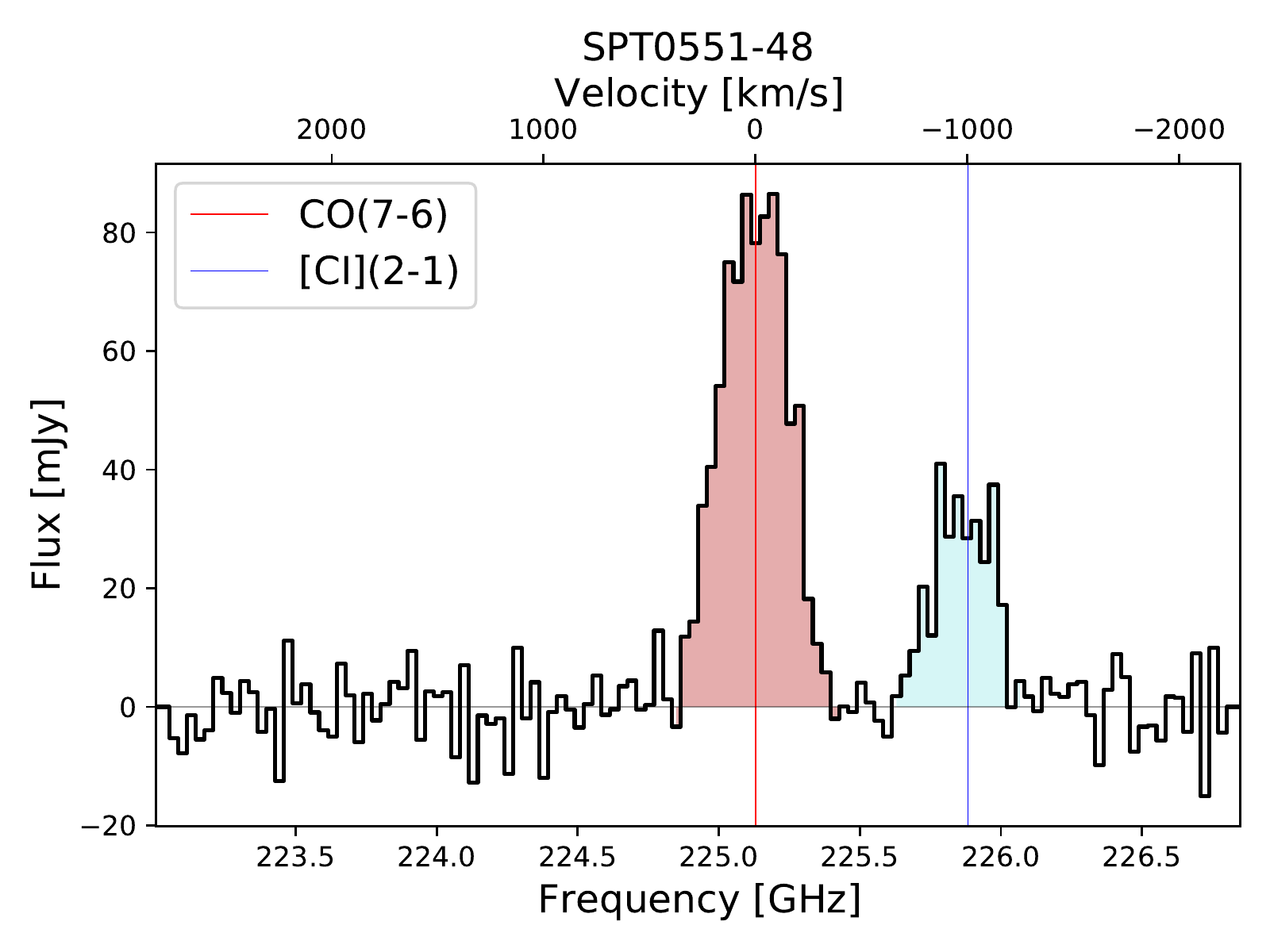} \\

\end{tabular}
\end{figure*}

\begin{figure*}[!]
\centering

\begin{tabular}{ccc}
\includegraphics[width=6cm]{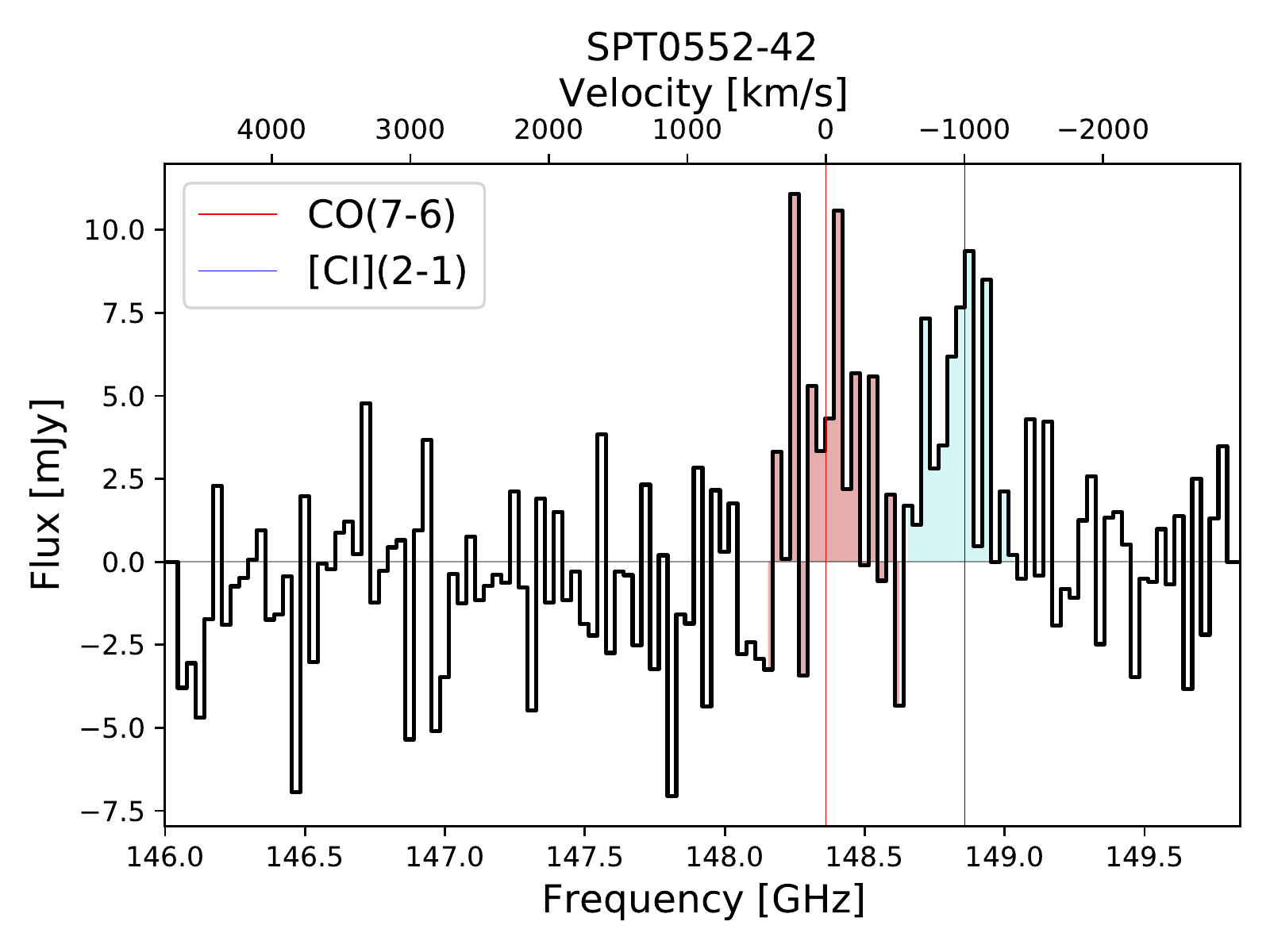} &  \includegraphics[width=6cm]{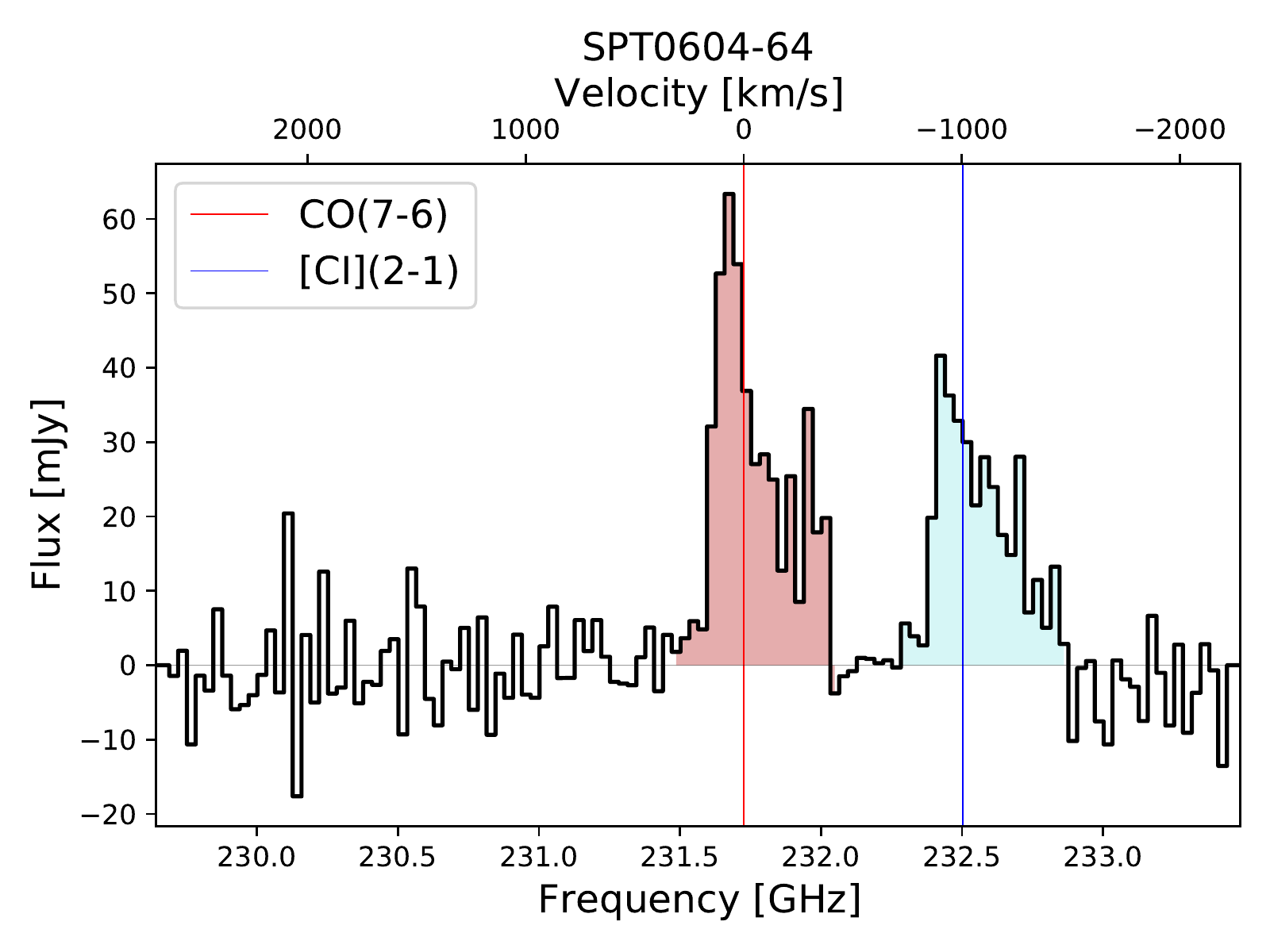} & \includegraphics[width=6cm]{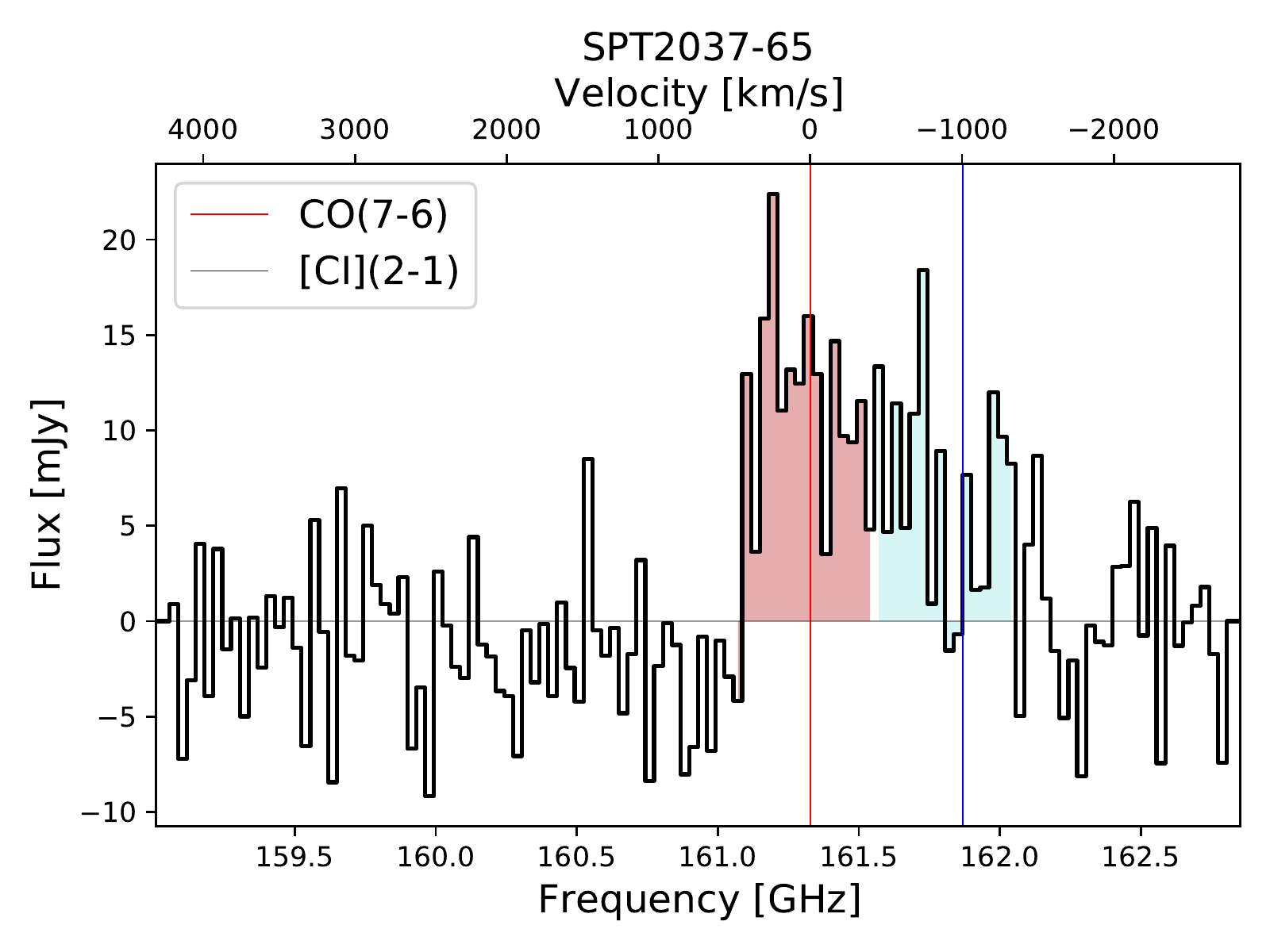} \\
\includegraphics[width=6cm]{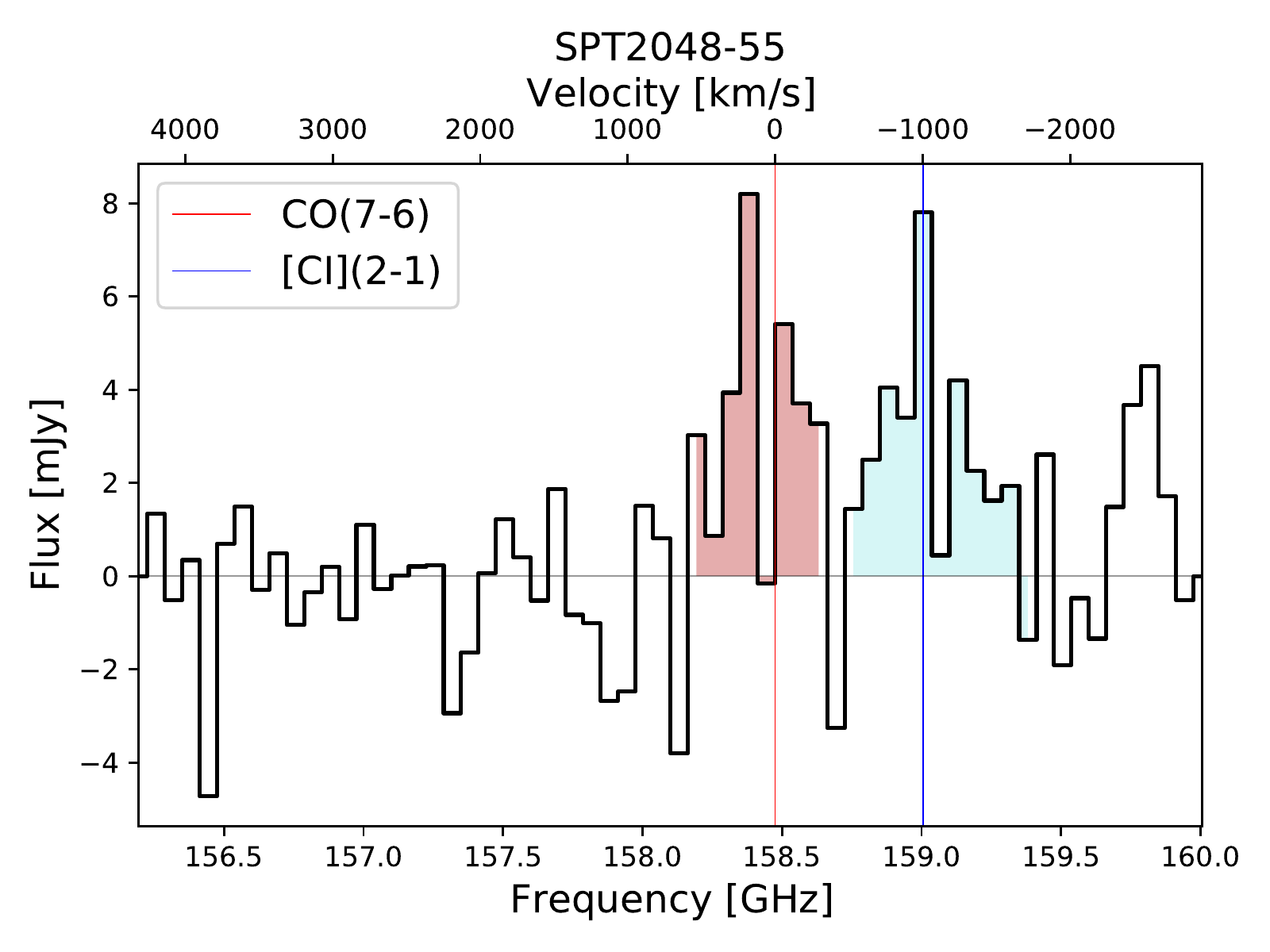} &  \includegraphics[width=6cm]{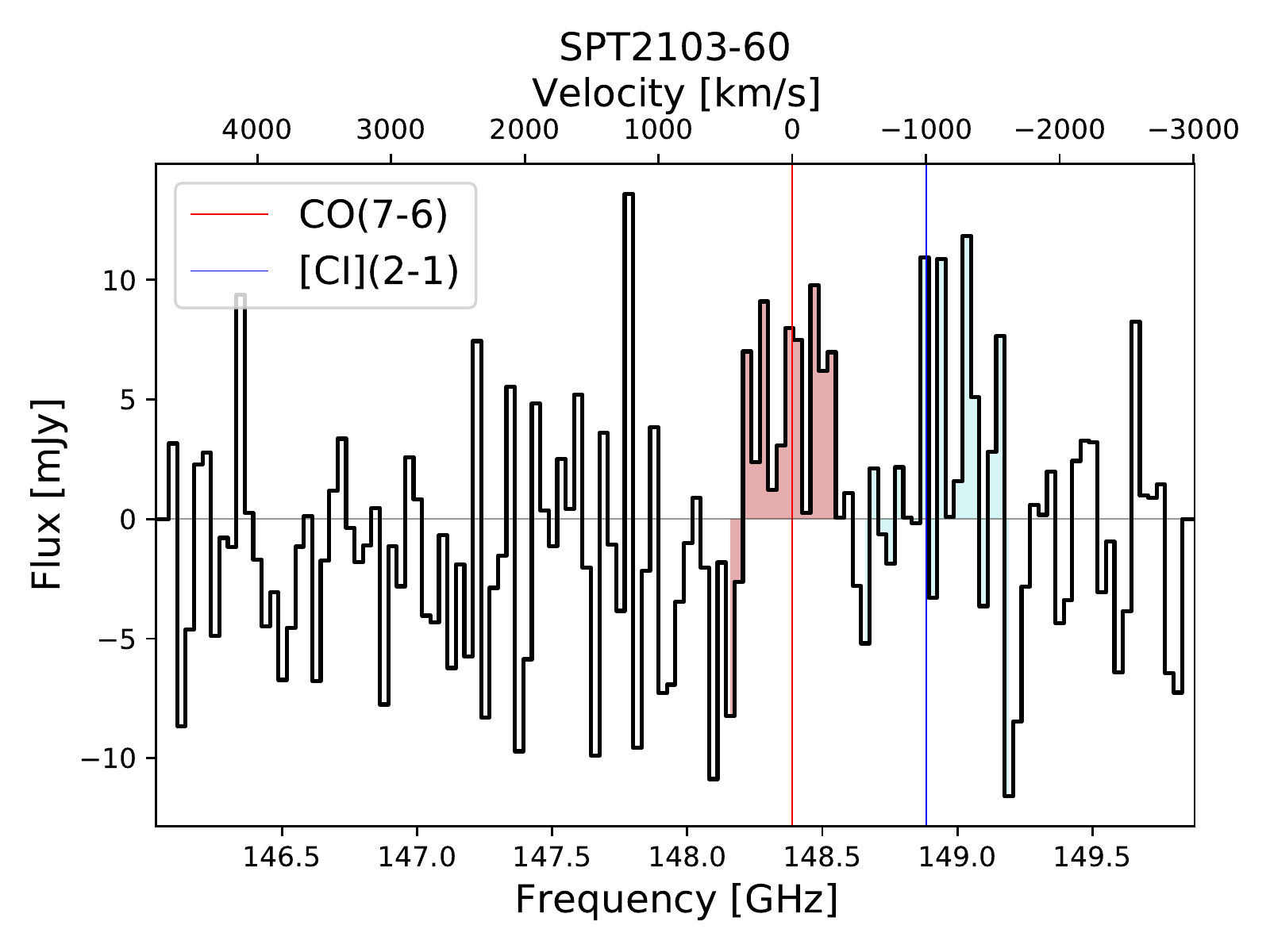} & \includegraphics[width=6cm]{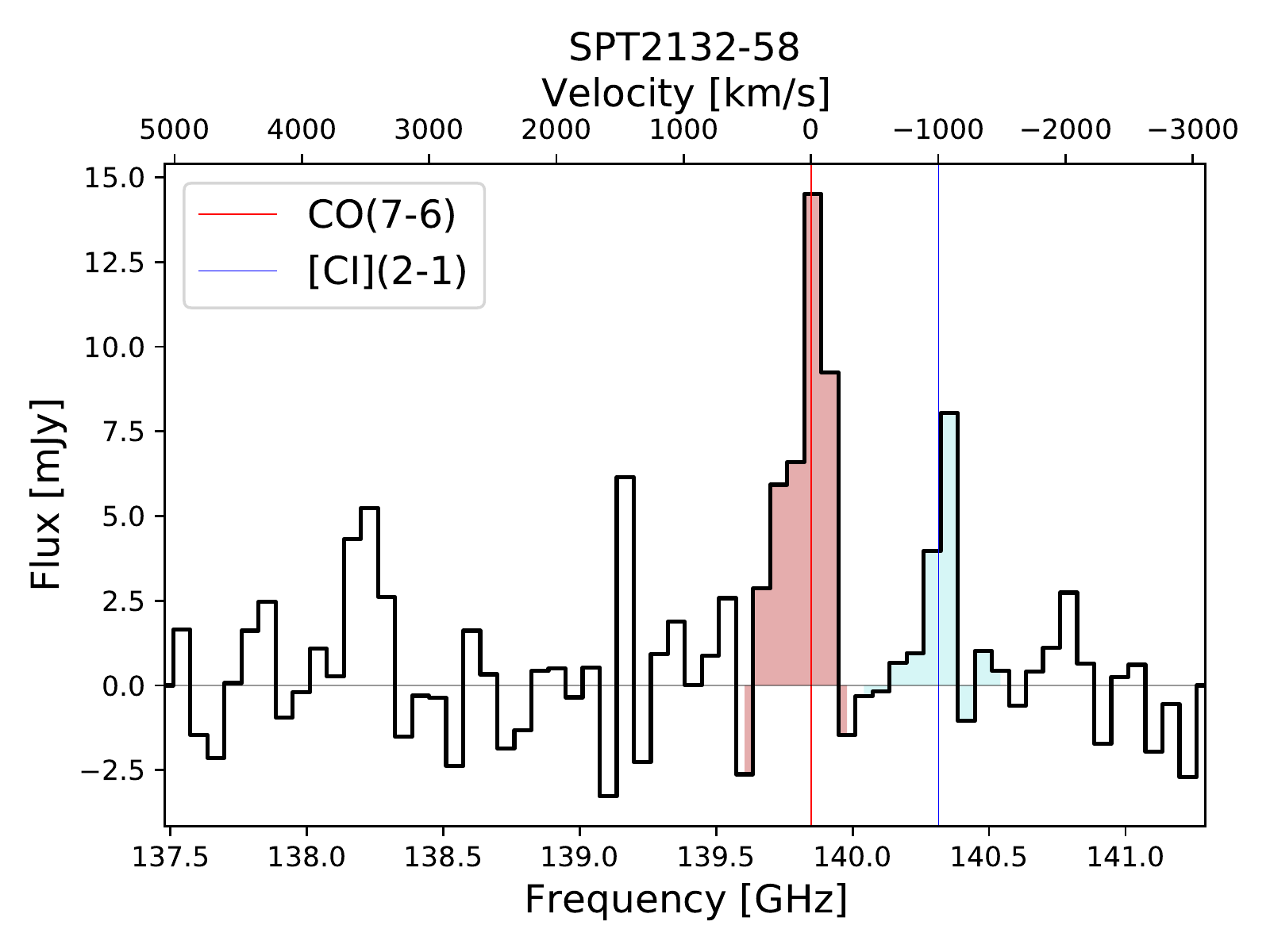} \\
\includegraphics[width=6cm]{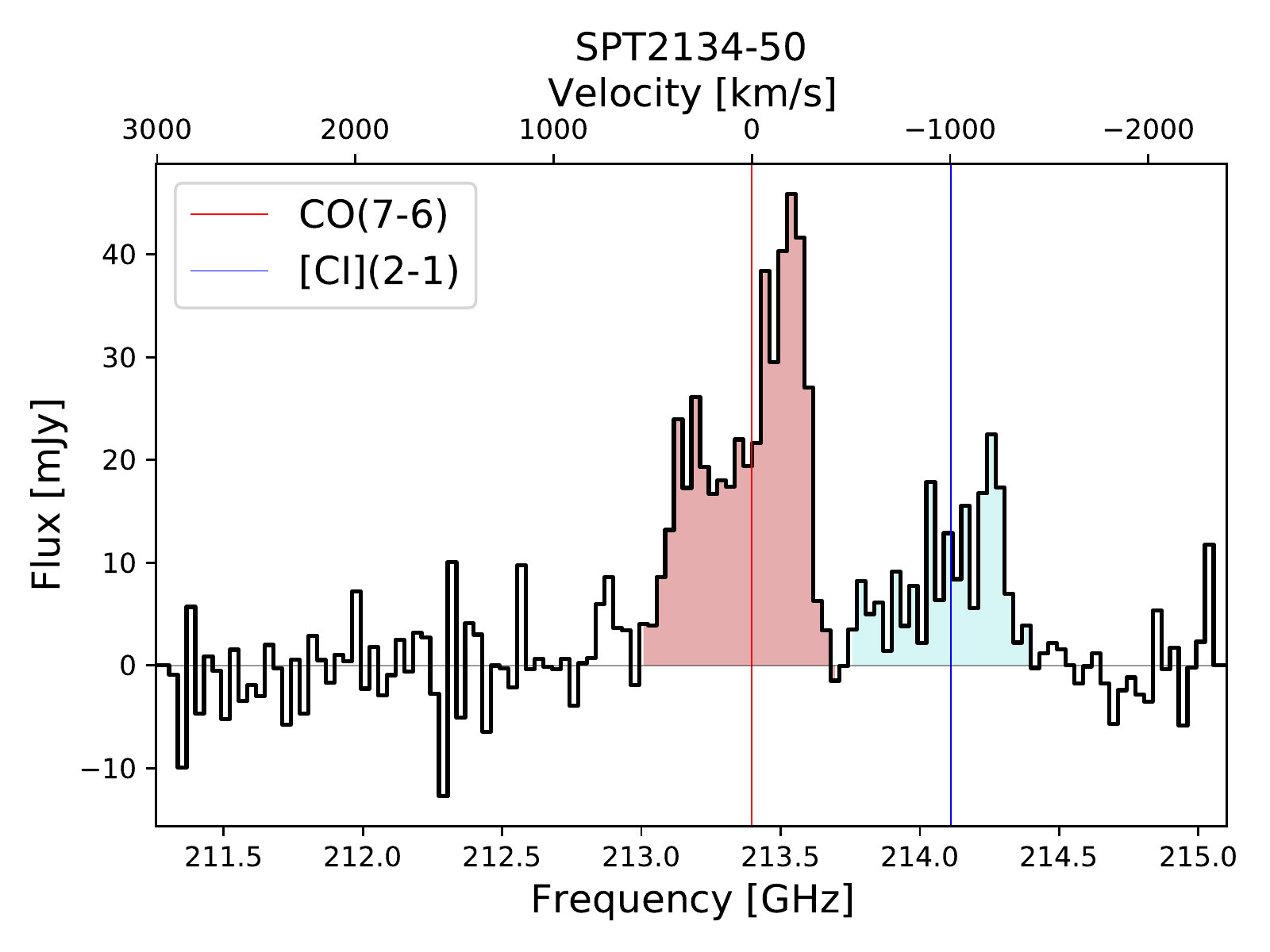} &  \includegraphics[width=6cm]{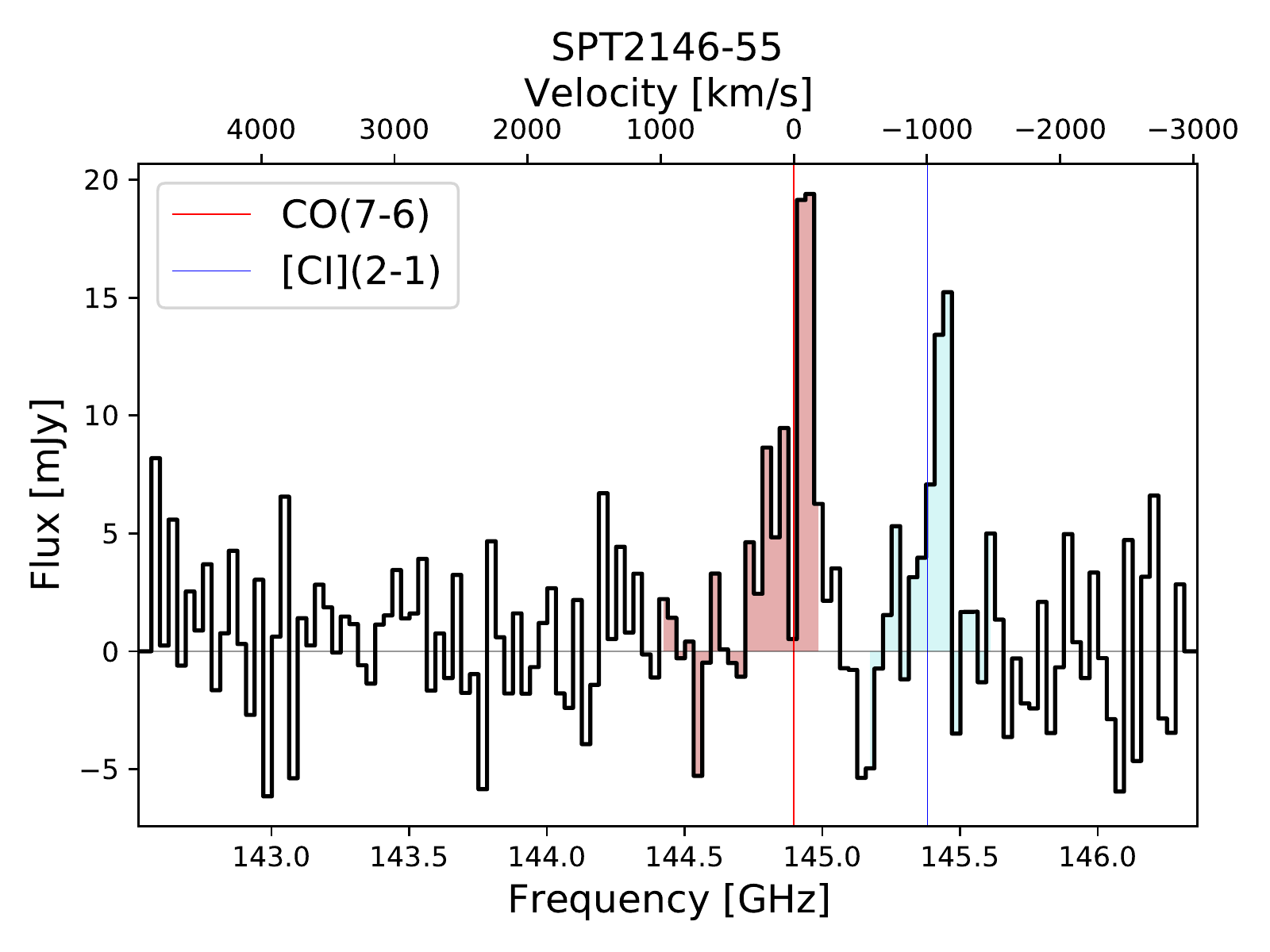} & \includegraphics[width=6cm]{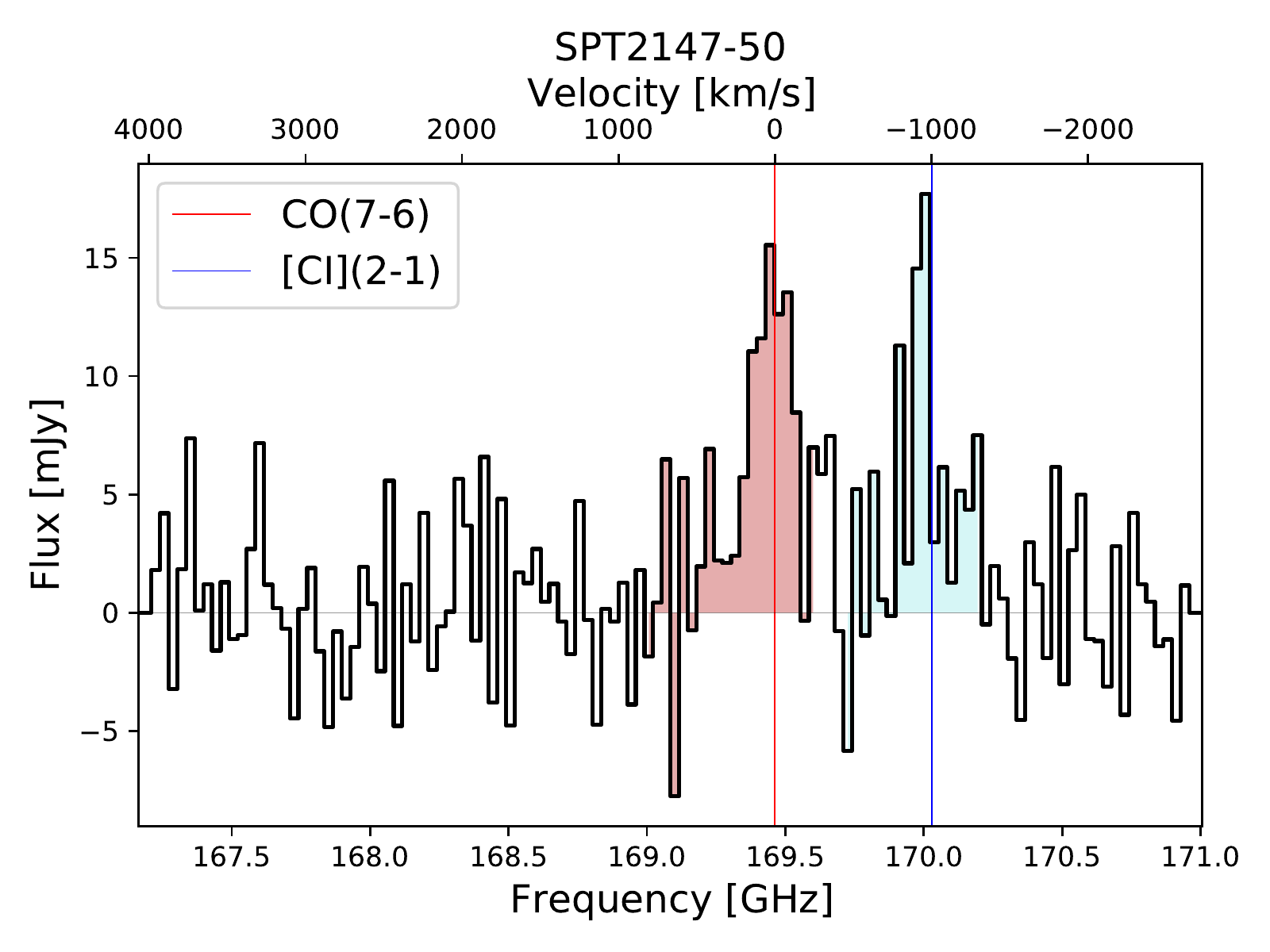} \\
\includegraphics[width=6cm]{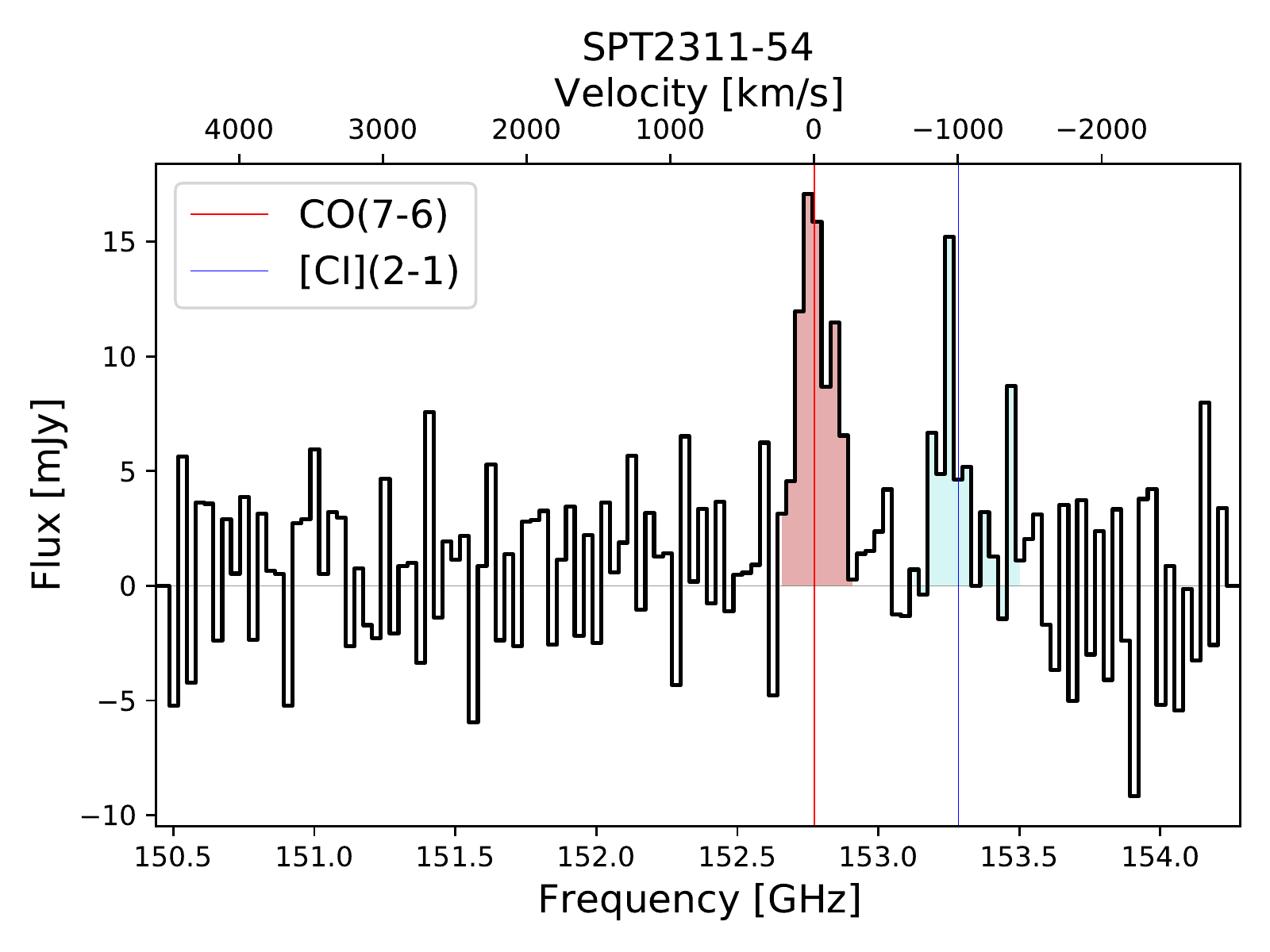} &  \includegraphics[width=6cm]{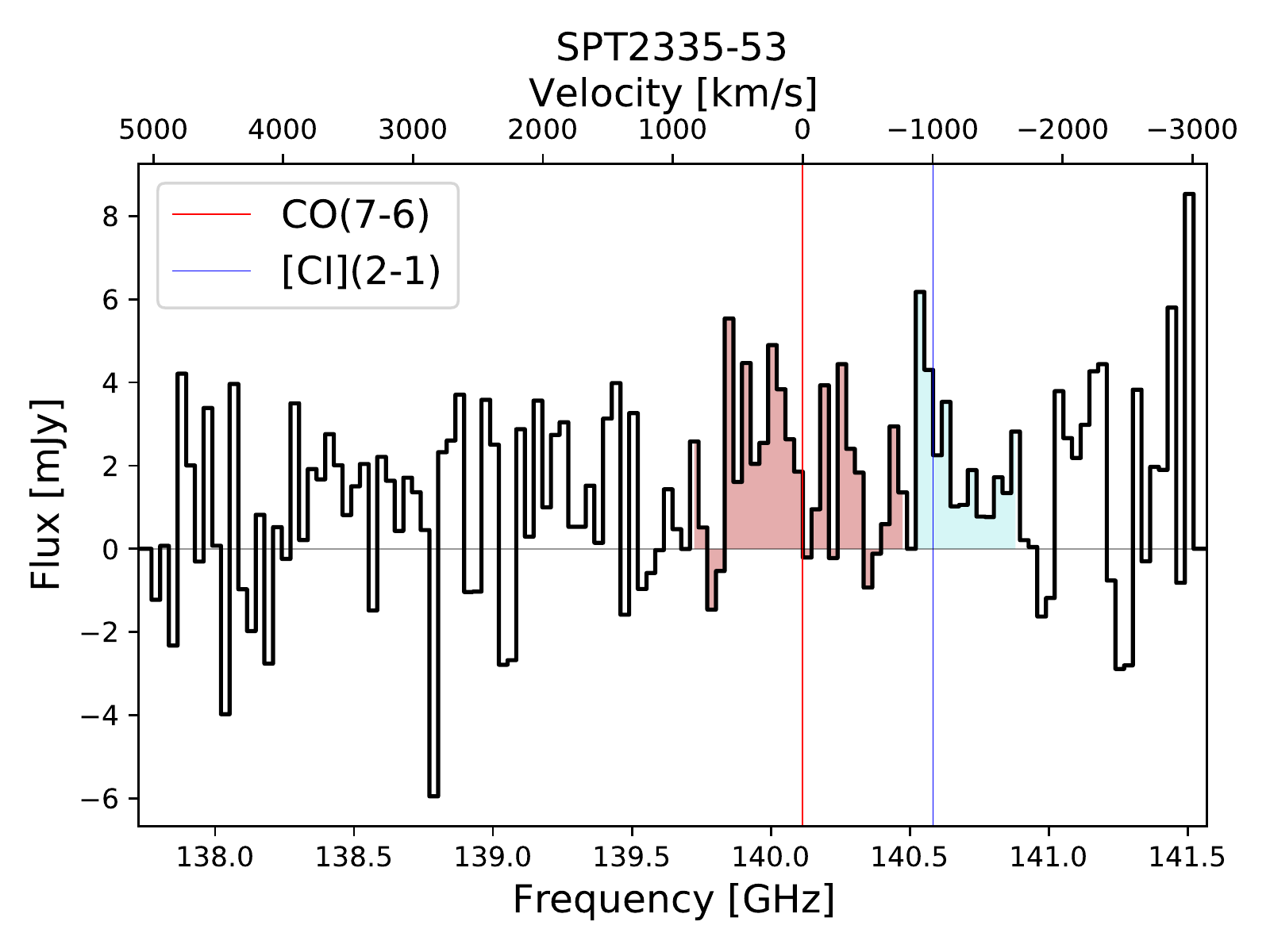} & \includegraphics[width=6cm]{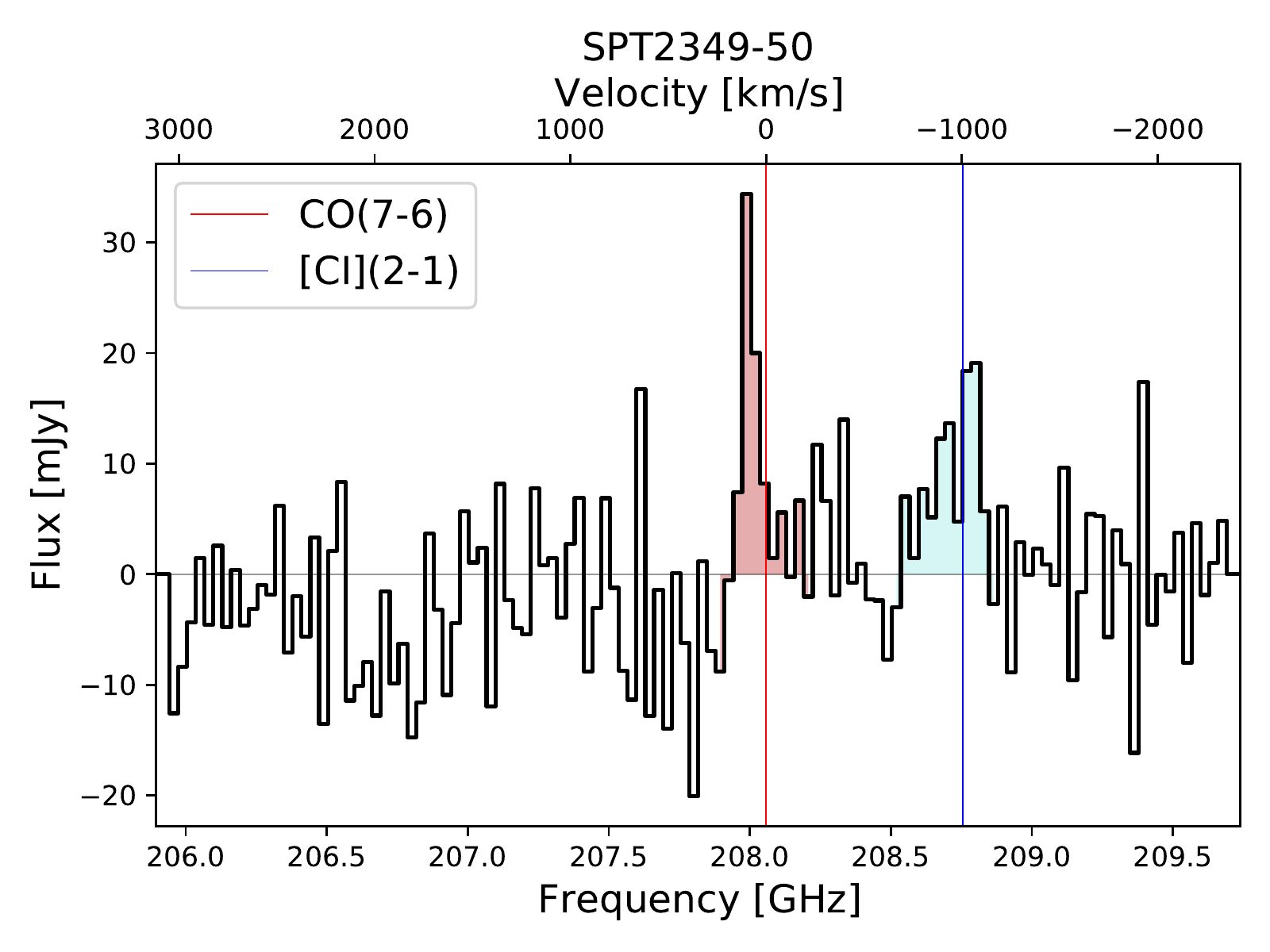} \\
\includegraphics[width=6cm]{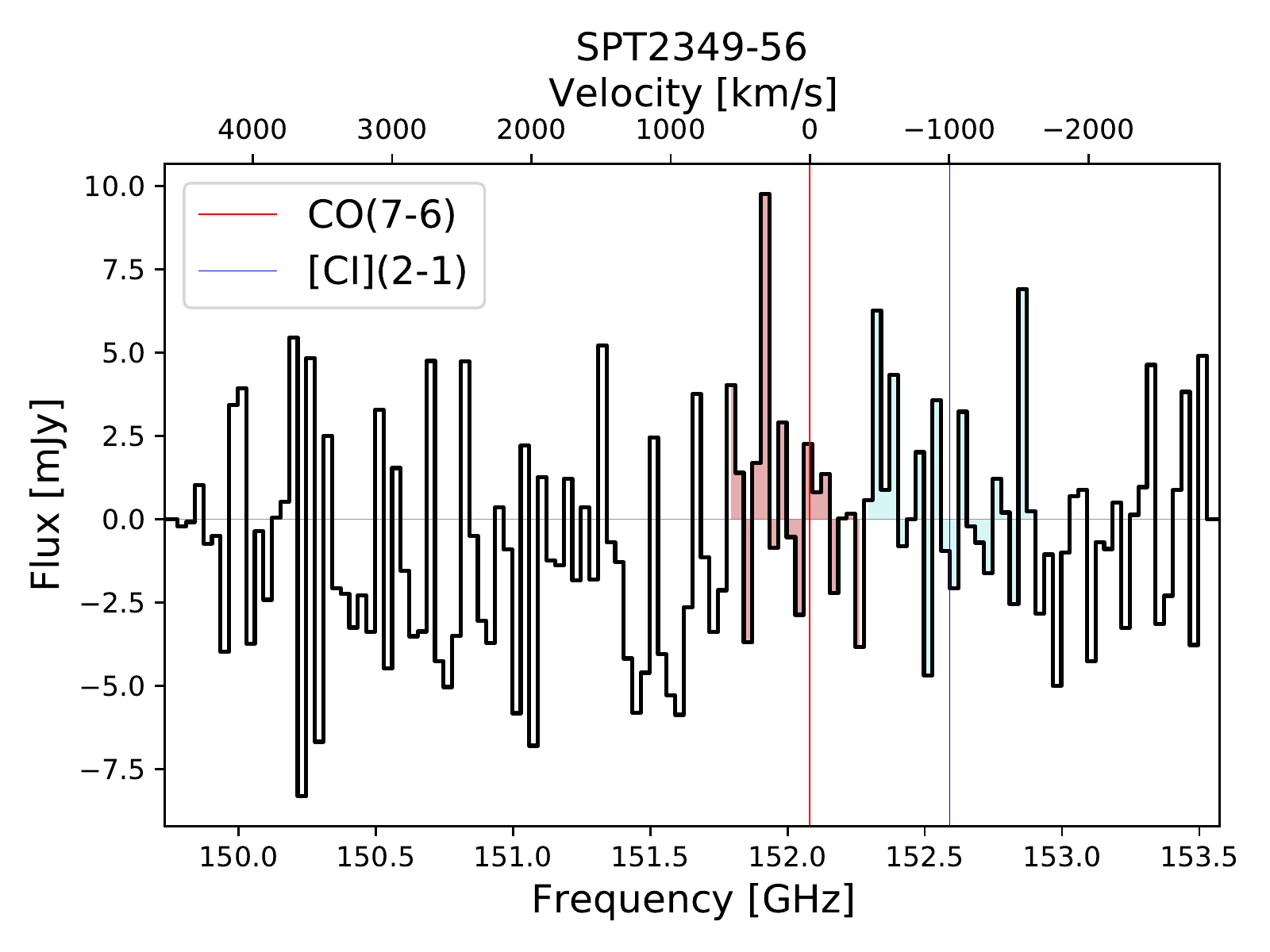} &  \includegraphics[width=6cm]{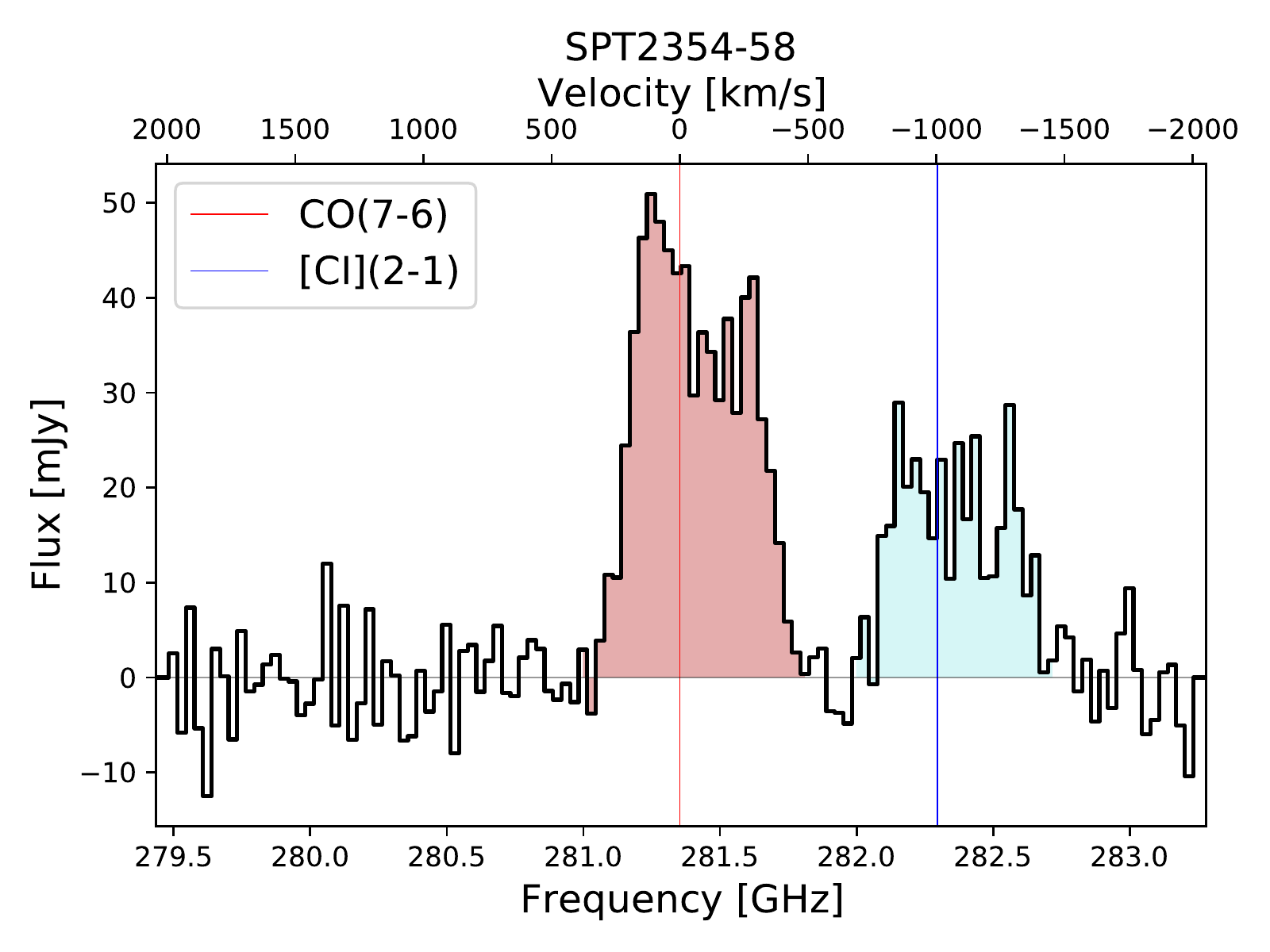} &  \\

\end{tabular}
\caption{\label{fig:ACA CI21 spectra} The [CI](2-1) and CO(7-6) spectra of our sample. The red and blue solid lines represent the CO(7-6) and [CI](2-1) frequencies based on the redshifts from \citet{Reuter20}. The velocity axis is centered on this CO(7-6) frequency. The red shaded region is the integration window of CO(7-6) line and the blue shaded region is the integration window of [CI](2-1) line. These windows are used to compute the integrated fluxes of the lines. }
\end{figure*}

\subsection{APEX [CI](2-1) sample and observations}\label{sec2.4}

In addition to our ACA sample, we also include a sample 9 DSFGs, for which we targeted the [CI](2-1) lines with APEX/SEPIA instrument (PIs: Béthermin and Strandet, project number: 097.A-0973). Six of the sources have tentative detections in the [CI](2-1) and/or CO(7-6) lines. For SPT0551-50, there was an ambiguity in the redshift, and thus APEX setup was not covering the [CI](2-1) observation window. The redshift has been constrained with the updated spectral scans presented in \citet{Reuter20}. We used the \texttt{CLASS} package of the \texttt{GILDAS} software\footnote{\url{https://www.iram.fr/IRAMFR/GILDAS/}} to reduce the data, and we used an automatic procedure described in Zhang et al., in prep., to assess the data quality and flag the bad data/baselines. The APEX/SEPIA spectra are presented in Fig.\,\ref{fig:APEX spectra}. 

Using an antenna gain conversion factor of 38 Jy K$^{-1}$, we obtain the APEX intensities. We then define a narrow integration window around the line peak and sum the intensities in this manually defined range. We estimate the noise by computing the standard deviation on the line-free channels multiplied by $\sqrt{\rm N}$, where N is the number of channels used to estimate the line flux.




\section{Data analysis}\label{sec3}

\subsection{Line imaging and performance}\label{sec:imaging and performance}

Initial calibrations of the data were done by the observatory using the standard ALMA pipeline based on the Common Astronomy Software Applications (\texttt{CASA}, \citealt{CASA}). We then image our data using the \texttt{tCLEAN} routine in \texttt{CASA}. All our sources are unresolved and thus we use a natural weighting function to optimise the S/N of our flux measurements. 

We make a first CLEANing using the \textit{cube} mode of  \texttt{tCLEAN}. From this first image, we can estimate the rms noise ($\sigma_{\rm rms}$) to determine the cleaning threshold. Additionally, we visualise the data to choose the relevant continuum model to use for its subtraction (constant or first-degree). We also check if the lines are clearly detected or if the data needs further re-binning. We then visually estimate the frequencies and the width of the line, which will be useful while subtracting the continuum emission. 

The next step is to obtain the line cubes, free of any continuum emission. For this, we use the \textit{uvcontsub} task of \texttt{CASA}. We can define the spectral windows and the frequency range containing our line emission, and the continuum model (constant or first-degree). This task then subtracts the continuum emission in the \textit{u-v} plane of our data cube and returns a continuum-free data cube.

Finally, we re-image the continuum-free line cubes with a cleaning threshold of 3\,$\sigma_{\rm noise}$, and rebin the data by a factor of 4 to further improve the S/N of each bin in our line images. From these line datacubes, we can estimate the noise level at the phase-centre by computing the standard deviation at every channel of the non-primary beam corrected line cubes, after masking the source. In Table\,\ref{tab:1_observation_details_ACA}, the mean sensitivity per channel ($\sigma_{\rm channel}$) is given for all our observations. The sensitivity for our sources varies from 1.6 -- 6.0 mJy, with an average sensitivity of 2.9\,mJy in a channel width of 0.031\,GHz.

\subsection{Spectrum extraction and Flux estimation}\label{sec:flux estimation}

We image the continuum emission of our sources by selecting the line-free spectral windows and using the multi-frequency synthesis mode (MFS, \citealt{Conway90}) of \texttt{tCLEAN} of CASA. This generates the continuum map of our sources. All our sources are gravitationally lensed and the lens models are presented in \citet{Spilker16}. Our sources are unresolved and well encompassed in the beam. We verify that the continuum position is in agreement with the source coordinates presented in \citet{Reuter20}. Assuming that the continuum and the line emission are co-spatial, we can extract the spectra at the position of the peak of continuum emission. We thus extract the spectrum from the line-cubes at this exact position of the continuum emission. The [CI](1-0) and [CI](2-1) spectra of our sample are presented in Fig.\,\ref{fig:ACA CI spectra} and Fig.\,\ref{fig:ACA CI21 spectra} respectively.



From the spectra, we can visually estimate the integration window of the line emission for our sources with sufficient signal of the line. This frequency range is then used as the integration range to obtain the integrated fluxes for each of the lines. In Fig.\,\ref{fig:ACA CI21 spectra}, the blue-shaded region represents the [CI](2-1) integration window, the red-shaded region represents the CO(7-6) integration window. The violet-shaded region in Fig.\,\ref{fig:ACA CI spectra} represents the [CI](1-0) integration window. For four of our sources, SPT0136-63, SPT0345-47, SPT2335-53 and SPT2349-56, the [CI](2-1) line is not visually detected, (i.e. no sufficient signal), and thus we use a width of 400\,km/s (approximately the median width of [CI](2-1) line of our sample) as a line integration width to estimate the upper-limits of the line fluxes. The [CI](2-1) line of two of our sources, SPT0150-59 and SPT0155-62 are blended with the CO(7-6) emission. In order to properly estimate the fluxes in this scenario, we estimate the width of the line by fitting the spectra using a two-Gaussian profile and estimating the width of each component (see Sect.\,\ref{App: Deblending}).

To estimate the integrated flux of these lines, we construct a moment-0 map using the \textit{immoments} function of CASA. The integration windows for these lines are represented in Fig.\,\ref{fig:ACA CI spectra} and \ref{fig:ACA CI21 spectra}. For the two sources with blended [CI](2-1) and CO(7-6) emission, we use the width estimated from the Gaussian fitting to use as the integration range in the moment-0 maps. From these moment-0 maps, we measure the integrated flux at the continuum source position. Since the source is unresolved, the flux is well-contained in a single synthesised beam. To estimate the uncertainties on the flux, we select a polygonal region near the source and measure the noise RMS in this region. The fluxes and their uncertainties for all the lines imaged for our sample is presented in Table\,\ref{tab:2_flux_catalogue_ACA}. 

We observed 8 sources in [CI](1-0), 6 of these sources were securely detected ($>5\,\sigma$) and 2 sources have tentative detection ($3>\sigma>5$). Of the 29-targets for [CI](2-1), 4 sources were undetected and are presented as $3\,\sigma$ upper-limits. 7 of the sources were tentatively detected ($3>\sigma>5$) and 18 sources were securely detected ($>5\,\sigma$). The non-detections are presented as $3\,\sigma$ upper-limits.

For the ancillary [CI](1-0) data, 7 of the sources are tentative detections (3 < $\sigma$ < 5) and 3 are non-detections ($<3\,\sigma$) from \citet{Bothwell17} and 4 sources with tentative detections, 1 secure detection ($>5\,\sigma$) and 6 non-detections from \citet{Reuter20}.

From these fluxes, we estimate the line luminosities using the formalism from \citet{Solomon05}
\begin{equation}\label{eq:line luminosity}
    \rm L_{\rm line} = 1.04 \times 10^{-3}\,S_\nu \Delta\upsilon\,D_L^2\,\nu_{\rm obs}\,\, [\rm L_{\odot}]
\end{equation}
where $S_\nu \Delta\upsilon$ is the integrated intensity in Jy km/s, obtained from the moment-0 maps, $D_L^2$ is the luminosity distance in Mpc and $\nu_{\rm obs}$ is the observed frequency of the line in GHz.


\begin{figure*}[h]
\centering

\begin{tabular}{cc}
\includegraphics[width=8cm]{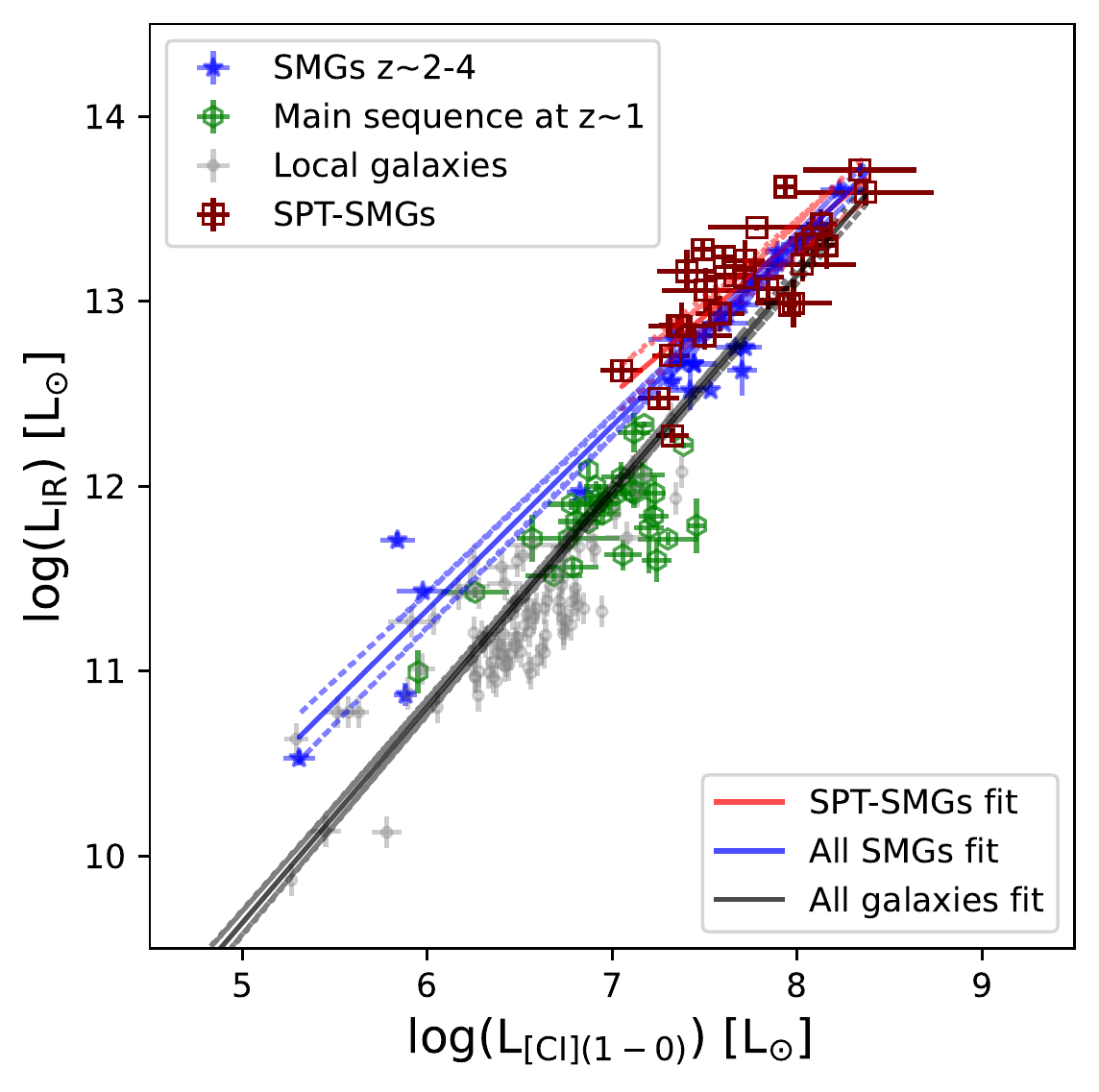} &  \includegraphics[width=8cm]{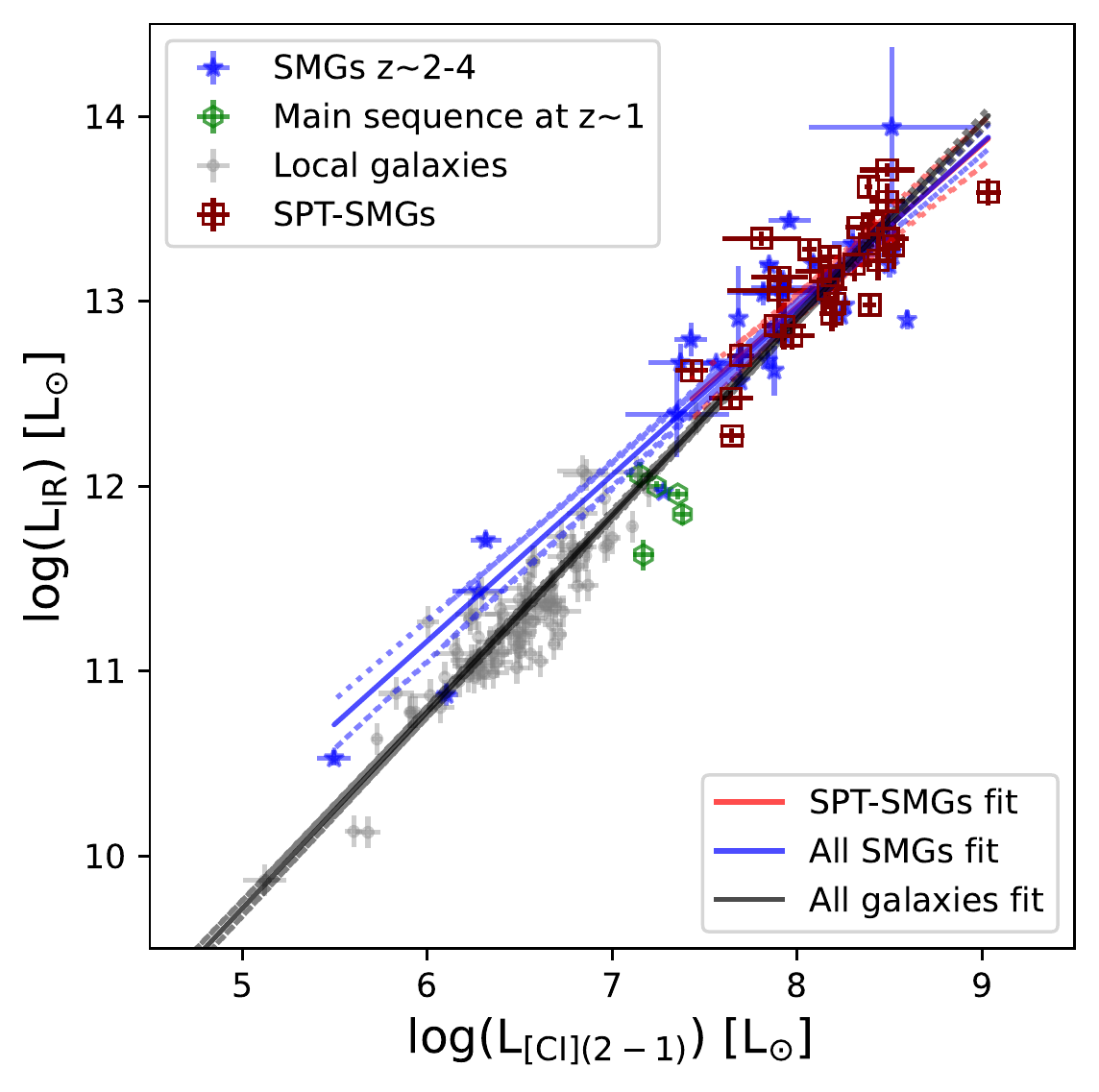} \\ 
\end{tabular}
\caption{\label{fig:LIR_LCI} IR luminosities versus [CI] line luminosities for our sample. The left panel shows the L$_{\rm IR}$ against L$_{[\rm CI](1-0)}$ for our sample and the right panel shows the L$_{\rm IR}$ against L$_{[\rm CI](2-1)}$ for our sample. Both the luminosities are corrected for magnification and are in the units of L$_{\odot}$. ACA-[CI] sample is represented by red squares. We compare our sample with the \citet{Valentino20} compilation. The blue stars represent the SMGs \citep{Walter11,Alaghband-Zadeh13,Bothwell17,Yang17,Andreani18,Canameras18,Nesvadba19,Dannerbauer19,Jin19}. The main-sequence galaxies at $z\sim$1 are represented by green hexagons \citep{Valentino18,Bourne19,Valentino20}. The local FTS samples of star-forming galaxies are represented by grey circles \citep{Veron-cetty10,Liu15,Kamenetzky14}. The fit for the entire compilation is represented as the grey solid line. The blue line represents the fit for the our sample combined with the literature SMGs and the red line represents the fit for our sample. The 1\,$\sigma$  limit of the fit represented as dashed lines. }
\end{figure*}


\begin{table*}
\centering
\caption{\label{tab:fitting LIR LCI}Fit parameters of the L$_{\rm IR}$ vs L$_{\rm [CI]}$ for all the samples in Fig.\,\ref{eq:line luminosity}.  }
\begin{tabular}{ccccc}
\hline
\hline
&&&&\\
Sample & Slope ($\beta$) & Intercept ($\alpha$) & Slope ($\beta$) & Intercept ($\alpha$) \\
&[CI](1-0)&[CI](1-0)&[CI](2-1)&[CI](2-1)\\
&dex&dex&dex&dex\\
&&&&\\
\hline
&&&&\\
SPT-SMGs & 0.90$\pm$0.17 & 6.14$\pm$1.28 & 0.88$\pm$0.14 & 5.94$\pm$1.11 \\
All SMGs & 1.00$\pm$0.05 & 5.34$\pm$0.41 & 0.90$\pm$0.05 & 5.77$\pm$0.42 \\
Local galaxies and main-sequence at $z\sim1$ & 0.98$\pm$0.04 & 4.95$\pm$0.23 & 1.02$\pm$0.03 & 4.63$\pm$0.19 \\
Full sample & 1.16$\pm$0.03 & 3.83$\pm$0.21 & 1.06$\pm$0.02 & 4.40$\pm$0.13 \\
\hline
\end{tabular}
\tablefoot{All the fit parameters are obtained from the \texttt{Linmix} linear-regression fitting module. The errors on the slopes and intercepts are obtained from the MCMC chain posteriors in the \texttt{Linmix} linear-regression fitting module.}
\end{table*}


\section{[CI] line properties}\label{CI line properties}

\subsection{Relation between [CI] and infrared luminosity}\label{CI IR luminosities}



Understanding the relation between the SFR and gas mass for different galaxy populations is crucial to decode the nature and evolution of galaxies. The relation between the L$_{\rm IR}$ and L$_{\rm [CI]}$ can be a proxy to the integrated Kennicutt-Schmidt (KS) law, as the bolometric L$_{\rm IR}$ between 8 and 1000\,$\mu$m (L$_{\rm IR}$) can provide a good estimate of the SFR \citep{Kennicutt98, Murphy11,Kennicutt12} and [CI] line can be used to estimate the total molecular gas \citep[e.g.][]{Weiss03,Papadopoulos04}. For our sample, we compute the [CI] luminosities using the relation in Eq.\,\ref{eq:line luminosity} with the integrated fluxes estimated from the moment-0 maps (see Sect.\,\ref{sec:flux estimation}). The L$_{\rm IR}$ for our sample are presented in \citet{Reuter20} and are based on the \textit{Herschel} and ground-based sub-mm photometry, providing a very reliable estimate of this quantity. We correct both the luminosities for their magnification obtained from the lens models presented in \citet{Spilker16} and assuming a median magnification of 5.5 for sources without reliable lens modelling.

In Fig.\,\ref{fig:LIR_LCI}, we compare the IR luminosity against the luminosities of both [CI] transitions for our sample. The left panel shows the L$_{\rm IR}$ versus the [CI](1-0) luminosity and the right panel presents the L$_{\rm IR}$ versus the [CI](2-1) luminosity. We fit the relation between the line and the IR luminosity using a linear regression model of \texttt{Linmix} package \citep{Kelly07}. The slopes and intercept of the fits for both the lines are tabulated in Table\,\ref{tab:fitting LIR LCI}. Additionally, we compare our sample with the SMGs, main sequence (MS) at $z\sim1$, and the local galaxy compilation presented in \citet{Valentino20} in both the IR-[CI] luminosity plots. We also fit a linear regression model for the SMGs combined with our sample (represented as all SMGs in Fig.\,\ref{fig:LIR_LCI}), for the local galaxies and the main-sequence galaxies, and for the entire galaxy population including our sample. 
For both the [CI]-lines, the IR versus [CI] luminosity relation exhibits a slight variation in slopes between the different populations shown in Fig.\,\ref{fig:LIR_LCI} and Table\,\ref{tab:fitting LIR LCI}. In the case of the L$_{\rm IR}$ versus L$_{\rm [CI](1-0)}$ relation, both our sample (SPT-SMGs, 0.90 $\pm$ 0.17) and our compilation of SMGs (1.00 $\pm$ 0.05) are compatible at 1\,$\sigma$ with a linear relation. However, our sample of SPT-SMGs span across a small range of luminosities, thereby making their slopes uncertain. The local and z$\sim$1 main-sequence galaxies also have a similar slope (0.98 $\pm$ 0.04) to the SPT-SMGs and the combined SMG sample, but the slope of the combined samples containing all the galaxies is higher (1.16$\pm$0.03). In the case of L$_{\rm IR}$ versus L$_{\rm [CI](2-1)}$ relation, the slopes are little flatter for SPT-SMGs (0.88$\pm$0.14) and all SMGs (0.9$\pm$0.05) compared to the linear slopes for the local galaxies (1.02$\pm$0.03) and the combined population (1.06$\pm$0.02). The slopes remain nearly linear across all populations.




We compare our observed slopes with the L$_{\rm FIR}$ or L$_{\rm IR}$ to L$^{\prime}_{\rm CO(1-0)}$ best-fit relation presented in \citet{Greve14}, \citet{Liu15} and \citet{Kamenetzky16}. Since CO (n$_{\rm crit} \rm CO(1-0) \sim 2.1 \times 10^3\,cm^{-3}$ and n$_{\rm crit} \rm CO(2-1) \sim 1.1 \times 10^4\,cm^{-3}$) and [CI] (n$_{\rm crit} \rm [CI](1-0) \sim 4.7 \times 10^2\,cm^{-3}$ and n$_{\rm crit} \rm [CI](2-1) \sim 1.2 \times 10^3\,cm^{-3}$) are both tracers of the cold gas, we could expect to find similar results. However, this is not trivial because [CI] traces lower density density gas than CO. In the case of \citet{Greve14}, they find a slope of 1.00 $\pm$ 0.05 for L$_{\rm IR}$ versus L$^{\prime}_{\rm CO(1-0)}$ and 1.05 $\pm$ 0.10 for L$_{\rm IR}$ versus L$^{\prime}_{\rm CO(2-1)}$ relation for a sample of local ULIRGs and $z>1$ DSFGs, which agrees with our slope of L$_{\rm IR}$ versus L$_{\rm [CI](1-0)}$ for the combined SMG sample and the sample of local galaxies and z$\sim$1 main-sequence galaxies. \citet{Kamenetzky16} found a slope of 1.27$\pm$0.04 for the L$_{\rm FIR}$ versus L$^{\prime}_{\rm CO(1-0)}$ relation in a sample of galaxies (AGN, main sequence, and ULIRGs). In the case of their subsample of ULIRGs, they found a lower slope of 1.15$\pm$0.09 than for the full sample as we found for [CI](1-0), but their slope is in 1.5\,$\sigma$ tension with unity.



The slope variations in relations between gas mass tracers ([CI](1-0) or low-J CO) and SFR tracers (L$_{\rm IR}$) for different populations such as starbursts and main-sequence galaxies, has been previously explored in e.g. \citet{Daddi10_sflaws, Genzel10}. \citet{Daddi10_sflaws} suggest that this could be due to the two populations having a different SFR-gas mass relation. Starburst are expected to have less gas for the same SFR and have thus a lower [CI] or CO luminosity. Since the low-L$_{\rm IR}$ galaxies are mainly main-sequence and the bright ones are more often starbursts \citep[e.g.][]{Sargent12}, this could lead to a steeper slope for the full sample containing both populations. 


In the case of L$_{\rm IR}$ versus L$_{\rm [CI](2-1)}$ relation, slopes of 0.90$\pm$0.05 for the combined SMGs and 0.88$\pm$0.14 for the SPT-SMGs are comparable to the mid-J CO versus L$_{\rm FIR}$ slopes (0.94 $\pm$ 0.11 for CO(3-2) for the ULIRG sample) found in \citet{Kamenetzky16}. We also find that the local and main-sequence galaxies have nearly the same [CI](1-0)/IR luminosity ratios, but lower [CI](2-1)/IR ratios compared to the SMGs and SPT-SMG sample. The difference in trends for starbursts and main-sequence, as seen in the case of [CI](1-0), may not be significant for [CI](2-1) due to the excitation of the CO and [CI] being higher for starbursts. For the higher energy transition, this compensates the lower gas content and the main-sequence and starburst have a similar relations. We cannot constrain the stellar-mass to SFR relation (position relative to the main sequence) for these galaxies due to the unavailability of stellar mass estimates based on the photometry. From the stellar mass estimates of \citet{Ma2015}, SPT-SMGs in thier analysis were found to be possibily representative of a starburst population. However, due to the sample size and the uncertainties in the stellar-mass estimates, it is difficult to disentangle whether these galaxies represent a starbursting population or an extended main-sequence population.




\subsection{[CI] excitation temperature}\label{exc temp}

The excitation temperature of [CI] can be estimated from the line luminosity ratio of the two [CI] transitions, $R_{CI} = \rm L^{\prime}_{[CI](2-1)}/L^{\prime}_{[CI](1-0)}$ for an optically thin scenario \citep{Stutzki97, Weiss03, Walter11}. The excitation temperature, $\rm T_{ex}$ is computed as :
\begin{equation*}
    \rm T_{ex} = \frac{38.8}{ln(\frac{2.11}{R_{CI}})}\,\,\,[K]. 
\end{equation*}

\begin{figure}
\centering

\includegraphics[width=9cm]{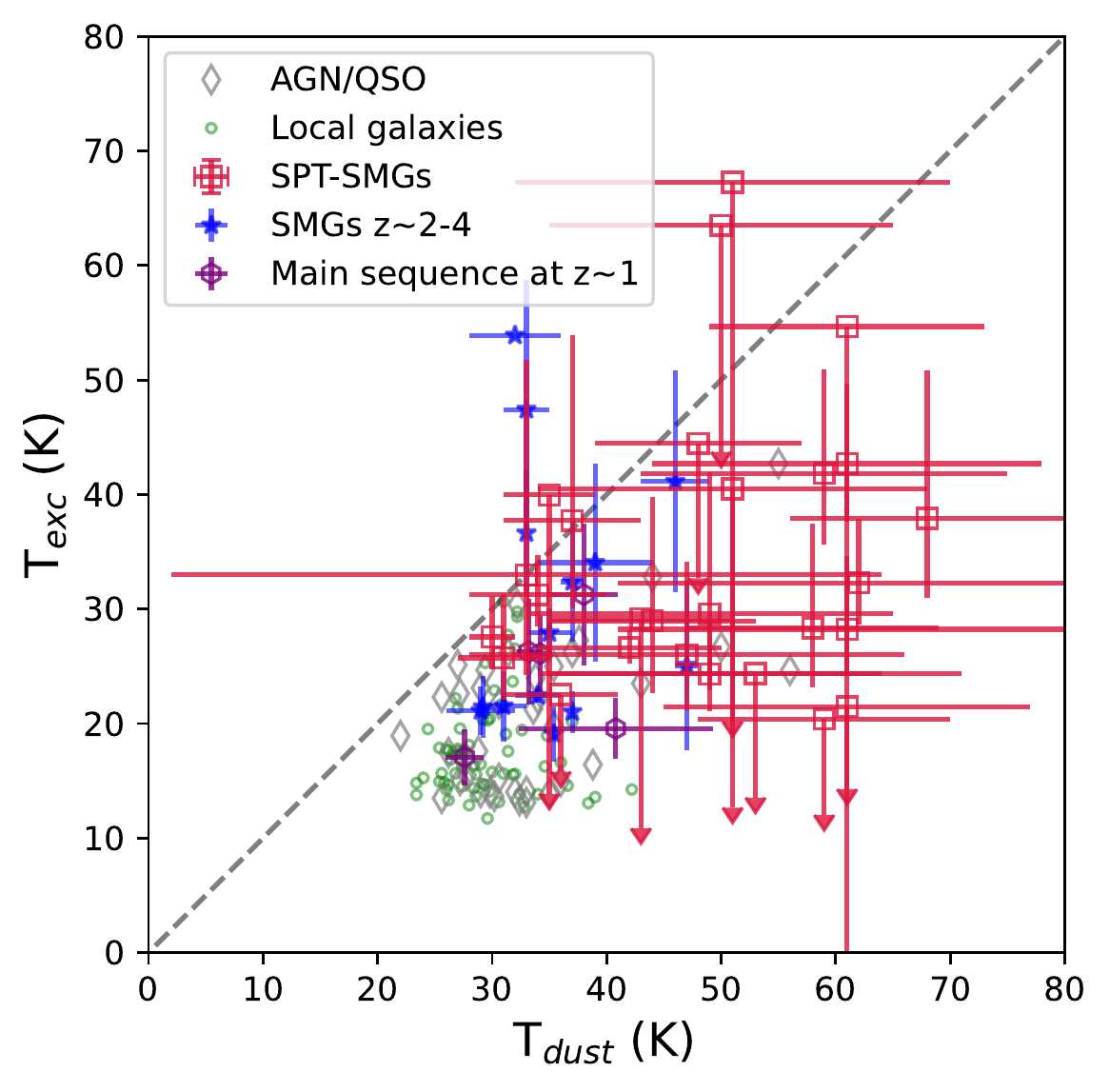}   

\caption{\label{fig:excittation_dust_temp}[CI] excitation versus dust temperature for our sample. Our sources are represented as red-squares. Sources with $1\,\sigma \rm to 3\,\sigma$ detections in either/both [CI] lines are represented as upper-limits. The dust temperatures for our sample are from the SED fits of \citet{Reuter20}. We also compare the excitation temperatures of SMGs (blue stars), main-sequence at $z\sim1$ (purple hexagons), local galaxies (green points) and AGNs (grey diamonds) from the \citet{Valentino20} compilation. The grey-dashed line corresponds to the 1:1 relation between the dust and [CI] excitation temperatures.}
\end{figure}

For the sources with at least tentatively detected in both the [CI] transitions, we can directly derive the excitation temperature and its uncertainties. To estimate the uncertainties of the excitation temperature, we make a simulation with the observed [CI]-fluxes. We generate a random Gaussian distribution with each of the [CI] fluxes with the width of the distribution derived from the error of the fluxes. For each of these points, we compute the excitation temperature and generate asymmetrical error bars on this distribution as the 16th and 84th percentile. Galaxies with only one [CI] transition tentatively detected are considered as upper-limits or lower-limits.

The excitation temperature of our sample varies from 17.7 to 64.2 K with a mean value of 34.5 $\pm$ 2.1 K. These temperatures are slightly higher than the mean temperatures reported in \citet{Valentino20} ($\big < \rm T_{\rm ex} \big> = 25.6 \pm 1.0$ K) and \citet{Nesvadba19} (T$_{\rm ex} = $ 21 -- 37 K) but comparable to the mean temperature reported in \citet{Walter11} ($\big <\rm  T_{\rm ex} \big> = 29.1 \pm 6.3$ K). Overall, our temperatures are slightly higher than the commonly adopted [CI] excitation temperature of T$_{ex} = $ 30 K \citep{Alaghband-Zadeh13, Bothwell17}, but compatible to a 2\,$\sigma$ level. 

\citet{Reuter20} presents the dust mass and dust temperatures for the SPT-SMGs by fitting the SED using a modified black-body law. They fix the Rayleigh-Jeans spectral slope, $\beta$ to 2, to mitigate degeneracies between redshift and dust temperature and reduce the number of free parameters. Additionally, they also define $\lambda_0$ as a function of T$_{\rm dust}$ using the empirical relation provided by \citet{Spilker16}.  Thus only 3 free parameters, photometric redshift, dust temperature ($\rm T_{\rm dust}$) and the overall SED normalisation are used in the SED fitting procedure.\footnote{We refer to \citet{Reuter20} for the details on the SED fitting and the estimation of dust temperatures and dust masses that will be used in this work.}

In Fig.\,\ref{fig:excittation_dust_temp}, we compare the [CI]-excitation temperature to the dust temperatures for our sample. In general, all our sources (with secure estimates of T$_{\rm ex}$), $\rm T_{\rm ex} \lesssim T_{\rm dust}$. These results agree with the SMG population from \citet{Nesvadba19} and with the compilation of galaxies presented in \citet{Valentino20}. Furthermore, we find similar results as \citet{Bothwell17}, $\rm T_{\rm kin} < T_{\rm dust}$ found for SPT-SMGs, assuming $\rm T_{\rm kin} = T_{\rm ex}$ at LTE. In Fig.\,\ref{fig:excittation_dust_temp}, we also plot the SMGs, main-sequence galaxies and local galaxies from the compilation sample presented in \citet{Valentino20}. Overall, our sources seem to have a higher dust and excitation temperature compared to the local galaxies and main-sequence, but comparable excitation temperatures to the SMGs.

The dust temperature could be used as a proxy to the excitation/gas temperature due to the gas to dust coupling, assuming an LTE condition \citep[e.g.][]{Narayanan11,CarilliWalter13, daCunha13}. From our results and the previous studies mentioned above, we find a lower gas excitation than the dust temperature systematically. This could be explained by the gas and dust not being in thermal equilibrium \citep{Canameras15, Nesvadba19}. Furthermore, the dust and [CI] emission may not be originating from the same phase of the ISM. The dust emission can be dominated by the central star-forming regions, whereas, [CI] could trace cooler, extended regions \citep[e.g.][]{Valentino18,Nesvadba19}. The SPT-SMGs have a higher dust temperature compared to the other populations in Fig.\,\ref{fig:excittation_dust_temp}. But, comparing the dust temperatures between different samples is potentially biased due to the differences in the parameterisation of the SED modelling and/or the different sampling of the dust SED. 

\citet{Cortzen20} compared the [CI] excitation temperatures of GN20, a starburst at $z\sim4$ to its dust temperature. They found a [CI] excitation temperature of $\rm T_{\rm ex} = 48.2^{+15.1}_{-9.2}\,K$ for the galaxy, compared to the dust temperature, $\rm T_{\rm dust} = 33 \pm 2\,K$ using an SED modelling with an optically thin regime. This strong outlier suggested that the dust could be optically thick thereby appearing deceptively cold. Hence, with the optically thick dust, they estimated $\rm T_{\rm dust} = 52 \pm 5\,K$, which was more agreeable with the excitation temperature. In our sample, we do not find candidates hinting at optically thick dust component, thus suggesting that GN20 could be an exception.

However, the recent works of \citet{Papadopoulos22} concluded that [CI] line excitation is subthermal and excitation temperatures cannot be derived with the line ratios assuming an LTE condition. Furthermore, they concluded that non-LTE [CI] line ratios could be used to constrain the dust SED models rather than estimating the [CI] excitation.




\begin{figure}
\centering

\includegraphics[width=8.5cm]{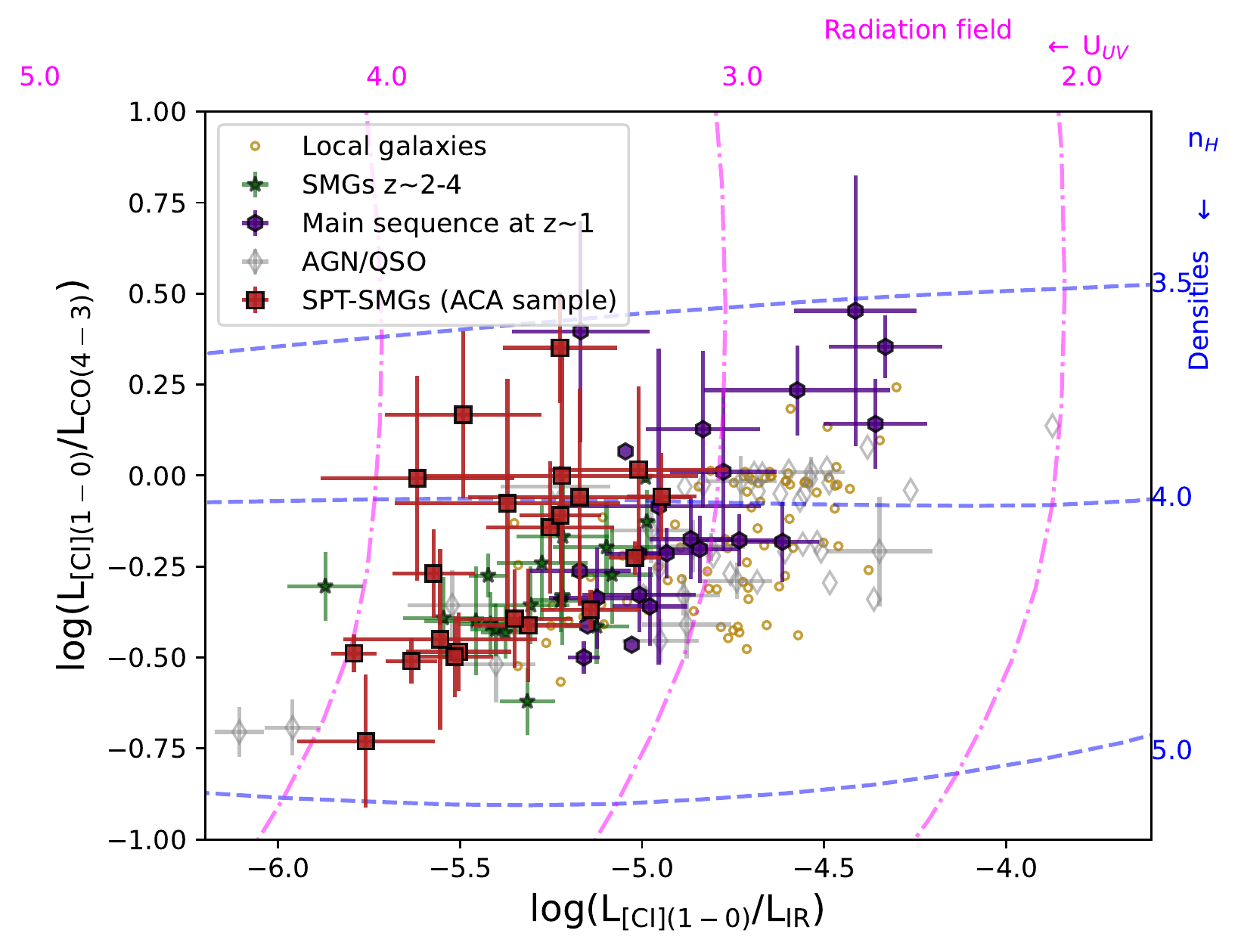}  \includegraphics[width=8.5cm]{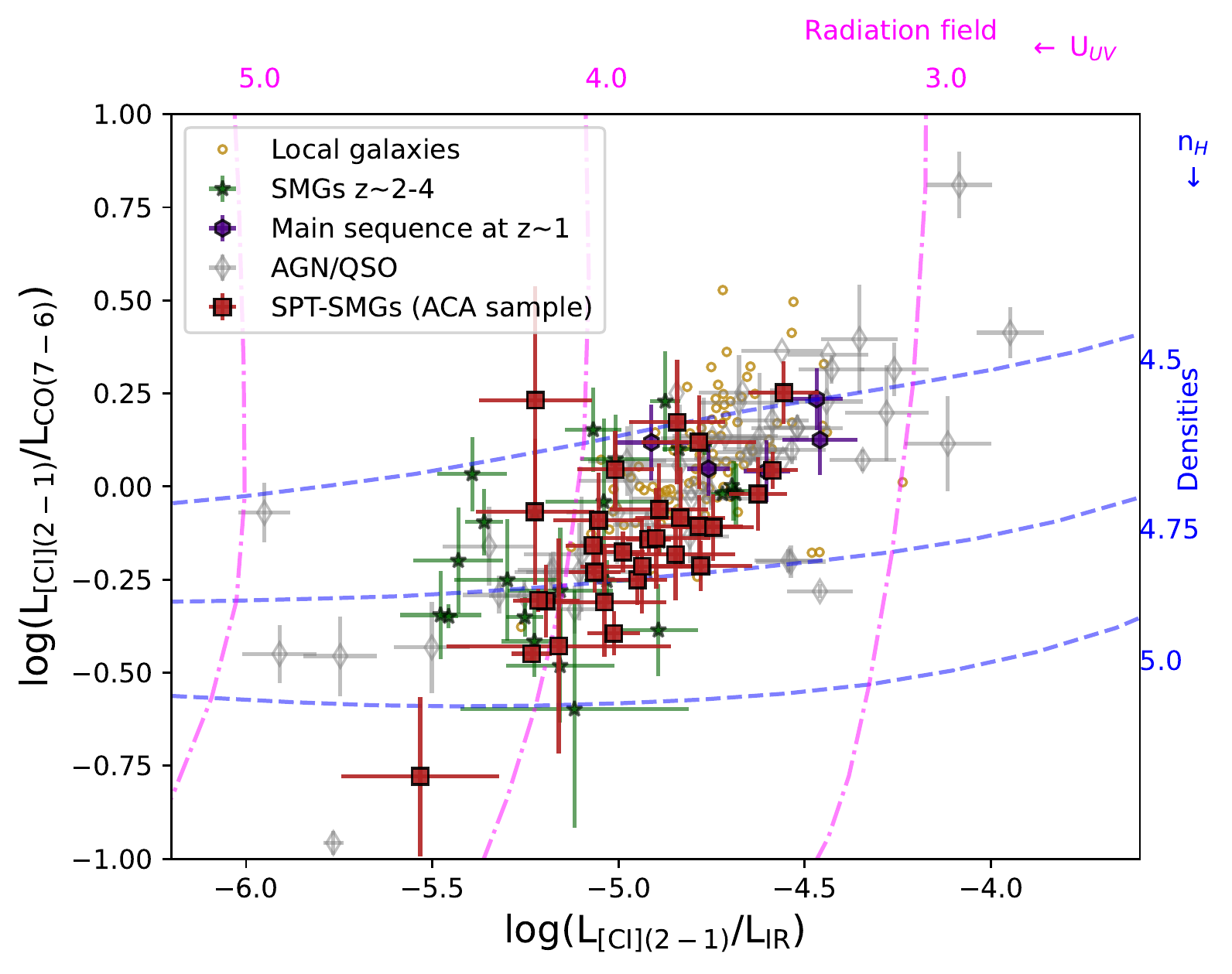} 

\caption{\label{fig:line ratio plots}  $\rm L_{\rm [CI](1-0)}/L_{\rm CO(4-3)}$ versus $\rm L_{\rm [CI](1-0)}/L_{\rm IR}$ in the top-panel and  $\rm L_{\rm [CI](2-1)}/L_{\rm CO(7-6)}$ versus $\rm L_{\rm [CI](2-1)}/L_{\rm IR}$ in the bottom panel for our sample. Our sources are represented as red-squares. The SMGs (darkgreen stars), Main-sequence at $z\sim1$ (indigo hexagons), local galaxies (golden empty circles) and AGNs (grey diamonds) from the \citet{Valentino20} compilation are plotted along with our sample. The iso-contours of density and radiation field intensity are plotted as grey dashed and grey dot-dashed lines, respectively. These isocontours are obtained from the PDR modelling of \citet{Kaufman06} from the PDR-toolbox \citep{pdrtoolbox}. The individual luminosities are not corrected for magnification.}
\end{figure}
\subsection{ISM diagnostics with line ratios}\label{ism diagnostics}

Line ratios can be used as a tracer of ISM properties of galaxies. A ratio of an extended gas tracer such as [CI] \citep{Papadopoulos04, Papadopoulos04b, Walter14, Bothwell17} and a relatively dense gas tracer such as high-J CO can be a proxy to the density of the ISM. We compute the $\rm L_{[CI](1-0)}/L_{CO(4-3)}$ and $\rm L_{[CI](2-1)}/L_{CO(7-6)}$ ratios for our sample. The choice of the CO transitions is due to the close spectral proximity of the lines to the [CI]-spectral frequencies at these redshifts. The CO(7-6) and [CI](2-1) lines can be imaged in the same spectral window at these redshifts. Similarly, the [CI](1-0) and CO(4-3) can also be observed in a single frequency range for some of these galaxies. We compare these [CI]-to-CO luminosity ratios with the [CI]-IR luminosity ratio. [CI] being an extended gas tracer can be a proxy to the gas mass and IR luminosity traces the star-formation, thus this ratio is a tracer of the radiation field of the ISM.


Figure\,\ref{fig:line ratio plots} shows the $\rm L_{[CI](1-0)}/L_{CO(4-3)}$ versus $\rm L_{[CI](1-0)}/L_{IR}$ in the top panel and $\rm L_{[CI](2-1)}/L_{CO(7-6)}$ versus $\rm L_{[CI](2-1)}/L_{IR}$ in the bottom panel. Additionally, we also compare the isocontours of the density and radiation field from the PDR modelling of \citet{Kaufman06} available from the PDR-toolbox \citep{pdrtoolbox} with our sample on Fig.\,\ref{fig:line ratio plots}. However, there are some caveats in comparing line ratios to PDR models. The line emission may not arise from the same region of the PDRs and they cannot fully reproduce the line excitation in starbursts without additional heating mechanisms \citep[e.g.][]{Papadopoulos12}. Additionally, a part of the IR emission can be arising from the HII regions, and hence may not be reproduced by PDRs. The IR emission arising from the warm dust may also be responsible for the heating mechanisms of the sources \citep{Valentino18}.



We compare our sample with the compilation presented in \citet{Valentino20}, consisting of SMGs, main-sequence galaxies at $z\sim1$ and local galaxies. In the $\rm L_{[CI](1-0)}/L_{CO(4-3)}$ versus $\rm L_{[CI](1-0)}/L_{IR}$ plot (top panel, Fig.\,\ref{fig:line ratio plots}), our sample has a lower $\rm L_{[CI](1-0)}/L_{IR}$ ratio compared to the local galaxies and the main-sequence at $z\sim1$ with an overlap. In terms of the $\rm L_{[CI](1-0)}/L_{CO(4-3)}$ ratios, they have similar values as main-sequence galaxies and the SMGs.

While comparing the $\rm L_{[CI](2-1)}/L_{CO(7-6)}$ versus $\rm L_{[CI](2-1)}/L_{IR}$ (bottom panel, Fig.\,\ref{fig:line ratio plots}), our sample has comparable $\rm L_{[CI](2-1)}/L_{IR}$ with the SMGs, main-sequence and local galaxies. In terms of the $\rm L_{[CI](2-1)}/L_{CO(7-6)}$ ratio, they are in the lower-end in comparison to the main-sequence and the local galaxies. The difference in trends between populations for the [CI]/CO ratio could arise due to the variation in CO-SLEDs at high-J CO lines for the different populations, but not in the mid-J CO lines \citep[e.g.][]{Casey14,Liu15,Daddi15,Yang17,Canameras18,Valentino20}. Furthermore, the galaxies with higher densities also have a stronger radiation field intensity.

To summarise, the differences in line ratios between the populations when compared to the PDR models suggest that our sample and the SMGs have comparable densities as the main-sequence and local galaxies, but lower radiation field intensities despite an overlap. In both these plots, our sources are diverse, with  radiation field intensities ranging from $10^{2.5}$ to $10^{4.5}$ Habing and densities ranging from 3.5 to 5.0 $\rm cm^{-3}$. However, all the ratios in Fig.\,\ref{fig:line ratio plots} are not corrected for magnification assuming the contribution of differential magnification might not be very significant ($<$ 24$\%$\,found in the analysis of SPT-SMGs from \citealt{Gururajan22}).

\begin{figure*}[ht]
\centering

\begin{tabular}{cc}
\includegraphics[width=8.5cm]{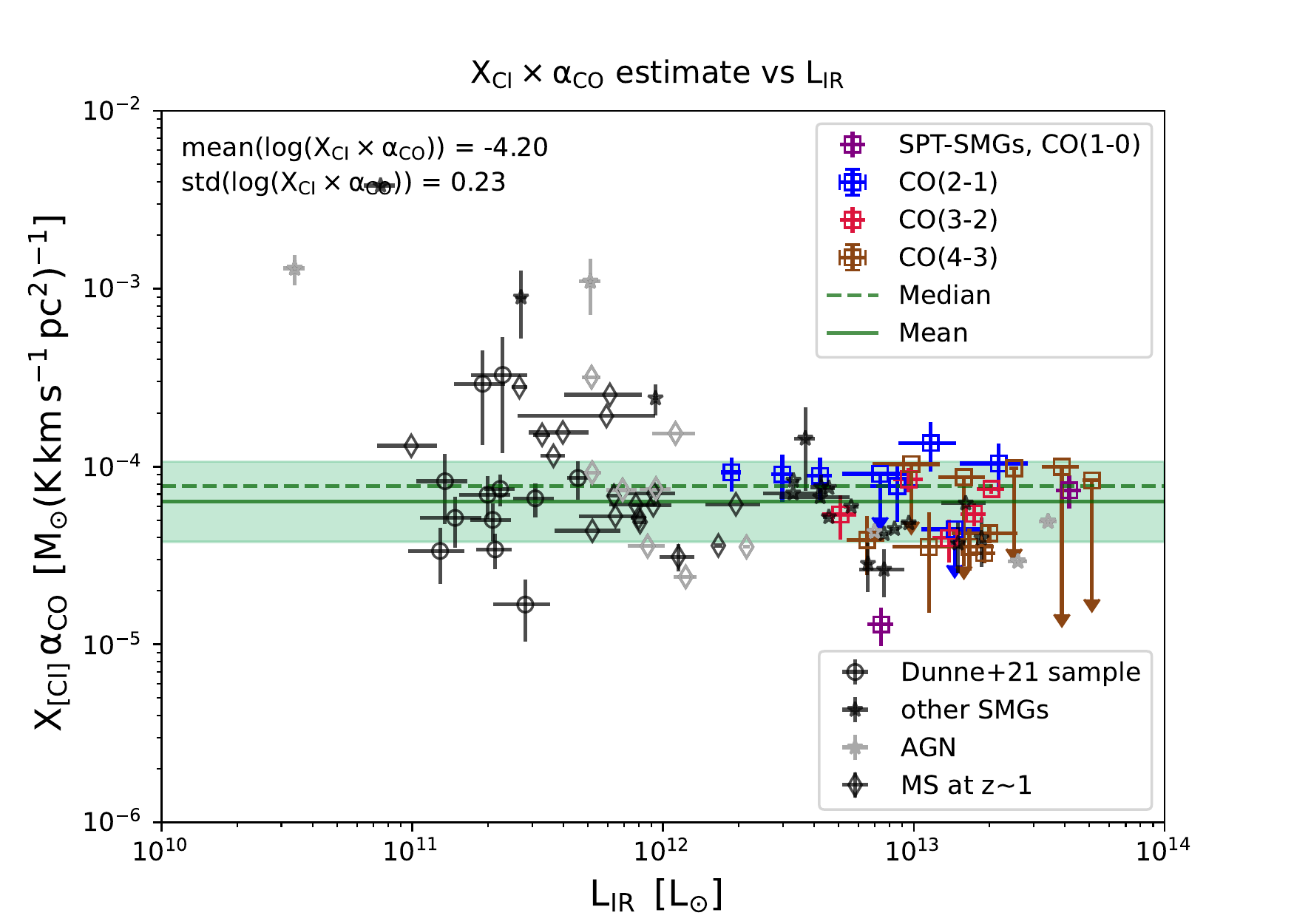} & 
\includegraphics[width=8.5cm]{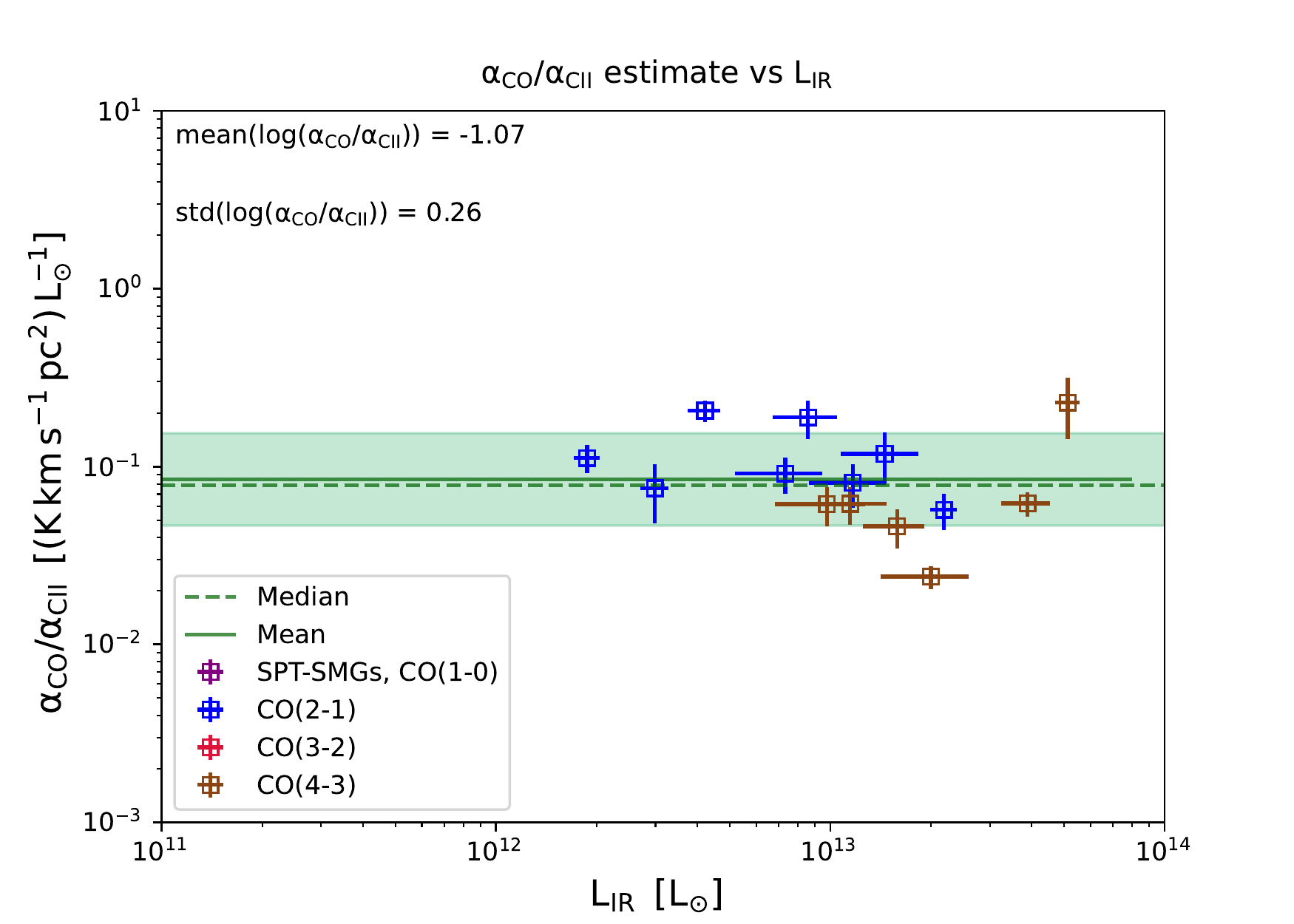}\\
\includegraphics[width=8.5cm]{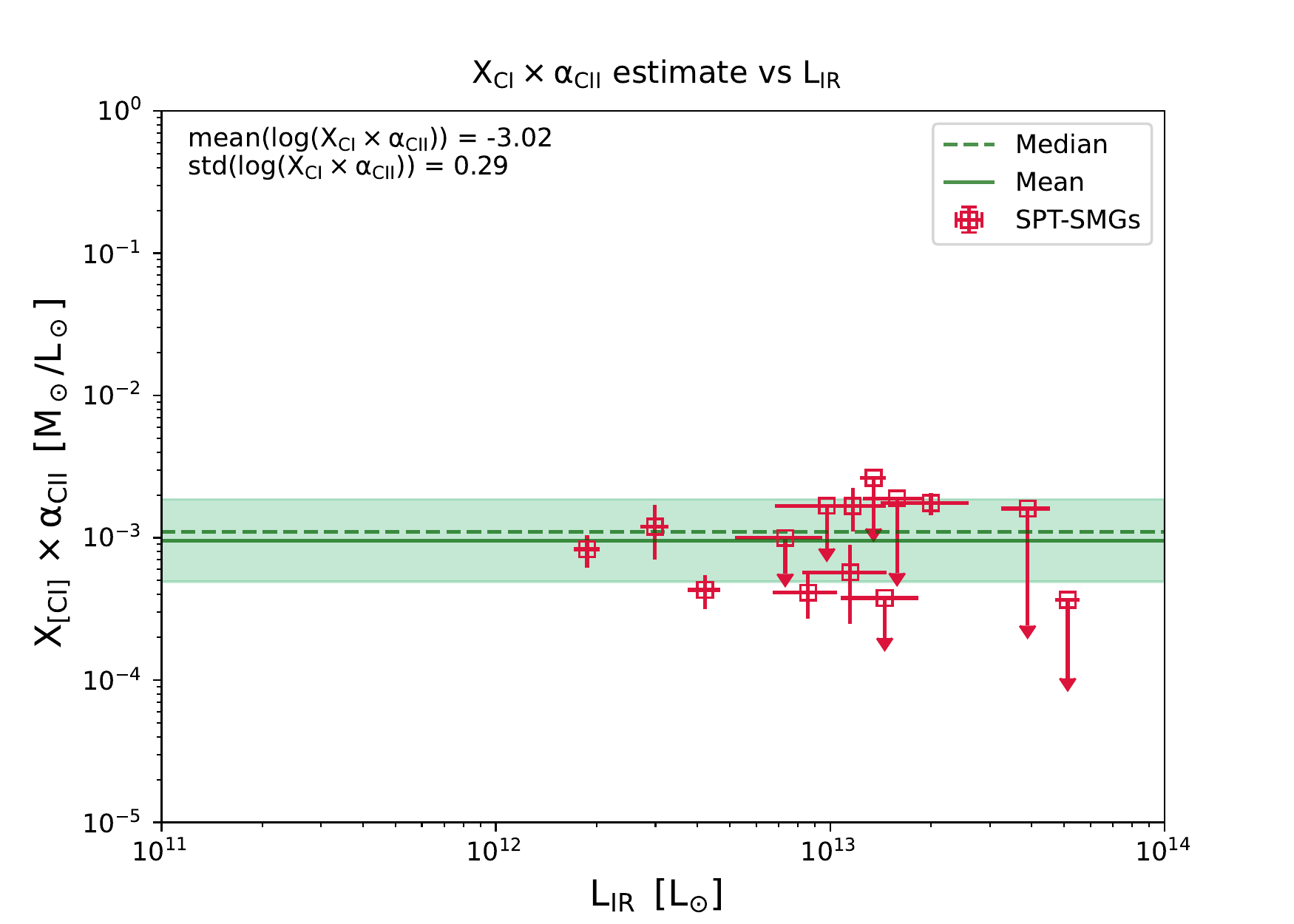} &
\includegraphics[width=8.5cm]{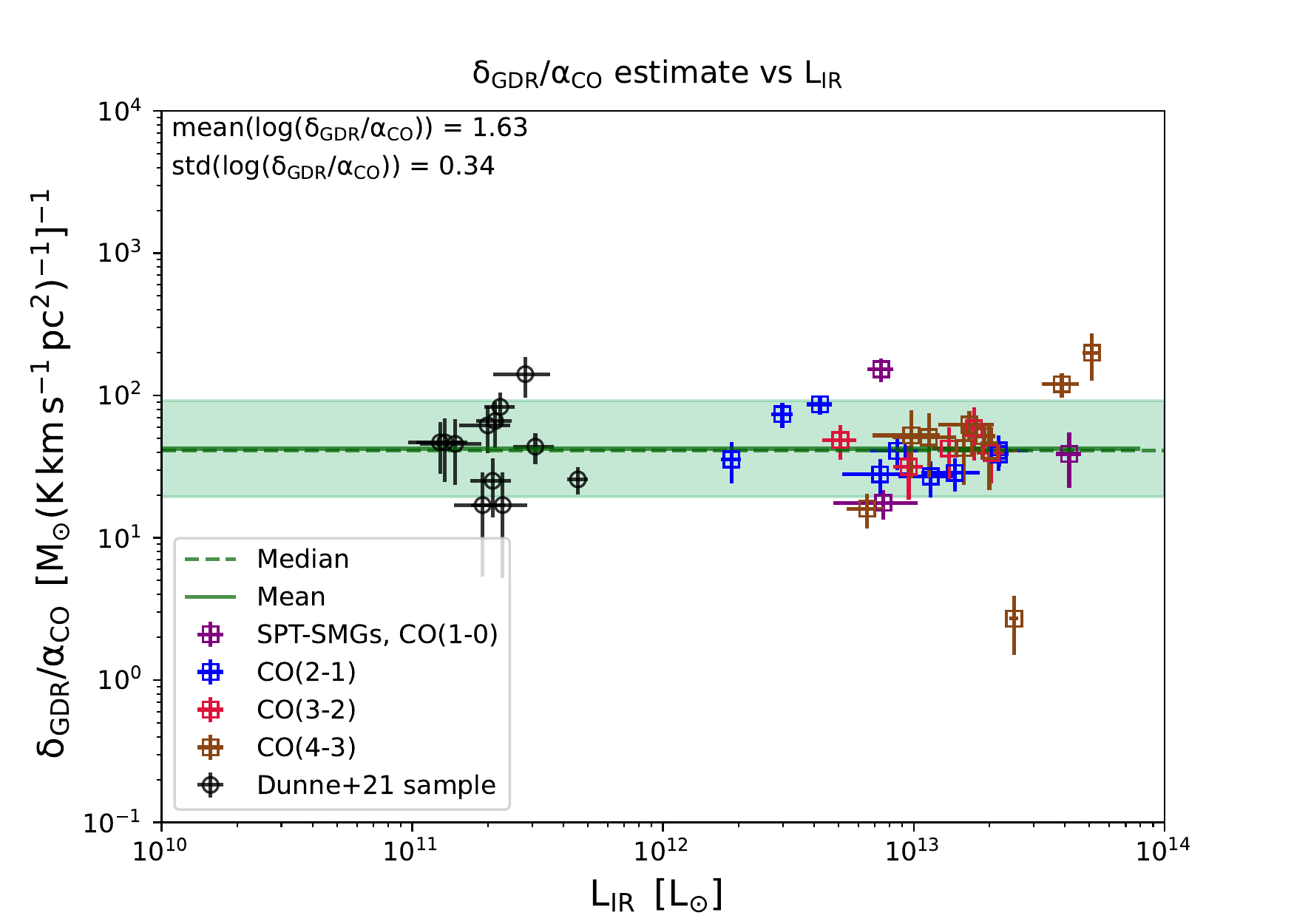} \\
\includegraphics[width=8.5cm]{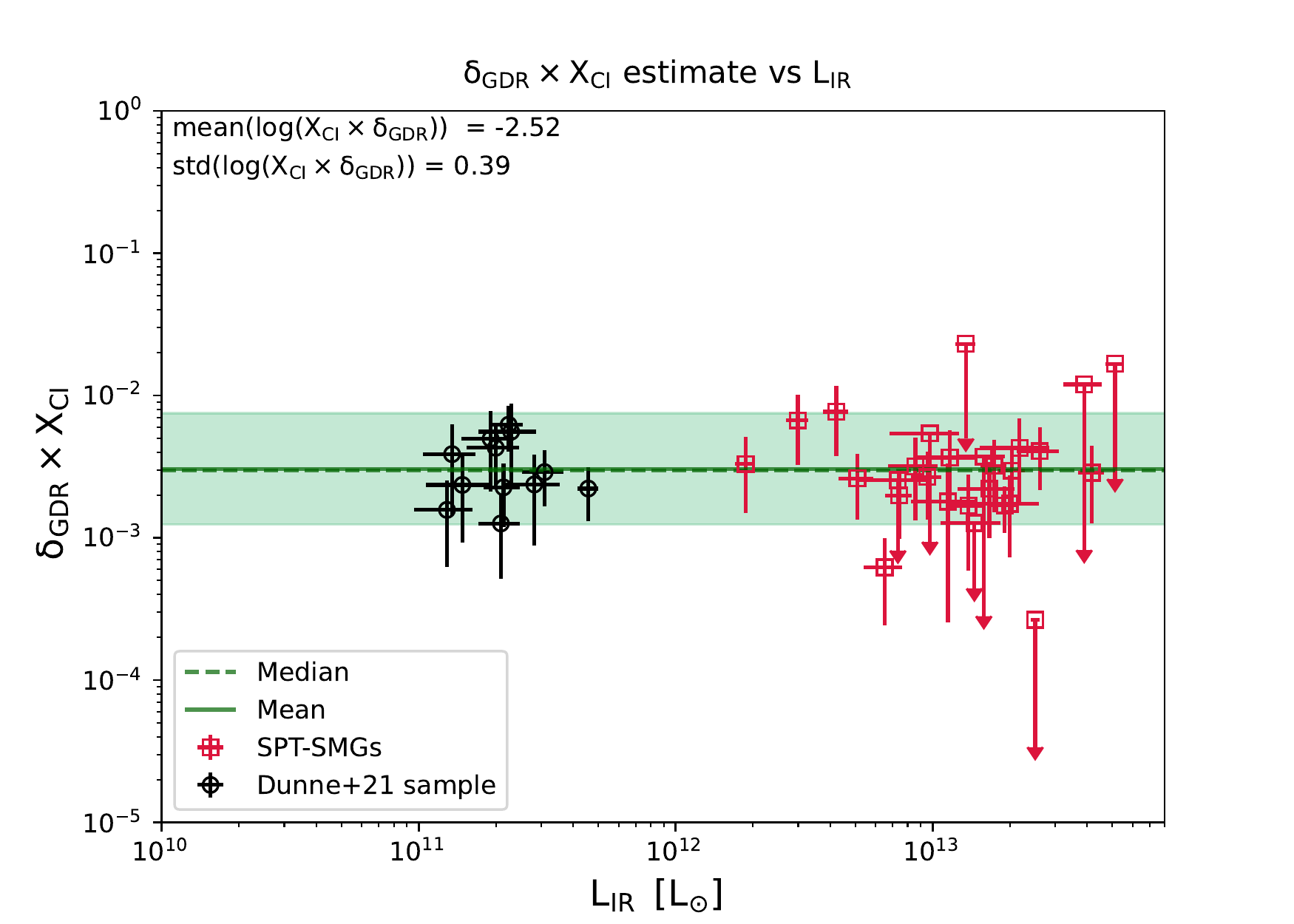} &
\includegraphics[width=8.5cm]{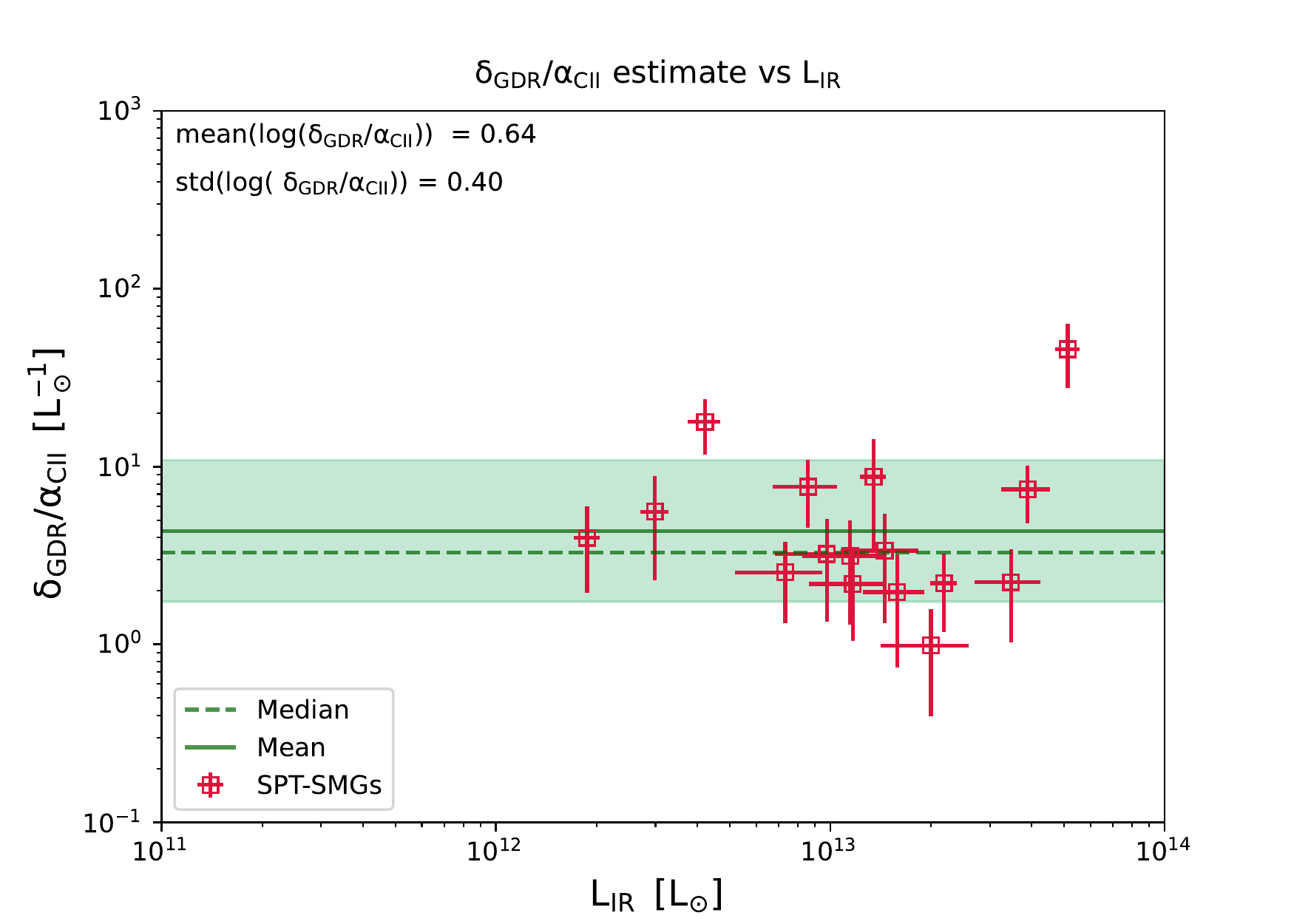} \\

\end{tabular}
\caption{\label{fig:mass comparisions} The \XCI $\times$ \aCO, \XCI $\times$ \aCII, and \XCI $\times$ \gdr (left column) and \aCO / \aCII , \gdr / \aCO, and \gdr / \aCII (right column) as a function of L$_{\rm IR}$ for our sample. Our sample is represented in squares, color-coded by the CO transition used to estimate the CO-based gas mass, CO(1-0) transition in purple, CO(2-1) in blue, CO(3-2) in red and CO(4-3) in brown in top and center rows. We also compare our sample with the sample presented in \citealt{Dunne21} - black circles, SMGs in black stars, main-sequence at $z\sim1$ in black diamonds and AGNs in grey diamonds from the \citet{Valentino20} compilation. The mean and the median y values are shown as green solid and dashed lines respectively for the left column. The green-shaded region represents the 1\,$\sigma$ region around the sample mean. }
\end{figure*}
\section{Comparison of [CI] with other gas tracers}\label{sec5}

\subsection{Estimation of the gas mass}\label{gas mass estimation}


We compare the ability of [CI] as an estimator of molecular gas mass of the galaxies against traditionally used gas-mass tracers such CO(1-0) emission line, the dust mass, and the [CII] line emission. We estimate the molecular gas mass using the four tracers to test their agreement/disagreement with each other.

\subsubsection{[CI]-based gas masses}

One of the methods of estimating of [CI]-based molecular gas mass was introduced by \citet{Papadopoulos04b} as :
\begin{equation}\label{eq:CIgasmass}
    \rm M(H_2)^{[CI]} = 1375.8\times10^{-12}\, \frac{D_L^2\,S_{\nu}\Delta\upsilon}{(1+z)\,A_{10}\,Q_{10}\,X_{CI}} \,\,[M_{\odot}], 
\end{equation}

where, $\rm D_L$ is the luminosity distance to the galaxy in Mpc, $\rm A_{10} = 0.793 \times 10^{-7}\,s^{-1}$ is the Einstein coefficient, $\rm S_{\nu}\Delta\upsilon$ is the integrated intensity in Jy\,km/s and $z$ is the redshift. The above equation can be re-written in terms of [CI] luminosity, $\rm L_{\rm [CI]}\,[L_\odot]$ with $\rm A_{10} = 0.793 \times 10^{-7}\,s^{-1}$ and $\rm \nu_{\rm rest, [CI]} = 492.16\,GHz$ as :

\begin{equation}\label{eq:CIgasmass_new}
    \rm M(H_2)^{[CI]} = 3.39\times10^{-2}\, \frac{\rm L_{\rm [CI]}}{Q_{10}\,X_{CI}} \,\,[M_{\odot}], 
\end{equation}

The main uncertainties in the gas mass estimated from [CI] arise from assumptions of the [CI]-excitation factor/partition function, $\rm Q_{10}$, and the [CI]/H$_{2}$ abundance ratio, $\rm X_{CI}$. 

The [CI] excitation factor depends on the excitation conditions in the gas such as the temperature and the critical densities. The excitation factor of level $\rm J = 1$ to $\rm J = 0$ can be derived as: \footnote{Please refer to the Appendix A of \citet{Papadopoulos04a} to obtain an extended in-depth derivation of the excitation factor. In particular, refer to equations, A8 and A15.}

\begin{equation}
    \rm Q_{10} = \frac{3e^{-T_1/T_{ex}}}{1+3e^{-T_1/T_{ex}}+5e^{-T_2/T_{ex}}},
\end{equation}

where T$_1 = 23.6$ K and T$_2 = 62.5$ K are the excitation energy levels of atomic carbon and T$_{\rm ex}$ is the excitation temperature of the ISM. Since we have estimated the excitation temperature for most of our sources, we can constrain the gas mass without assuming a value for the [CI] excitation factor. Our sample has a median $\rm Q_{10}$ value of 0.45 $\pm$ 0.01, which is in agreement with the median value of 0.43 for the sample presented in \citet{Dunne21}, and the values used in \citet{Papadopoulos04a,Alaghband-Zadeh13,Nesvadba19}. These values are lower than the Q$_{10}$ value of 0.6 assumed in \citet{Bothwell17}, thereby giving us a higher mass estimate. Thus, in our sample, the only unknown for the [CI]-based gas mass estimates would be the [CI]/H$_{2}$ abundance ratio, $\rm X_{CI}$.



\subsubsection{CO-based gas masses}
The total molecular gas mass from CO(1-0) line observations can be estimated using the following relation :
\begin{equation}\label{eq:COgasmass}
    \rm M(H_2)^{CO} = \alpha_{CO}\,L^{\prime}_{CO(1-0)}\,\,[M_\odot],
\end{equation}
where $\rm L^{\prime}_{CO(1-0)}$ is the line luminosity of CO(1-0) in K km/s pc$^{2}$ and $\alpha_{\rm CO}$ is the CO-to-H$_2$ conversion factor with the units, $\rm M_\odot [K\,km/s\, pc^2]^{-1}$. 

The main uncertainty for the CO-based molecular gas mass estimates arises from the $\alpha_{\rm CO}$ factor. In addition, since the CO(1-0) transition cannot be observed for most of these galaxies, we use CO-SLEDs to derive the CO(1-0) fluxes from the other observed low-J and mid-J transitions. 
3 of our sources have CO(1-0) and 8 of our sources have CO(2-1) line fluxes from the ATCA observations presented in \citet{Aravena16}. We also have CO(3-2) and CO(4-3) fluxes from the ALMA band-3 spectral scans \citep{Reuter20}. To compute the gas mass, we use the lowest-J transition available for our sources, up to the CO(4-3) transition. The line luminosity ratios to obtain the CO(1-0) luminosities are from \citet{Spilker14,Harrington21} for the SMGs. We use R$_{21} = 0.88 \pm 0.07$, R$_{32} = 0.69 \pm 0.12$ and R$_{43} = 0.52 \pm 0.14$. 

The $\alpha_{\rm CO}$ factor is debated in the literature and can vary from 0.8 for ULIRGs or starburst-like environment \citep[e.g.][]{DownesSolomon98} to higher value of $\sim 4.4$ for Milky Way-like galaxies \citep[e.g.][]{Solomon87, Strong96, Abdo10} and have also been independently estimated for SMGs \citep[e.g.][]{Spilker15, Calistro-Rivera18}. The choice of $\alpha_{\rm CO}$ could thus give a gas mass varying nearly by a factor of 5. This factor, therefore, contributes to the main uncertainty for the gas mass estimated from CO luminosity.

\subsubsection{Dust-based gas masses}


The gas mass can be estimated from the dust mass assuming a gas-to-dust ratio, 
\begin{equation}\label{eq:dustgasmass}
    \rm M(H_2)^{dust} = \delta_{\rm GDR}\,M_{\rm dust}\,\,[M_\odot],
\end{equation}
where $\rm M_{\rm dust}$ is the dust mass of the galaxy in $\rm M_\odot$ units and $\delta_{\rm GDR}$ is the assumed gas-to-dust ratio. To have a good estimate of the dust mass, we need to model the SED of the galaxy, which in turn depends on certain factors. Constraining the peak of the SED from space-based observations (e.g. \textit{Herschel}) along with a good sampling of the Rayleigh-Jeans tail from millimetric ground-based observations can help improve the SED fitting. Additionally, constraining the dust emissivity index as a function of the dust continuum can reduce the number of free parameters in the SED fitting and provide better constraint on the dust mass. The gas-to-dust ratio is known to vary with metallicity \citep[e.g.][]{Magdis11,Leroy11,Popping22a,Popping22b}. The assumed gas-to-dust ratio is therefore the main factor contributing to the uncertainties in the dust mass derived gas mass. 

\subsubsection{[CII]-based gas masses}

\citet{Zanella18} proposed the following calibration relation to estimate the [CII]-based molecular gas mass : 
\begin{equation}\label{eq:CIIgasmass}
    \rm M(H_2)^{[CII]} = \alpha_{[CII]}\,L_{[CII]}\,\,[M_\odot],
\end{equation}
where \aCII is the [CII]-to-H$_2$ conversion factor with the units $\rm M_\odot \,L_\odot^{-1}$ and L$_{[\rm CII]}$ is the line luminosity in $\rm L_\odot$. They reported a median value of \aCII $\sim 31 \rm M_\odot/L_\odot$ and a median absolute deviation of 0.2 dex. \citep{Zanella18} also found that \aCII remains unaffected by metallicity or star-formation modes. This is in contrast to the other three tracers we consider where the main limitations arise from the conversion factors such as \aCO, \XCI and \gdr.

\subsection{Cross-calibration of the various gas conversion factors}\label{cross calib}

From our four gas mass tracers, we have four unknown factors. Since the absolute value of all these tracers depend on various factors and are highly debated, we cross-calibrate these tracers to test the agreement between them. Assuming that the gas mass estimated from each of these tracers are equivalent, we can obtain the following relations : 

For $\alpha_{\rm CO}$ and $\rm X_{CI}$, from equation Eq.\,\ref{eq:CIgasmass_new} and Eq.\,\ref{eq:COgasmass}:
\begin{equation*}
    \rm M(H_2)^{CO} = M(H_2)^{[CI]} 
\end{equation*}
\begin{equation}\label{eq:XCIalphaCO}
    \rm \Rightarrow \frac{3.39\times10^{-2}\,L_{\rm [CI]}}{Q_{10}\,L^\prime_{CO(1-0)}} = X_{CI}\times\alpha_{CO}
\end{equation}

For $\delta_{\rm GDR}$ and $\rm X_{CI}$, from equation Eq.\,\ref{eq:CIgasmass_new} and Eq.\,\ref{eq:dustgasmass} :
\begin{equation*}
    \rm M(H_2)^{dust} = M(H_2)^{[CI]} 
\end{equation*}

\begin{equation}
    \rm \Rightarrow \frac{3.39\times10^{-2}\,L_{\rm [CI]}}{Q_{10}\,M_{dust}} = X_{CI}\times\delta_{GDR}
\end{equation}

and for $\delta_{\rm GDR}$ and $\alpha_{\rm CO}$, from equation Eq.\,\ref{eq:dustgasmass} and Eq.\,\ref{eq:COgasmass}:

\begin{equation*}
    \rm M(H_2)^{dust} = M(H_2)^{CO}
\end{equation*}
\begin{equation}
    \rm \Rightarrow \frac{L^\prime_{CO(1-0)}}{M_{dust}} = \frac{\delta_{GDR}}{\alpha_{CO}}
\end{equation}
For $\alpha_{\rm [CII]}$ and $\rm X_{CI}$, from equation Eq.\,\ref{eq:CIgasmass_new} and Eq.\,\ref{eq:CIIgasmass}:
\begin{equation*}
    \rm M(H_2)^{[CII]} = M(H_2)^{[CI]} 
\end{equation*}
\begin{equation}\label{eq:XCIalphaCII}
    \rm \Rightarrow \frac{3.39\times10^{-2}\,L_{\rm [CI]}}{Q_{10}\,L_{[CII]}} = X_{CI}\times\alpha_{[CII]}
\end{equation}

For $\delta_{\rm GDR}$ and $\alpha_{\rm [CII]}$ , from equation Eq.\,\ref{eq:CIIgasmass} and Eq.\,\ref{eq:dustgasmass} :
\begin{equation*}
    \rm M(H_2)^{dust} = M(H_2)^{[CII]} 
\end{equation*}

\begin{equation}\label{eq:gdralphaCII}
    \rm \Rightarrow \frac{L_{[CII]}}{M_{dust}} = \frac{\delta_{GDR}}{\alpha_{[CII]}}
\end{equation}

and for $\alpha_{\rm [CII]}$ and $\alpha_{\rm CO}$, from equation Eq.\,\ref{eq:CIIgasmass} and Eq.\,\ref{eq:COgasmass}:

\begin{equation*}
    \rm M(H_2)^{[CII]} = M(H_2)^{CO}
\end{equation*}
\begin{equation}\label{eq:alphaCOalphaCII}
    \rm \Rightarrow \frac{L_{[CII]}}{L^\prime_{CO(1-0)}} = \frac{\alpha_{CO}}{\alpha_{[CII]}}
\end{equation}
These above equations only provide a ratio/product of the unknowns, $\rm \alpha_{CO}$, $\rm X_{CI}$, $\rm \delta_{GDR}$ and \aCII in terms of all the other observables. In this way, we do not have to assume an absolute value for any quantity since all four unknowns have their limitations. We can obtain a mean value for the cross-calibration of our sample and test its agreement with different values found in the literature.

In Fig.\,\ref{fig:mass comparisions}, we plot the six ratios as a function of L$_{\rm IR}$. We do not correct the fluxes or dust mass for magnification as we compute the ratio, although there might be small biases due to differential magnification. For the plots comparing the $\alpha_{\rm CO}$, we further indicate the CO transition used to compute the $\rm L^\prime_{CO(1-0)}$ to further check if there is an agreement between the different transitions. 

In the left column of Fig.\,\ref{fig:mass comparisions}, we compare the \XCI $\times$ \aCO, \XCI $\times$ \aCII, and \XCI $\times$ \gdr against the L$_{\rm IR}$. In the right column of Fig.\,\ref{fig:mass comparisions}, we compare \aCO / \aCII , \gdr / \aCO, and \gdr / \aCII against the L$_{\rm IR}$. We also include the sample of nearby ($z\sim0.3$) galaxies presented in \citet{Dunne21} and the compilation of SMGs, main-sequence and local galaxies presented in \citet{Valentino20} when the data are available. 

We compute the internal scatter of the sample and probe any underlying trends with \texttt{Linmix} \citep{Kelly07}. It is Bayesian linear regression fitting model, which can fit data with errors along with the upper-limits. The intercept, the slope, and the internal scatter of the sample from the fit are summarized in Table\,\ref{tab:results_cross_calib_linmix}.

Comparing the dependence of these tracers on L$_{\rm IR}$, the slope is compatible with 0 and thus we can compute directly a constant conversion factor. We can also see that the scatter is small, the maximum scatter being 0.4\,dex for \gdr/\aCII.  The four estimators are thus remarkably consistent for this population. In the case of $\rm X_{CI} \times \alpha_{CO}$ versus L$_{\rm IR}$ and \aCO/\aCII versus L$_{\rm IR}$ there is a good agreement with our estimates using the different CO lines. 

On comparing the \XCI $\times $ \aCO relation for our sources with the literature, we find a similar ratio as the sample presented in \citet{Dunne21}. Some of the sources from the \citet{Valentino20} have a higher value of $\rm X_{CI} \times \alpha_{CO}$, and most of these extreme sources are dominated by AGNs. In the case of $\delta_{GDR} /  \alpha_{CO}$ and  $\rm X_{CI} \times \delta_{GDR}$ versus L$_{\rm IR}$ we see a tight correlation of these tracers and a good agreement with the \citet{Dunne21} sample. In general, we do not see any trend of these tracers with L$_{\rm IR}$ for our sample, and all the tracers are in reasonably tight correlation with less than 0.41 dex scatter.

In Fig.\,\ref{fig:mass comparisions_temperature}, we compare these tracers against the dust temperature. In the $\rm X_{CI} \times \alpha_{CO}$ versus T$_{\rm dust}$, we do not see a any trend with temperature, and for $\rm X_{CI} \times \delta_{GDR}$, \aCO/\aCII and \gdr/\aCII we do not see a significant trend ($< 1\,\sigma$). On the other hand, we see a possible trend ($\sim 2.5\,\sigma$) for $\rm \delta_{GDR}$/$\rm \alpha_{CO}$ with T$_{\rm dust}$. This could originate from the degeneracy between the dust temperature and the dust mass in the SED modelling. One way to break this is by better sampling the SED and improving the modelling.


\section{Discussion}\label{sec:discussion mass comparison}

\subsection{Origin of the scatter between the various tracers}\label{scatter}

We found a small scatter for all the cross-calibration relations on Fig.\,\ref{fig:mass comparisions}. To understand the origin of this scatter, we make a simulation to test the contribution of measurement uncertainty on this scatter. This procedure is described in Appendix.\,\ref{Sec: scatter simulation}. Figure\,\ref{fig:origin of scatter} shows the comparison of the mock data with measurement error (in red) with the observed data (in black) for all our sources. For our cross-calibration relations, a significant fraction of this scatter can be reproduced by measurement uncertainties, and the Kolmogorov-Smirnov (KS) test shows that the distribution of the measurements and of a simulation, assuming a perfectly tight relation and only measurements uncertainties, are similar. However, even if we cannot detect it, there is certainly an intrinsic scatter between the tracers.

\subsection{Understanding the cross-calibration of tracers}\label{discussion cross calib}
Overall, the various tracers agree with each other within a $\lesssim$ 0.4\,dex scatter and we do not see any trend in comparing them with the L$_{\rm IR}$. The mean values of the cross-calibration are tabulated in Table\,\ref{tab:gas mass calibration}. Here we discuss the impact of the assumed $\alpha_{\rm CO}$ on the other tracers. We discuss two hypotheses: ULIRG-like value, (0.8, \citealt{DownesSolomon98, Engel10}) and Milky-Way like value (3.4, \citealt{Papadopoulos12, Bolatto13, Harrington21, Jarugula21}).


\begin{table}[]
    \centering
    \caption{Mean value of the cross-calibration of gas mass tracers.}\label{tab:gas mass calibration}
    \begin{tabular}{cc}
    \hline
    \hline
    &\\
    Cross-calibration of tracers & Value \\
    &\\
    \hline
    &\\
    \XCI $\times$ \aCO ($\times 10^{-5}$)& 6.31$\pm$0.67\\
    \XCI $\times$ \aCII ($\times 10^{-5}$) & 95.5 $\pm$17.1 \\
    \XCI $\times$ \gdr ($\times 10^{-5}$) & 302.0$\pm$52.2\\
    \aCO/\aCII & 0.08$\pm$0.01\\
    \gdr/\aCO & 42.66 $\pm$6.43\\
    \gdr/\aCII & 4.36$\pm$1.07\\
    \hline
    \end{tabular}
    \tablefoot{The mean value was estimated using \texttt{Linmix} which accounts for errors and upper-limits.}
    
\end{table}

Assuming an $\alpha_{\rm CO} = 3.4$, we can compute a value of X$_{\rm CI} = 1.86 \pm 0.20 \times 10^{-5}$. The value of X$_{\rm CI}$ is lower than the commonly adopted value of X$_{\rm CI} = 3 \times 10^{-5}$ \citep[e.g.][]{Papadopoulos04a, Bothwell17}. However, it is in agreement with [CI] abundance found in the Milky way \citep{Frerking89}, the sample presented in \citealt{Dunne21} (X$_{\rm CI} = (1.6 \pm 0.3) \times 10^{-5}$) and for the main-sequence galaxies in \citealt{Valentino18} (1.6-1.9$\times 10^{-5}$). 

We find a corresponding value of $\delta_{\rm GDR} = 145 \pm 22$ with $\alpha_{\rm CO} = 3.4$, and is higher than the commonly adopted value of $\delta_{\rm GDR} = 100$ for massive galaxies with near-solar metallicities \citep[e.g.][]{Leroy11, RemyR14}. \citet{Zavala22} also found a similar value of $\delta_{\rm GDR} = 105 \pm 40$ for a massive compact DSFG at $z = 6$. The high $\delta_{\rm GDR}$ value of our sample is in better agreement with the $\delta_{\rm GDR} = 129 \pm 57$ presented for the \citet{Dunne21} sample. With an \aCO $=3.4$, we find a corresponding value of \aCII $ = 40 \pm 6\, \rm M_\odot / L_\odot$. This is higher than the values calibrated in \citet{Zanella18}, \aCII $=30\, \rm M_\odot / L_\odot$ but is compatible at a 2\,$\sigma$ level.


Assuming an $\alpha_{\rm CO} = 0.8$, we can compute a higher value of X$_{\rm CI} = 7.9 \pm 0.80 \times 10^{-5}$. Similar values are adopted in \citet{Walter11} for a sample of QSOs, and they adopt an $\alpha_{\rm CO} = 0.8$ to derive the X$_{\rm CI}$ values. There are some plausible explanations for these X$_{\rm CI}$ values in the literature. \citet{Izumi20} find a similarly high value of X$_{\rm CI}$ $ \sim 7 \times 10^{-5}$. They suggest that these high [CI] abundances could result from an X-ray dominated region (XDR) of the ISM. The higher values of  X$_{\rm CI}$ can be found in denser regions \citep[e.g.][]{Bisbas21} and/or with higher metallicities \citep[e.g.][]{Heintz20}. Additionally, the destruction of CO due FUV photons in regions of high star-formation can further increase the [CI]-abundances \citep[e.g.][]{Bisbas21}. 

With this value, we derive a low $\delta_{\rm GDR} = 34 \pm 5$ for our sample. A higher gas phase metallicity can also lead to low gas-to-dust ratios \citep[e.g.][]{Hunt05,Leroy11,Saintonge13,Popping22a,Popping22b}. Thus, if our sample was dominated by super-solar metallicity galaxies (e.g. galaxies presented in \citealt{DeBreuck19}, Litke et al. submitted), such low values of $\delta_{\rm GDR}$ could be plausible. Using the models from \citet{Popping17}, on the gas-to-dust ratio and metallicity relation, we obtain a \gdr $\sim$ 300 for a galaxy with solar-metallicity at $z\sim4$ (as seen in \citealt{DeBreuck19}). Although the model demonstrates a decreasing \gdr with increasing metallicities, the \gdr predicted remains high. This could arise from model being inefficient in reproducing the dust production from metals at high $z$. We find similar values on comparing with the observations of \citealt{Popping22a}, \gdr $\sim$ 400. This value also includes the contribution of atomic hydrogen gas and is estimated using quasar absorption sightlines through the neutral ISM. These differences in methodology and selection could contribute to the much higher \gdr found in that work. In a near future, we should be able to probe the dependencies of the carbon abundance (X$_{\rm CI}$) and $\delta_{\rm GDR}$ on the metallicity with JWST.

Using an \aCO value of 0.8, we derive a low \aCII $=9.4 \pm 1.5\, \rm M_\odot / L_\odot$. This value is compatible with the \aCII $= 7^{+4}_{-1} \,\rm M_\odot / L_\odot$ reported by \citet{Rizzo21} for a sample of 5 SPT-SMGs. \citet{Vizgan22} estimated the \aCII values for a simulated sample of galaxies and found a median value of \aCII $= 18\, \rm M_\odot / L_\odot$ with a median absolute deviation of $10\, \rm M_\odot / L_\odot$. This is slightly higher than the \aCII values we derive with a ULIRG-like \aCO. They also find that the \aCII values can depend on the mass of the system with lower-values predicted for more-massive galaxies ($\rm M*>10^9\, M_\odot$). Additionally, they find the peak of the \aCII distribution for their sample at $\sim 15\, \rm M_\odot / L_\odot$ and find many galaxies with \aCII $<10\,\rm M_\odot / L_\odot$. \citet{Sommovigo21} find a variation in \aCII between normal star-forming galaxies and starbursts ($\lesssim 10\,\rm M_\odot / L_\odot$ \aCII in starbursts). 

\begin{figure}[ht]
\centering

\includegraphics[width=9cm]{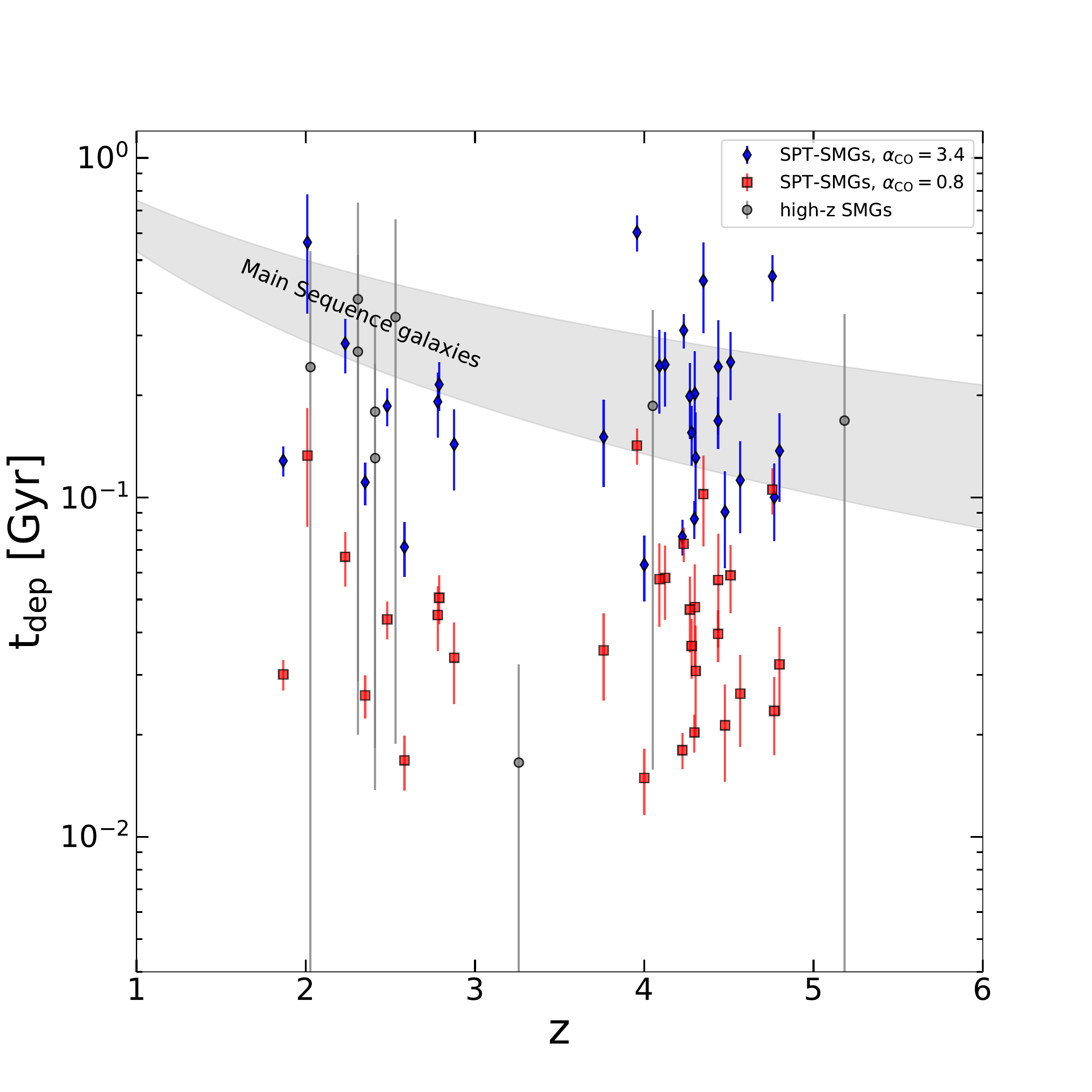}   

\caption{\label{fig:depletion timescale}Depletion timescale as a function of redshift for our sample. We compare the depletion timescale of our sample (blue diamonds and red squares) along with the SMGs (grey circles) from \citet{Carilli10,Walter12,Ivison13,Fu12,Fu13, Alaghband-Zadeh13}. The blue diamonds represent the depletion timescales corresponding to gas mass calculated using $\alpha_{\rm CO} = 3.4$ and the red squares are computed using $\alpha_{\rm CO} = 0.8$. The main-sequence in the shaded-grey region follows the relation presented in \citet{Saintonge13}, with $\alpha$ values ranging from -1.5 \citep{Dave12} to -1.0 \citep{Magnelli13}.}
\end{figure}

Additional constraints like dynamical mass estimates and metallicity can help get a better estimate of the absolute values of these tracers. Comparing the gas mass estimates with the dynamical mass can help us predict the values of these tracers \citep[e.g.][]{Bertemes18, Zavala22, Gururajan22}, however such data are available only for a small number of sources. We also make an important assumption that all these tracers trace the same region of molecular gas content of the galaxy, which has been debated in the literature. 

In Fig.\,\ref{fig:mass comparisions}, we do not see a trend with L$_{\rm IR}$ for the various tracers. However, in the y-axis, we compare either a product or ratio of these tracers with L$_{\rm IR}$. In this scenario, we may miss a systematic effect with L$_{\rm IR}$ if the tracers have an opposite (product) or similar (ratio) trend with L$_{\rm IR}$, and compensate each other. This can also be interpreted as a counter argument, that is, if one of the tracer is shown to not have a trend with L$_{\rm IR}$, then the other two tracers will also have no systematic effect versus L$_{\rm IR}$.

We discuss the possibilities of two values of \aCO and their implications on \XCI and \gdr in the above section. A third scenario could be the possibility of \aCO varying with L$_{\rm IR}$, such that \aCO decreases from 3.4 at low L$_{\rm IR}$ to 0.8 at high L$_{\rm IR}$. Since we do not see a trend with L$_{\rm IR}$ in Fig.\,\ref{fig:mass comparisions}, this could imply that \XCI, \gdr and \aCII should vary by the same factor in the following range of IR luminosities. Such a variation of the exact same factor in all four tracers is unlikely, and the various gas tracers are thus not likely to have a significant trend with L$_{\rm IR}$.



\subsection{Depletion timescales}\label{depletion timescale}

We finally compute the depletion timescale (t$_{\rm dep}$) of our sample as $\rm M_{\rm gas}/SFR$ [Gyr]. We use the gas mass estimated with CO line using an $\alpha_{\rm CO} = 3.4$ and find a mean t$_{\rm dep}$ of 212 $\pm$ 26 Myr with a range of 63.3 -- 603.9 Myr for our sample. We also compute the depletion timescale with a lower \aCO $=0.8$ and find a mean depletion time to 49.75 $\pm$ 6.07 Myr and the range varying from 14.9 -- 142.1 Myr.  In Fig.\,\ref{fig:depletion timescale}, we plot the evolution of depletion time with redshift $z$ for our sample, along with the SMGs in literature (\citealt{Carilli10,Walter12,Ivison13,Fu12,Fu13, Alaghband-Zadeh13}). The grey shaded region represents the depletion timescale evolution with redshift for main-sequence galaxies from \citet{Saintonge13}, t$_{\rm dep} = 1.5(1+z)^\alpha$, where $\alpha$ varies from -1.5 \citep{Dave12} to -1.0 \citep{Magnelli13}.

We do not see a clear trend in the evolution of t$_{\rm dep}$ with $z$ for our sample unlike the main-sequence. With $\alpha_{\rm CO} = 3.4$, our sample has a depletion timescale similar to the main-sequence at the redshift range, with 
8/29 sources having shorter depletion time than the main sequence, and 3/29 sources having longer depletion time than the main-sequence. Although these sources have a high SFR, their depletion time is similar to that of main-sequence galaxies. This could be a hint towards these systems having a large gas reservoir \citep[e.g.][]{Tacconi10,Daddi10a,Saintonge13,Dessauges-Zavadsky15,Bethermin15} with rapid accretion  of cold gas from the cosmic web \citep[e.g.][]{ Dekel09,Kleiner17,Kretschmer20,Chun20}. But, if we adopt a lower-value of $\alpha_{\rm CO} = 0.8$, these objects would fall much below the main-sequence with a short depletion timescale.




\section{Summary and conclusions}\label{conclusions}

We present a sample of 29-DSFGs in the redshift range 1.867 -- 4.799 and the flux catalogue of [CI](1-0), [CI](2-1) and [CII] lines for these galaxies, combining the ancillary observations presented in \citet{Bothwell17}, \citet{Reuter20} and \citet{Gullberg15}. The main conclusions of the work are presented below:
\begin{itemize}
    \item We compare the IR-luminosity to the [CI]-luminosity of our sample. This is a proxy to the integrated Kennicutt-Schmidt law, with the L$_{\rm IR}$ tracing the SFR and L$_{\rm CI}$ tracing the gas mass. In the case of L$_{\rm IR}$ versus L$_{\rm [CI](1-0)}$, we find that the slope of the relation for our sample is in agreement with the L$_{\rm FIR}$ versus L$^{\prime}_{\rm CO(1-0)}$ slopes presented in \citet{Greve14, Liu15, Kamenetzky16}. On combining our sample with local galaxies, main-sequence and other SMGs, we find a super-linear relation, suggesting that there may be a different trend between the starbursting SMGs and the local/main-sequence sample \citep{Daddi10_sflaws, Genzel10}. The relation between [CI](2-1) and IR luminosity does not show a difference in slope for SMGs and local/main-sequence galaxies which could be attributed to a higher excitation of [CI](2-1) in starbursts. 
    \item Comparing the $\rm L_{[CI](1-0)}/L_{CO(4-3)}$ versus $\rm L_{[CI](1-0)}/L_{IR}$ ratio of our sample with the compilation presented in \citet{Valentino20}, we find that our sample has comparable densities and radiation field intensities to the other SMGs. On comparing with the main-sequence and the local galaxies, they have higher intensities, but with a strong overlap. 
    \item We compute the [CI] excitation temperature for our sample, and it ranges from 17.7 -- 64.2 K with a mean sample value of 34.5 $\pm$ 2.1 K. On comparing [CI] excitation temperatures to the dust temperature, $\rm T_{\rm ex}/T_{\rm dust} <  1$. We do not find any candidates for cold/optically thick dust. 
    \item Comparing our molecular mass estimates with four tracers, [CI](1-0), CO(1-0), dust, and [CII] we provide a cross-calibration for the uncertain parameters X$_{\rm CI}$, $\alpha_{\rm CO}$, $\delta_{\rm GDR}$, and \aCII. Overall, there is a good agreement between all these tracers and the scatter between \aCO, \XCI,\gdr, and \aCII (Fig.\,\ref{fig:mass comparisions}) can be reproduced by measurement uncertainties.
    \item Higher values of $\alpha_{\rm CO}$, ($\sim3.4$) predict a lower [CI]-abundance, a higher dust-to-gas ratio and value of [CII]-to-H$_2$ conversion factor similar to \citet{Zanella18}. This could be possible in low-metallicity regimes. ULIRG-like values of $\alpha_{\rm CO}$, give a higher [CI] abundance and a low \aCII and very low \gdr which can be the case for metal rich systems. 
\end{itemize}

\begin{acknowledgements}
We thank Francesco Valentino for providing the data associated with his [CI] data compilation. We thank the referee for their valuable comments. ALMA is a partnership of ESO (representing its member states), NSF (USA) and NINS (Japan), together with NRC (Canada), MOST and ASIAA (Taiwan) and KASI (Republic of Korea), in cooperation with the Republic of Chile. The Joint ALMA Observatory is operated by ESO, AUI/NRAO and NAOJ. The National Radio Astronomy Observatory is a facility of the National Science Foundation operated under cooperative agreement by Associated Universities, Inc. This publication is based on data acquired with the Atacama Pathfinder Experiment (APEX). APEX is a collaboration between the Max-Planck-Institut fur Radioastronomie, the European Southern Observatory and the Onsala Space Observatory. Based on observations made with ESO Telescopes at the La Silla Paranal Observatory under programme ID 097.A-0973. This work was supported by the Programme National “Physique et Chimie du Milieu Interstellaire” (PCMI) of CNRS/INSU with INC/INP co-funded by CEA and CNES. This work was supported by the Programme National Cosmology et Galaxies (PNCG) of CNRS/INSU with INP and IN2P3, co-funded by CEA and CNES. MA acknowledges support from FONDECYT grant 1211951, ANID+PCI+INSTITUTO MAX PLANCK DE ASTRONOMIA MPG 190030, ANID+PCI+REDES 190194 and ANID BASAL project FB210003. TRG acknowledges support from the Carlsberg Foundation (grant no CF20-0534). The Cosmic Dawn Center (DAWN) is funded by the Danish National Research Foundation under grant No. 140.
\end{acknowledgements}
\bibliographystyle{aa} 
\bibliography{bibliography.bib}
\begin{appendix}

\section{Gas mass tracers versus dust temperatures}
We compare the cross-calibration relations \XCI $\times$ \aCO, \XCI $\times$ \aCII, \XCI $\times$ \gdr, \aCO/\aCII, \gdr/\aCO and \gdr/\aCII with the dust temperature for our sample. Overall, we do not find a significant ($>3\,\sigma$) for these tracers. The mild trend ($\sim 2.5\,\sigma$) we find for \gdr/\aCO could be linked to the T$_{\rm dust}$ - M$_{\rm dust}$ degeneracy in the SED fitting.


\begin{table*}[]
    \centering
    \caption{Slope, intercept and scatter from the cross-calibration relation}\label{tab:results_cross_calib_linmix}
    \begin{tabular}{cccc}
    \hline
    \hline
    &&&\\
    Data& Slope & Intercept & Scatter \\
    &&&\\
    \hline
    &&&\\
    \multicolumn{4}{c}{Versus $\rm L_{\rm IR}$}\\
    &&&\\
    \hline
    &&&\\
    log(\XCI $\times$ \aCO)& -0.04$\pm$0.17& -5.59$\pm$2.22&0.23$\pm$0.05\\
    log(\XCI $\times$ \aCII)  &  0.11$\pm$0.27 & -4.4$\pm$3.6&0.29$\pm$0.13\\
    log(\XCI $\times$ \gdr)  & -0.02$\pm$0.21& -2.23$\pm$2.78&0.39$\pm$0.09 \\
    log(\aCO/\aCII) & -0.16$\pm$0.24& 1.1$\pm$3.1&0.26$\pm$1.0\\
    log(\gdr/\aCO) & 0.04$\pm$0.21 & 1.1$\pm$2.8 & 0.34$\pm$0.07 \\
    log(\gdr/\aCII) & 0.07$\pm$0.34 & -0.27$\pm$4.55&0.41$\pm$0.14\\
    \hline
    &&&\\
    \multicolumn{4}{c}{Versus $\rm T_{\rm dust}$}\\
    &&&\\
    \hline
    &&&\\
    \XCI $\times$ \aCO & ($-0.74 \pm 1.62)\times10^{-6}$& ($9.0\pm5.4)\times10^{-5}$&($2.6\pm0.8)\times10^{-5}$\\
    \XCI $\times$ \aCII  &  $(-4.57\pm6.02)\times10^{-5}$ & $(2.4\pm1.6)\times10^{-3}$&$(5.4\pm3.5)\times10^{-4}$\\
    \XCI $\times$ \gdr  & $(2.26\pm8.16)\times10^{-5}$& $(-0.28 \pm2.8)\times10^{-3}$& $(9\pm4)\times10^{-4}$\\
    \aCO/\aCII & $(1.3\pm1.9)\times10^{-3}$& $(3.1\pm8.8)\times10^{-2}$&0.058$\pm$0.022\\
    \gdr/\aCO & 1.31$\pm$0.52 & -12.6$\pm$22.6&21$\pm$6\\
    \gdr/\aCII & 0.05$\pm$0.06 & 0.82$\pm$3.24&1.8$\pm$1.4\\
    \hline
    \end{tabular}

\end{table*}

\begin{figure*}[ht]
\centering

\begin{tabular}{cc}
\includegraphics[width=8.5cm]{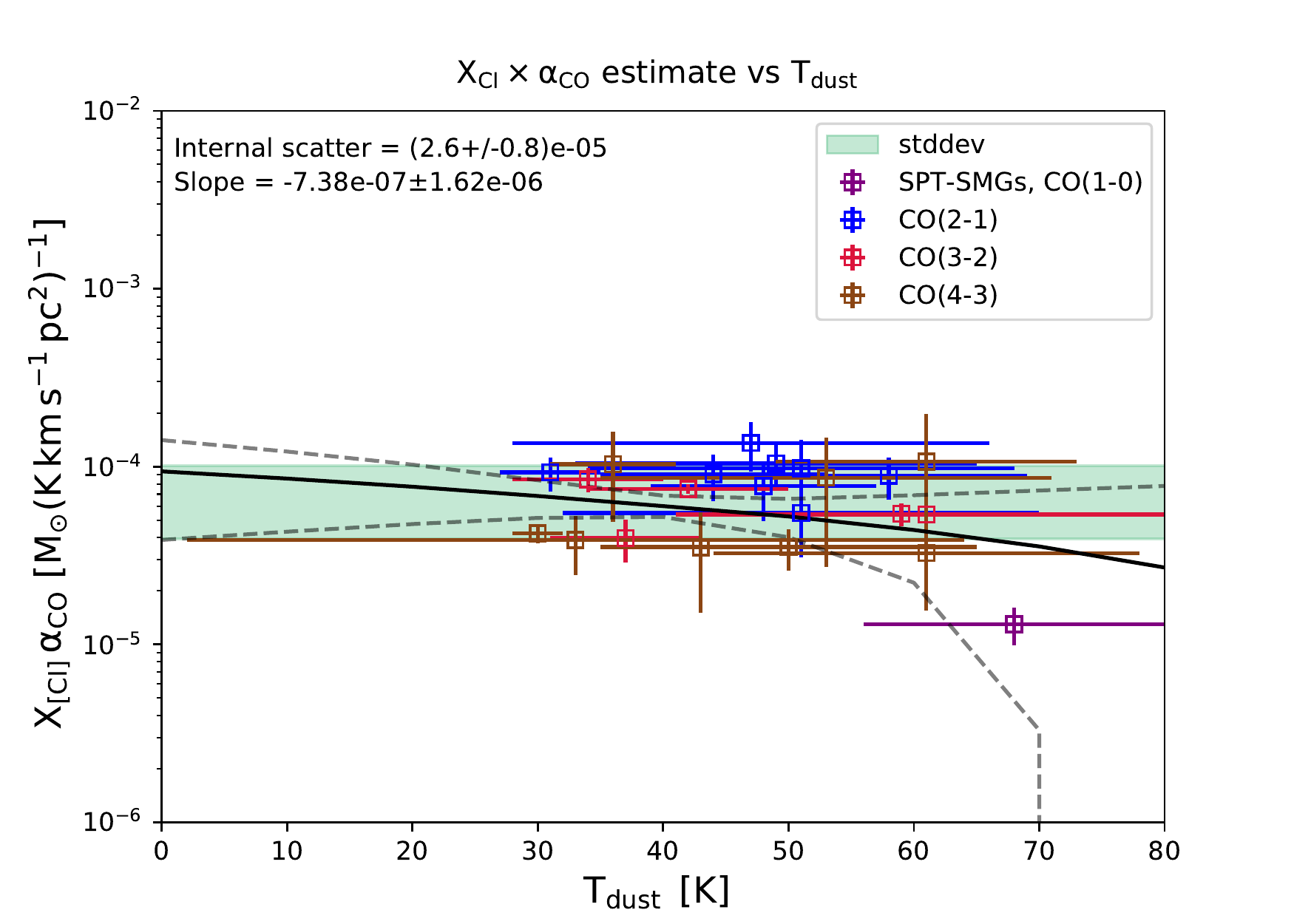} & 
\includegraphics[width=8.5cm]{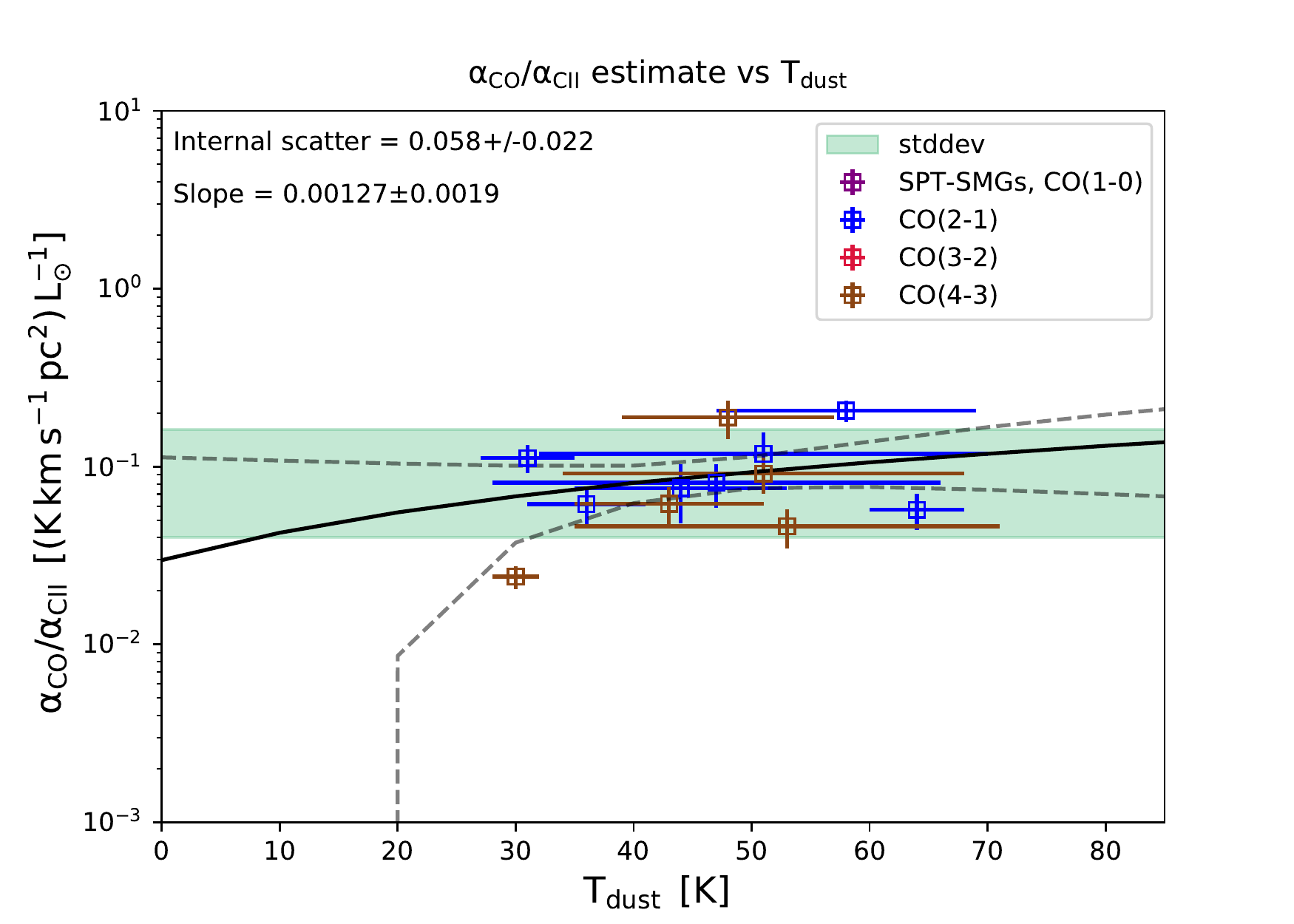}\\
\includegraphics[width=8.5cm]{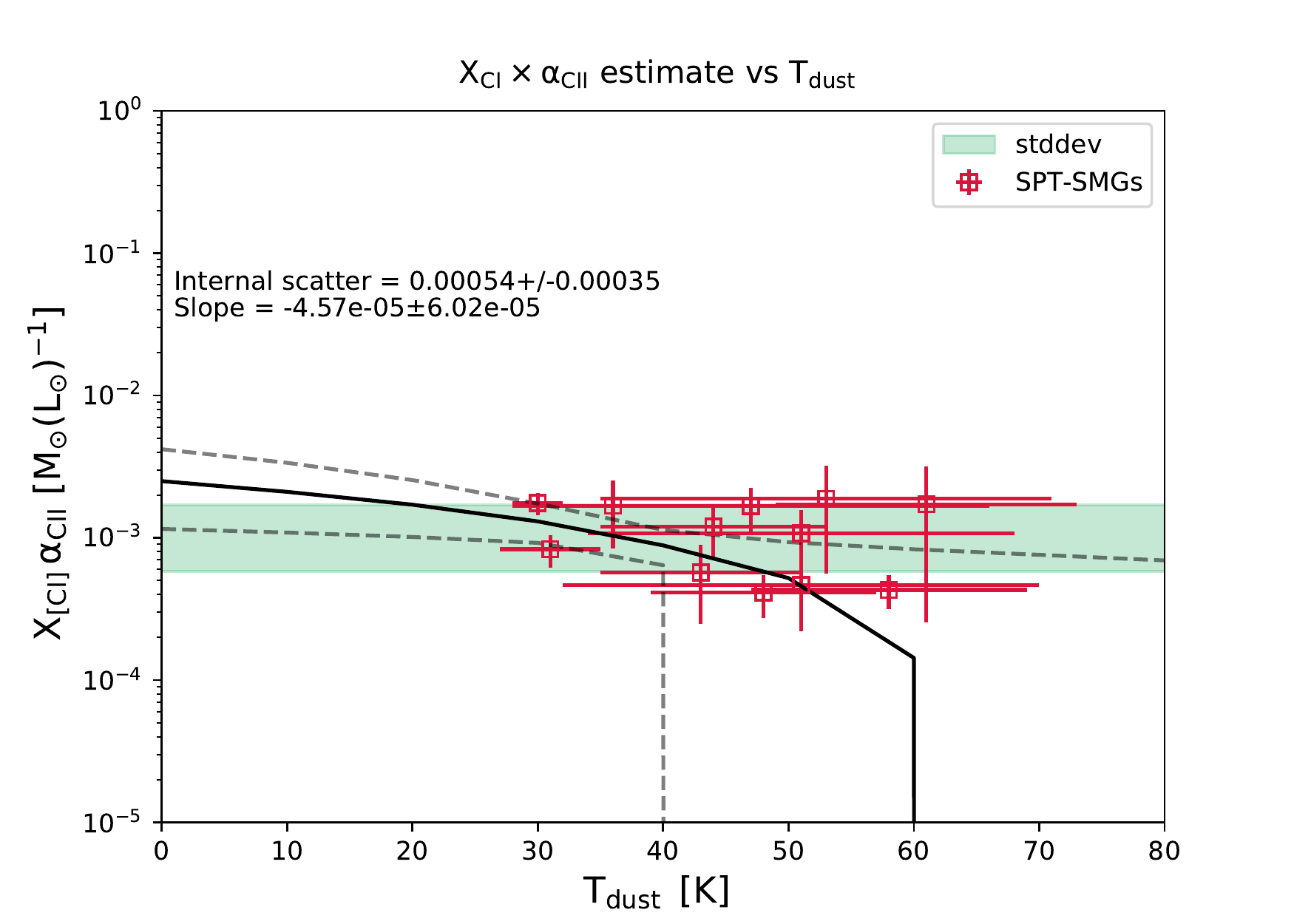} &
\includegraphics[width=8.5cm]{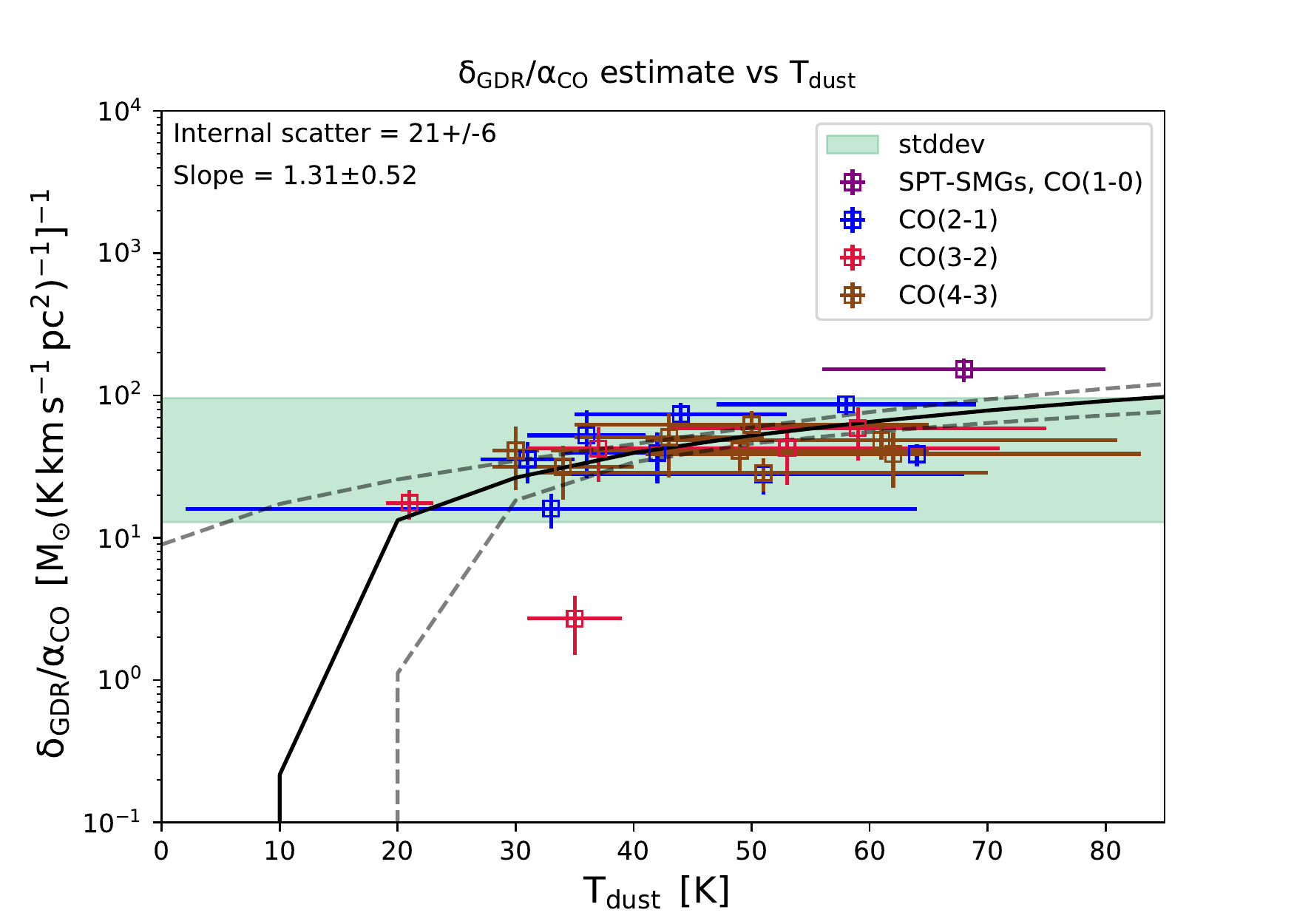} \\
\includegraphics[width=8.5cm]{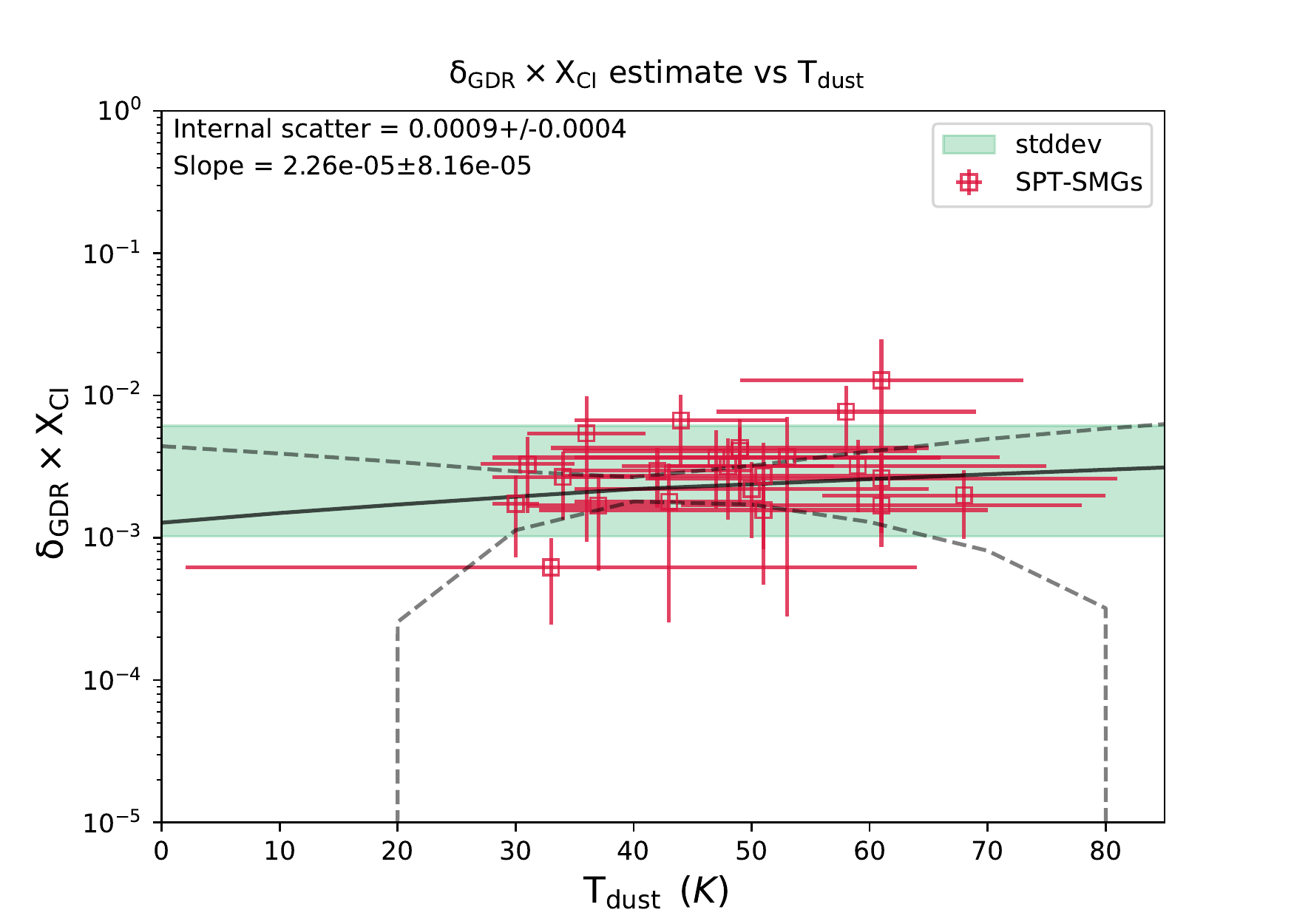} &
\includegraphics[width=8.5cm]{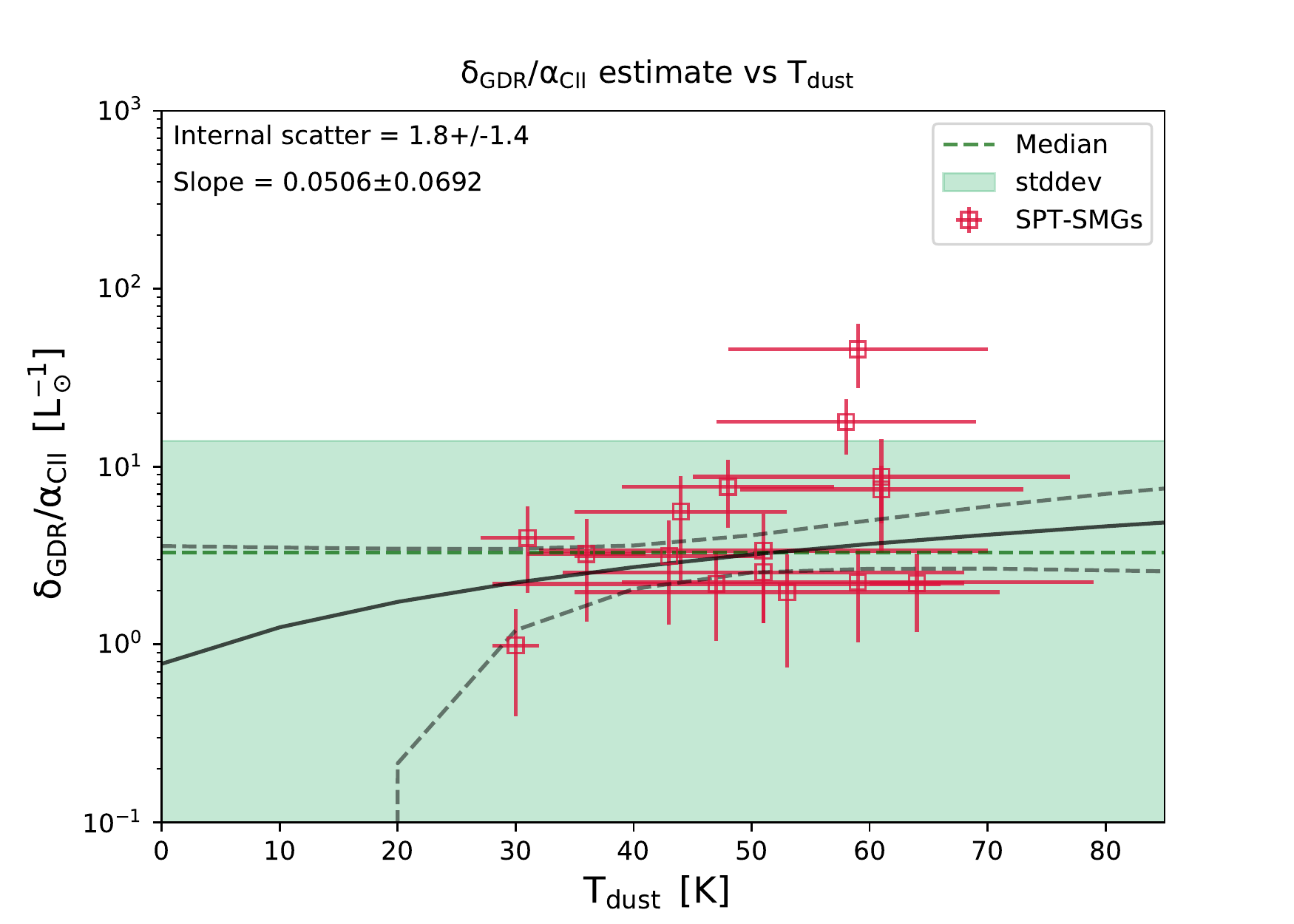} \\

\end{tabular}
\caption{\label{fig:mass comparisions_temperature} The \XCI $\times$ \aCO, \XCI $\times$ \aCII, and \XCI $\times$ \gdr (left column) and \aCO / \aCII , \gdr / \aCO, and \gdr / \aCII (right column) as a function of T$_{\rm dust}$ for our sample. Our sample is represented in squares, color-coded by the CO transition used to estimate the CO-based gas mass, CO(1-0) transition in purple, CO(2-1) in blue, CO(3-2) in red and CO(4-3) in brown in top and center rows. The green-shaded region represents the 1\,$\sigma$ region around the sample mean. we also plot the linear regression best-fit and the 1\,$\sigma$ region from \texttt{Linmix} as the black solid and black-dashed line respectively. }
\end{figure*}

\section{APEX Spectra}
\begin{figure*}[h]
\centering

\begin{tabular}{ccc}
\includegraphics[width=6cm]{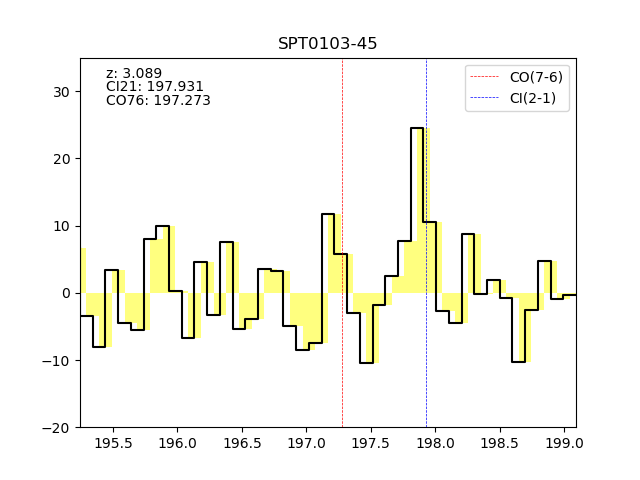} &  \includegraphics[width=6cm]{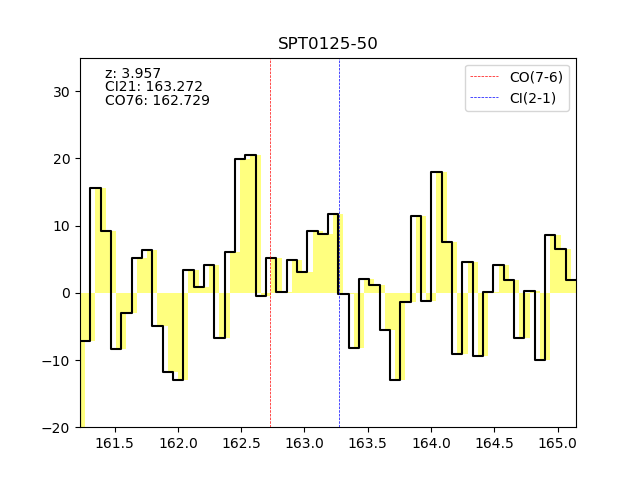} & \includegraphics[width=6cm]{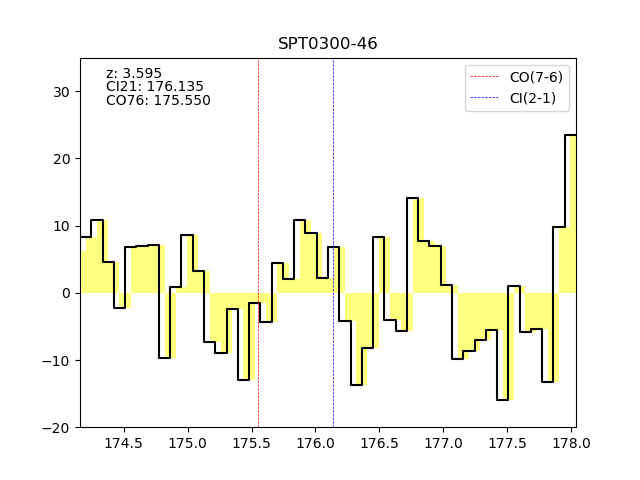} \\
\includegraphics[width=6cm]{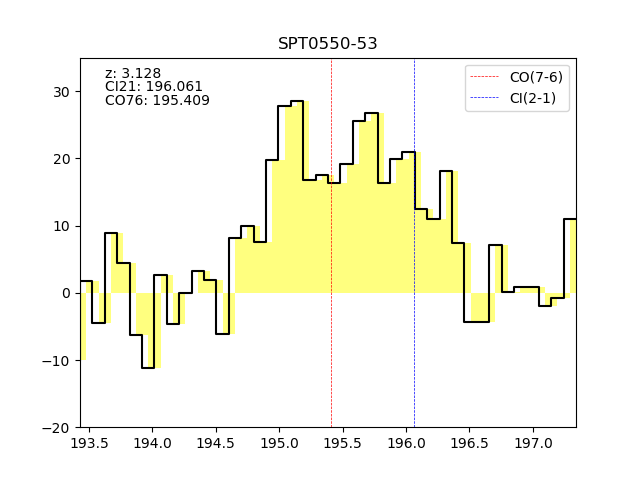} &  \includegraphics[width=6cm]{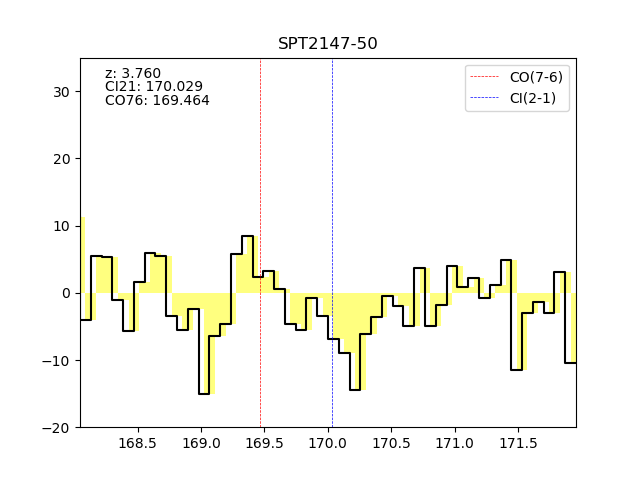} & \includegraphics[width=6cm]{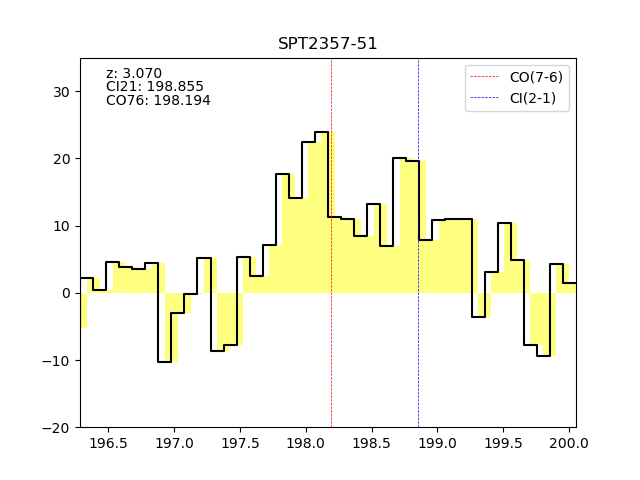} \\

\end{tabular}
\caption{\label{fig:APEX spectra} The spectra of the APEX sample. The red-dotted line indicates the CO(7-6) line and the blue-dotted line indicates the [CI](2-1) line.}
\end{figure*}

The spectra of our APEX observations of the sources are shown in Fig.\,\ref{fig:APEX spectra}.

\section{Deblending [CI](2-1) and CO(7-6) lines}\label{App: Deblending}

\begin{figure}[h]

\includegraphics[width=9cm]{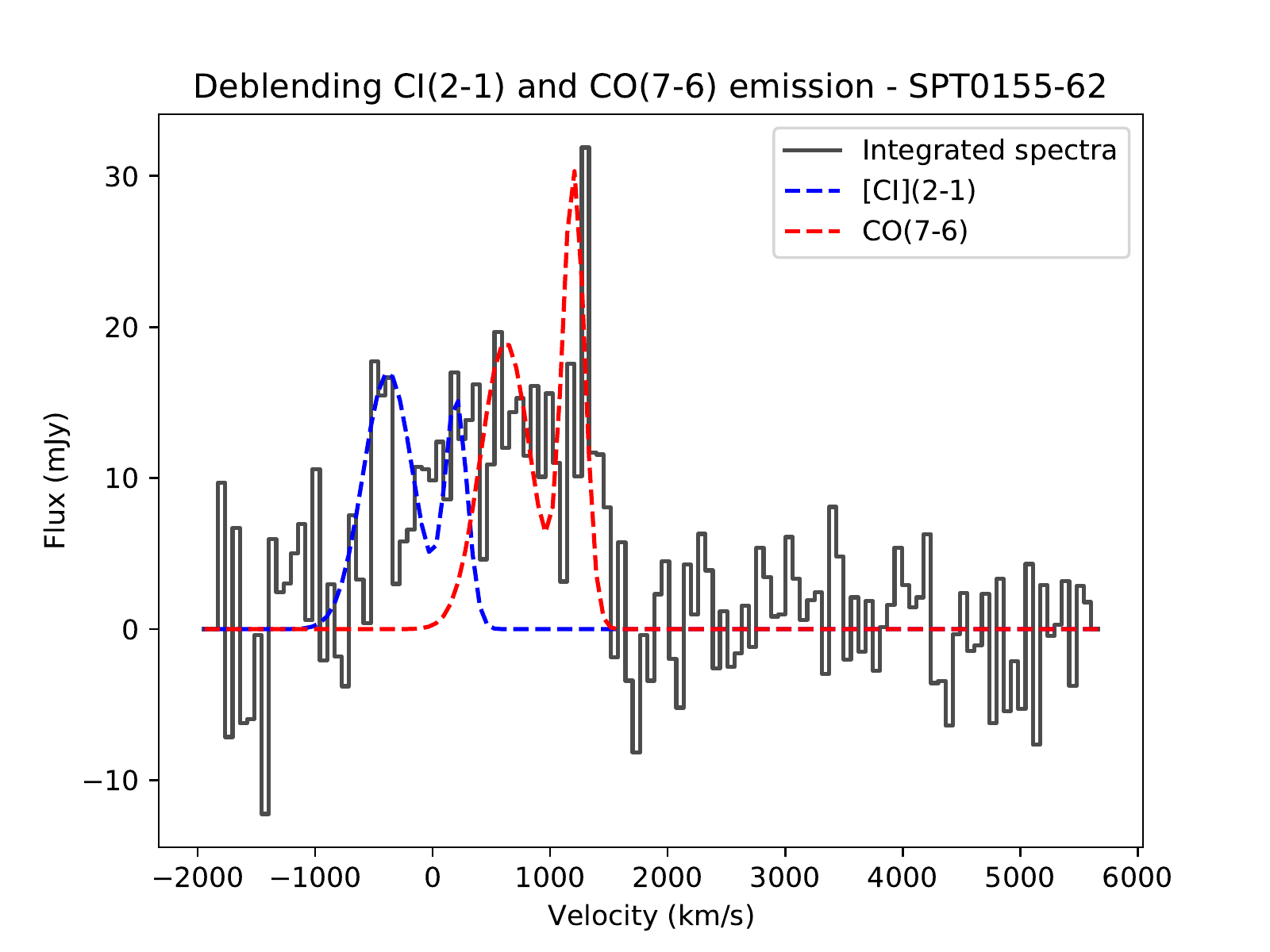}  \includegraphics[width=9cm]{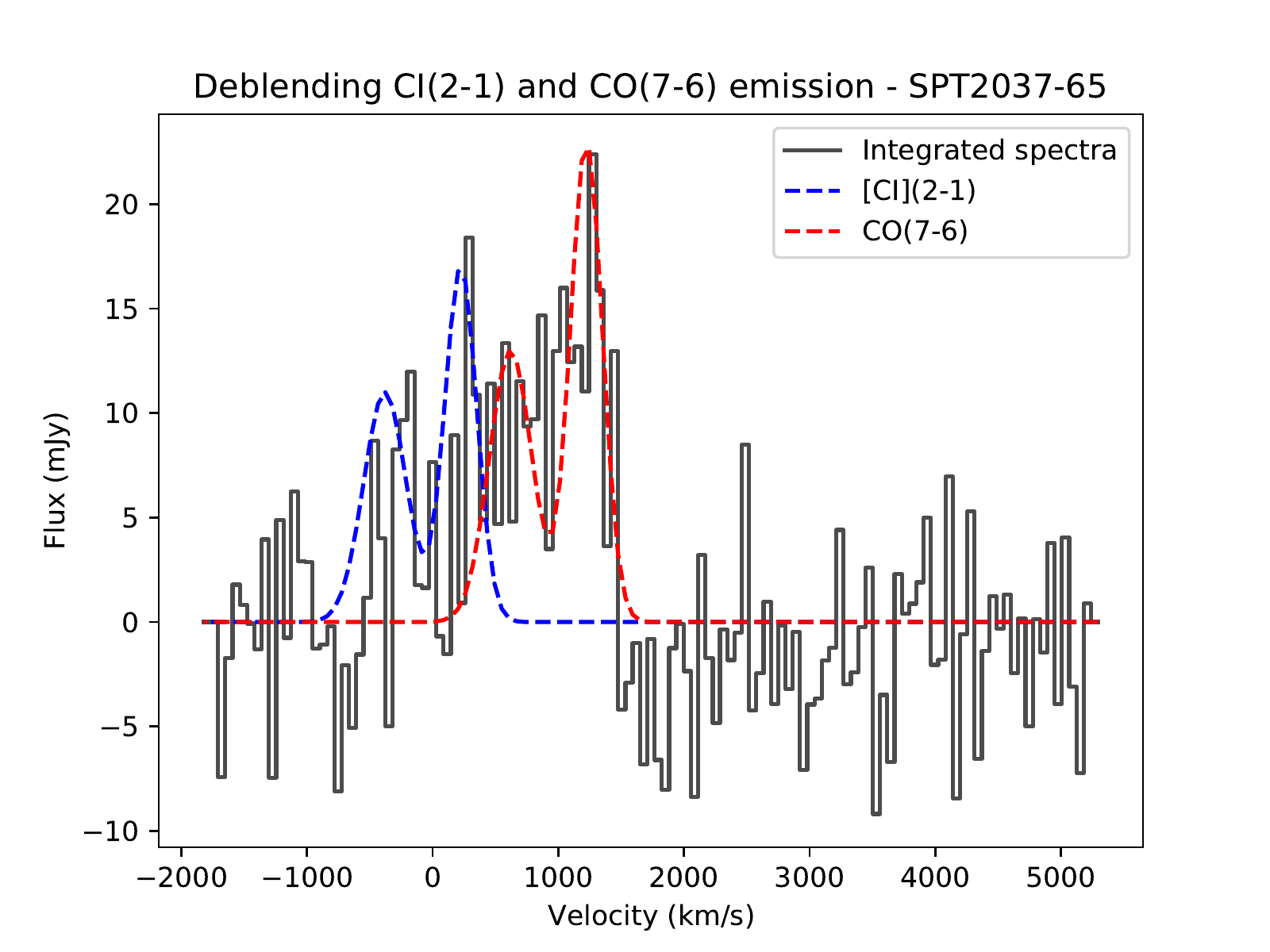}

\caption{\label{fig:deblending} Deblending the CO(7-6) and [CI](2-1) lines for sources SPT0155-62 (top-panel) and SPT2037-65 (bottom-panel). The spectra is represented by the black solid line. The curve-fit derived gaussian fits for CO(7-6) is represented as red dashed line and [CI](2-1) is represented as blue dashed line.  }
\end{figure}

In Fig.\,\ref{fig:ACA CI21 spectra}, the spectra of sources SPT0155-62 and SPT2037-65 have blended CO(7-6) and [CI](2-1) emission. We use a multiple-Gaussian fit in-order to estimate the fraction of flux in the blended region. For SPT0155-62, we fit 4 Gaussians with 6 free parameters. We use for amplitudes, 2 central velocities and 2 line widths, thereby forcing the same width for each of these lines. We thus use the combined width of these two components for each of these lines as the integration window for the moment-0 maps. Additionally, we can also compute the integrated flux from our fitting procedure. The line fluxes of CO(7-6) and [CI](2-1) estimated by this method are 12.1 $\pm$ 1.9 and 8.9 $\pm$1.6 Jy km/s. The difference between these fluxes and that estimated by the moment-0 maps is $< 1\,\sigma$. We use the same model for SPT2037-65 and estimate CO(7-6) flux as 9.1 $\pm$ 1.7 Jy km/s and [CI](2-1) as 7.1 $\pm$ 1.4 Jy km/s. The difference between the moment map estimated flux is $< 1.2\,\sigma$ for this source. Hence we proceed to use the fluxes estimated by the moment-0 map for uniformity.

\newpage
\section{Origin of scatter between various tracers}\label{Sec: scatter simulation}

In Fig.\,\ref{fig:mass comparisions}, we see a scatter while comparing our tracers with the L$_{\rm IR}$. Although there is no strong trend and the scatter is less than $<0.41$\,dex, the factor driving this is unclear. In order to probe the contribution of measurement uncertainties to this scatter, we make a test to simulate our data with additional uncertainties to see if they can reproduce the scatter. 

To do so, we derive a random Gaussian distribution of one of the tracers, example $\alpha_{\rm CO}$, in terms of another tracer example X$_{\rm CI}$. In other words, for the X$_{\rm CI} \times \alpha_{\rm CO}$ relation against the L$_{\rm IR}$ (Fig.\,\ref{fig:mass comparisions}, left column, top row), the $\alpha_{CO}$ is traced by the observable quantity, $\rm L^{\prime}_{CO(1-0)}$ and X$_{\rm CI}$ is traced by $\rm [1375.8\,D_L^2\,S_\nu \Delta\upsilon] / [(1+z)Q_{10}\,A_{10}]$. We generate a random Gaussian for $\rm L^{\prime}_{CO(1-0)}$ as a function of the X$_{\rm CI}$ tracer and the median of the X$_{\rm CI} \times \alpha_{\rm CO}$ relation. We allow the sigma of this distribution to the be combined error of the two quantities. This therefore gives us a large array of $\rm L^{\prime}_{CO(1-0)}$ values with additional noise. We then use every $\rm L^{\prime}_{CO(1-0)}$ value to compute  X$_{\rm CI} \times \alpha_{\rm CO}$ relation using Eq.\,\ref{eq:XCIalphaCO} for each of our source.

\begin{table*}[]
    \centering
    \caption{KS test between the observed data and the simulated data with noise}\label{tab:KS test results}
    \begin{tabular}{ccc}
    \hline
    \hline
    &&\\
    Data & Deviation & p-Value \\
    &&\\
    \hline
    &&\\
    \XCI $\times$ \aCO & 0.18& 0.39\\
    \XCI $\times$ \aCII  &  0.30 & 0.13\\
    \XCI $\times$ \gdr  & 0.18& 0.35 \\
    \aCO/\aCII & 0.21& 0.48\\
    \gdr/\aCO & 0.15 & 0.50\\
    \gdr/\aCII & 0.32 & 0.06\\
    \hline
    \end{tabular}

\end{table*}

\begin{figure*}\label{fig:scatter simulation}
\centering

\begin{tabular}{cc}

\includegraphics[width=9cm]{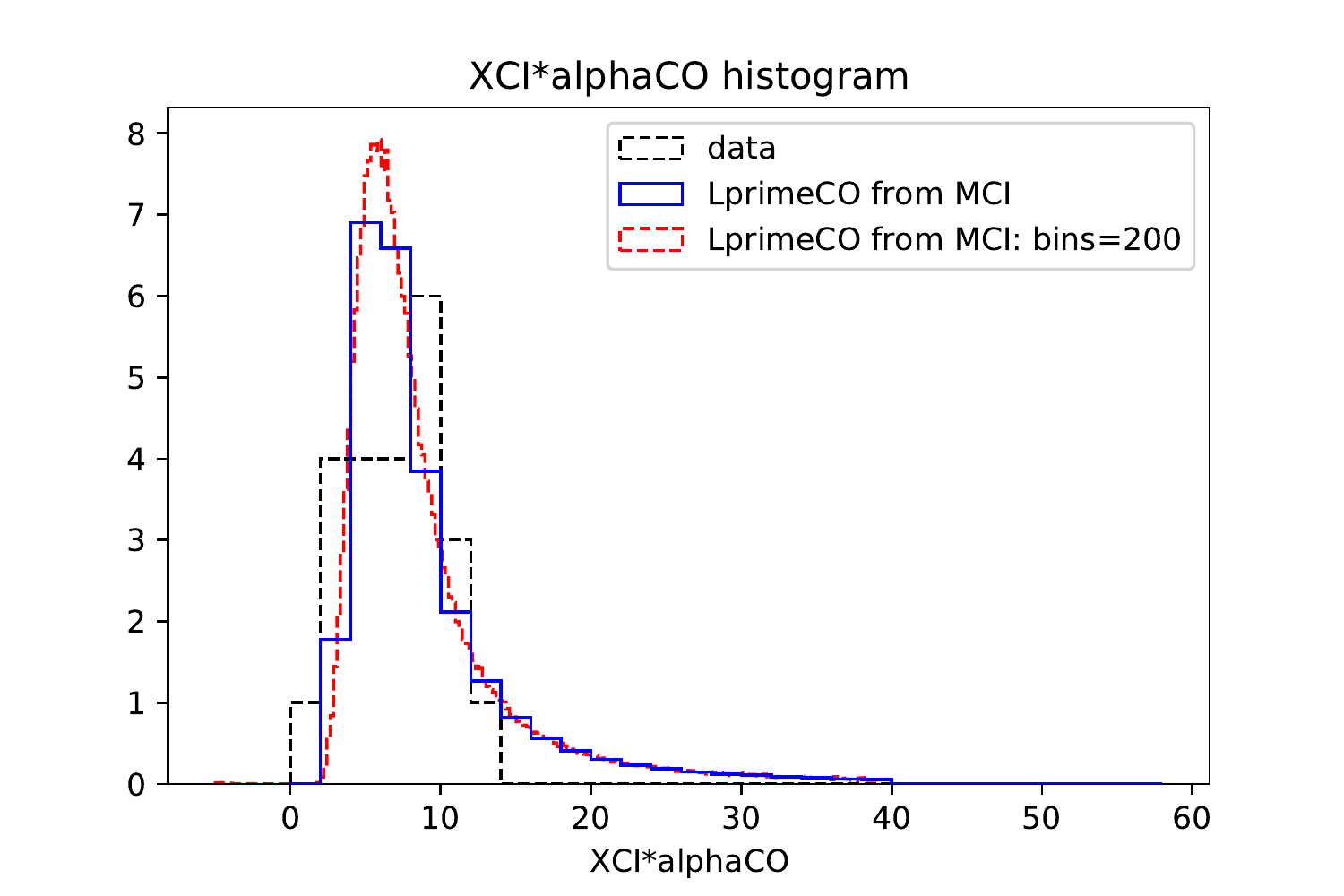} &
\includegraphics[width=9cm]{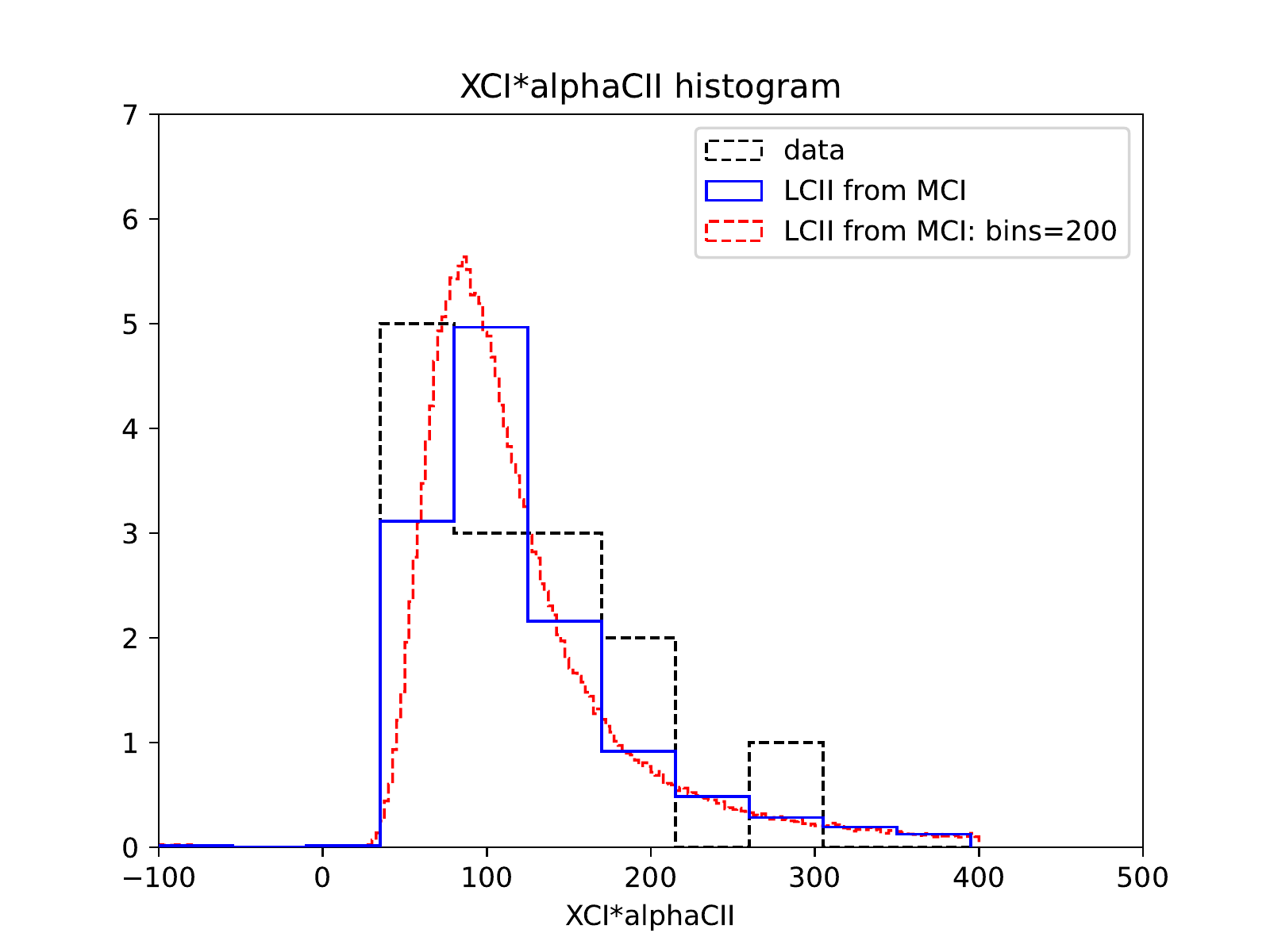} \\
\includegraphics[width=9cm]{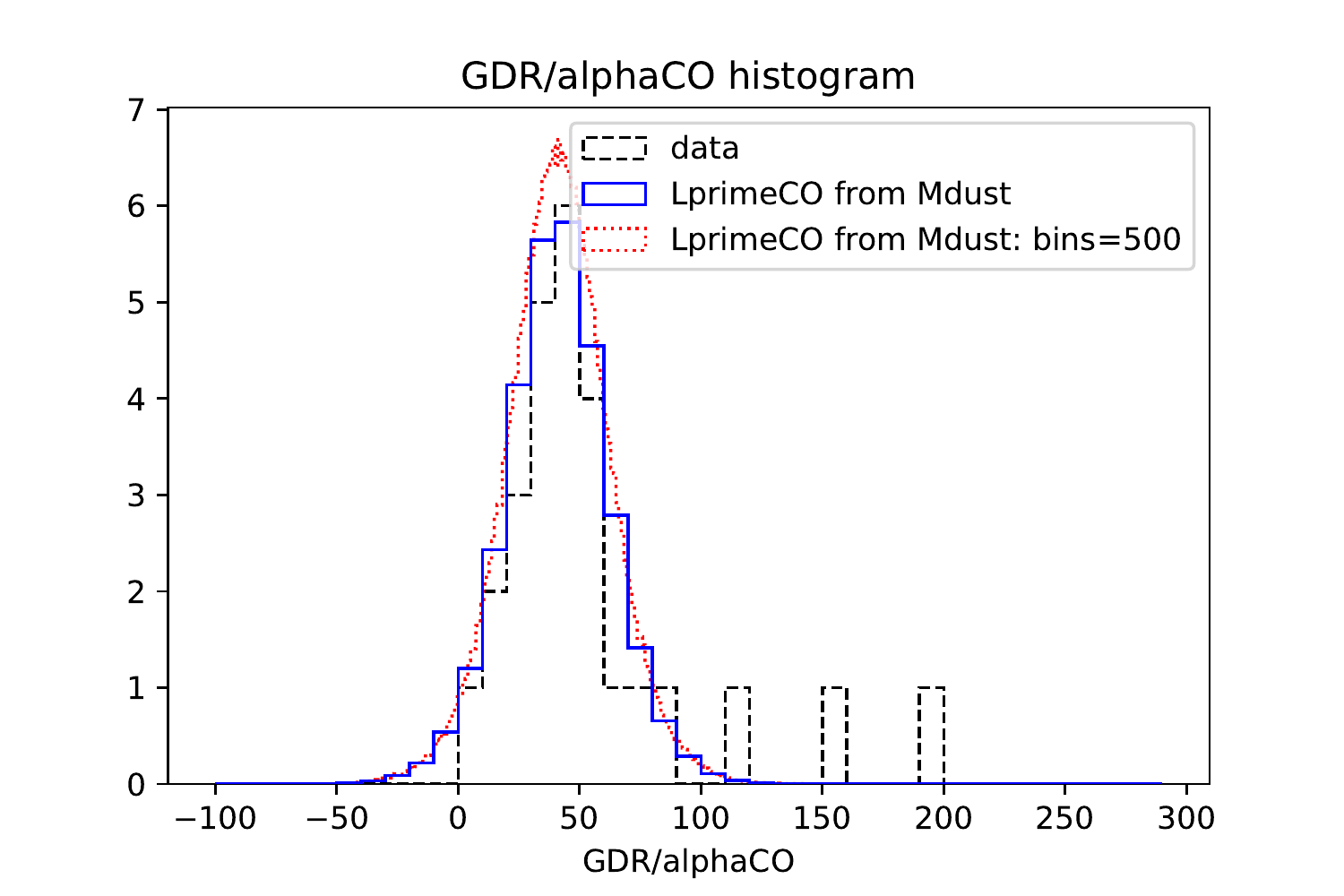} &
\includegraphics[width=9cm]{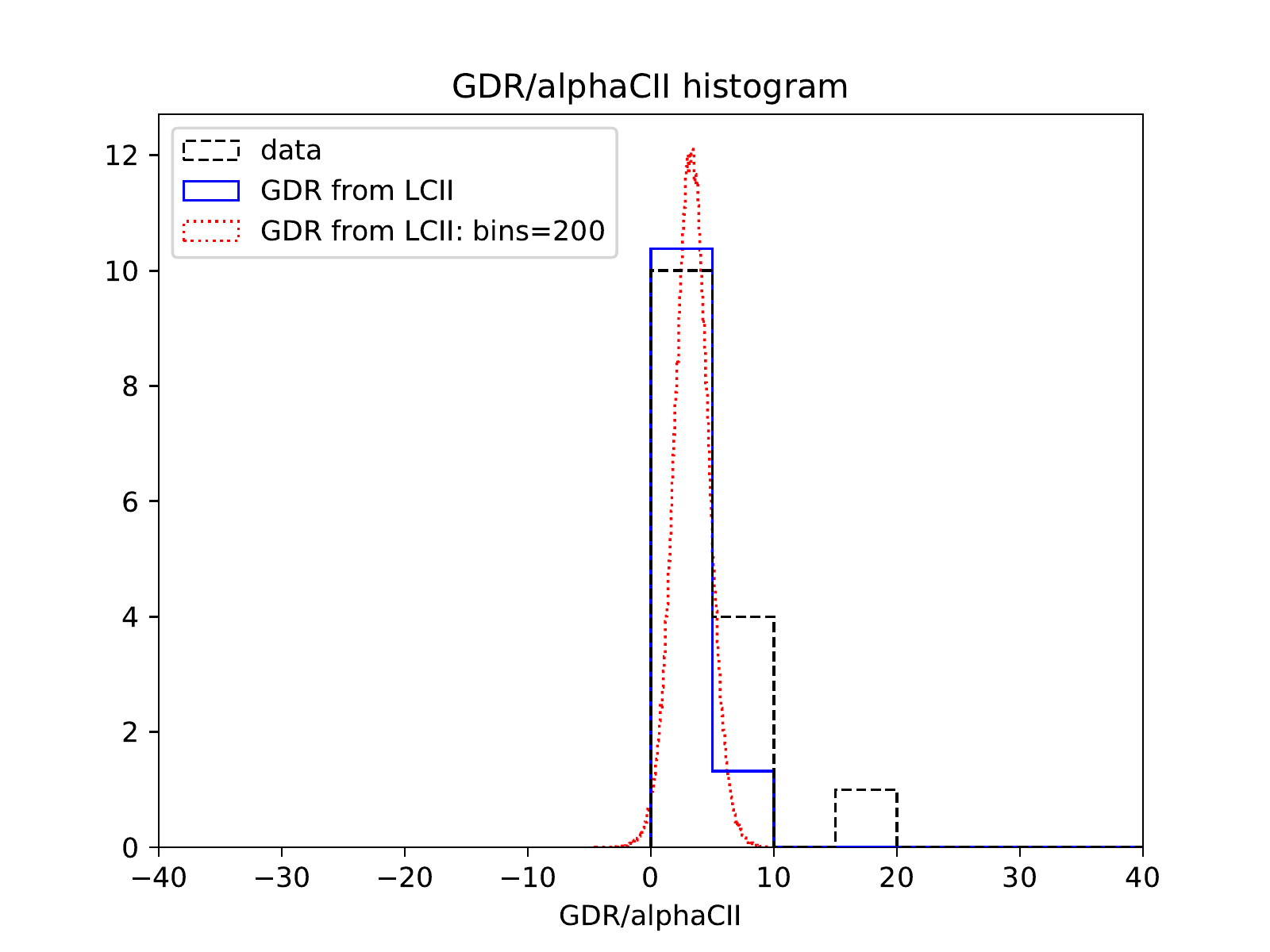} \\
\includegraphics[width=9cm]{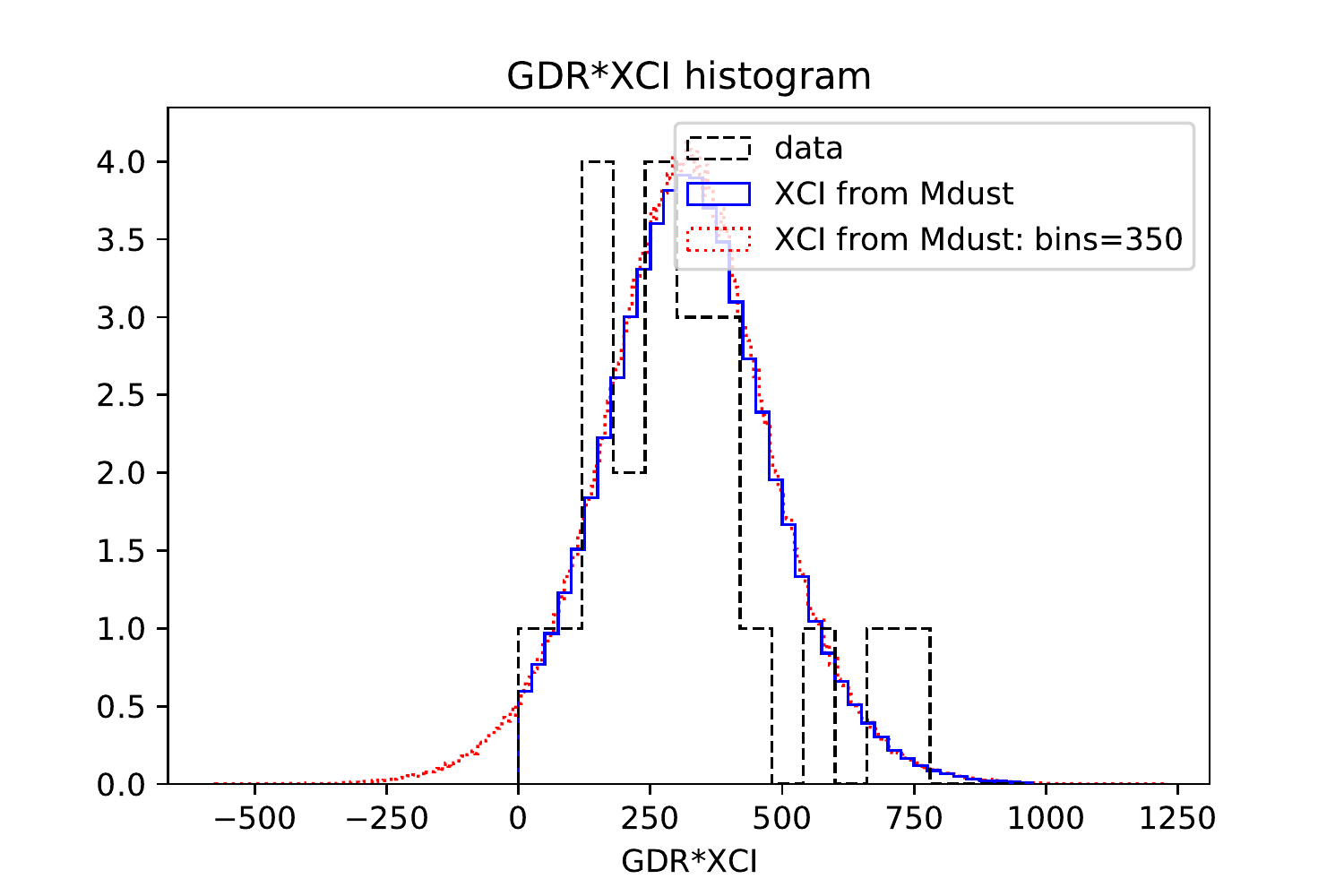} &
\includegraphics[width=9cm]{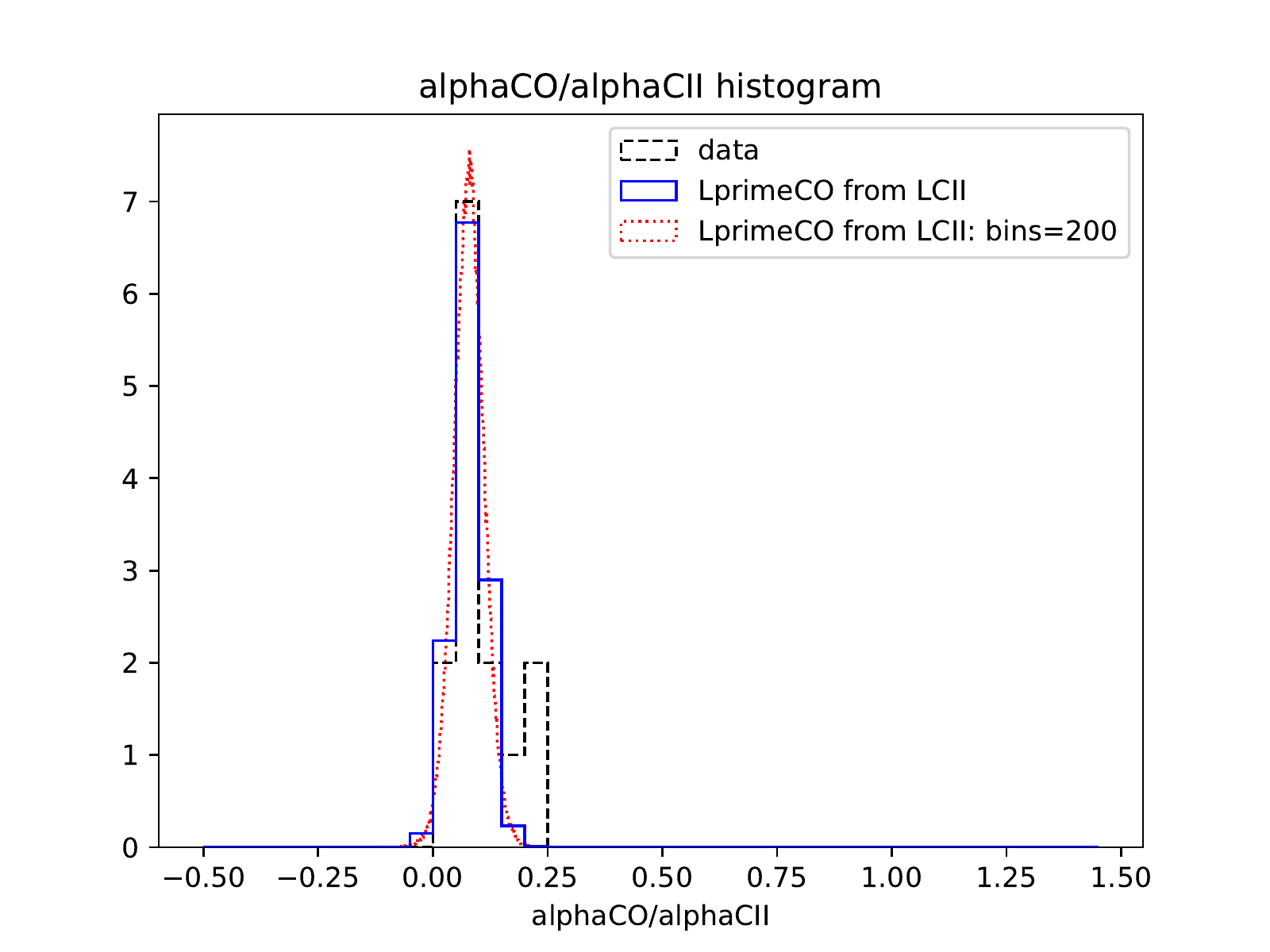} \\

\end{tabular}

\caption{\label{fig:origin of scatter} The cross-calibration with the tracers, $\rm X_{\rm CI}\times \alpha_{\rm CO}$ (left) and \XCI $\times$ \aCII (right), $\delta_{\rm GDR}/\alpha_{\rm CO}$ (left) and \gdr/\aCII (right), and the $\rm X_{\rm CI}\times \alpha_{\rm CO}$ (left) and \aCO/\aCII (right) are compared against the simulation described in Sect.\,\ref{Sec: scatter simulation} in the first, second and third row respectively. The resulting data, including the measurement uncertainty is represented as the blue-solid line (binned the same as the original data), and red-dotted line. }
\end{figure*}

In Fig.\,\ref{fig:scatter simulation}, we compare the observed ratios between tracers, along with simulated ratios assuming no scatter but including instrumental noise. The KS tests comparing our observed distribution and the simulated distribution (Table\,\ref{tab:KS test results}) return p-values higher than 0.05 for all the combination of tracers. Thus, we do not find any evidence of an intrinsic scatter, which is probably undetected due to our small sample size and the large measurement uncertainties.


\end{appendix}

\end{document}